\documentclass[11pt,twoside,final]{huthesis}

\usepackage{epsfig,bm,epsf,float}

\begin{document}


\hyphenation{me-ta-sta-ble me-ta-sta-bles}
\hyphenation{re-tro-mir-ror}

%

\newcommand{\twoStwoS}{\mbox{$2S$-$2S$}}
\newcommand{\percc}{\mbox{cm$^{-3}$}}
\newcommand{\upproton}{\mbox{$\uparrow\!\!\!\!\!\!\!\!\:\:-$}}
\newcommand{\downproton}{\mbox{$\downarrow\!\!\!\!\!\!\!\!\:\:-$}}

\newcommand{\vecmu}{{\mbox{\boldmath $\mu$}}}
\newcommand{\ethr}{{\epsilon_t}}
\newcommand{\kb}{{k_B}}
\newcommand{\veff}{{V_{eff}}}
\newcommand{\etal}{{\em et al.}}
\newcommand{\erf}{{\rm erf}}
\newcommand{\oneStwoS}{\mbox{$1S$-$2S$}}
\newcommand{\oneSoneS}{\mbox{$1S$-$1S$}}
\newcommand{\Hdown}{\mbox{$H\!\!\downarrow$}}
\newcommand{\Hup}{\mbox{$H\!\!\uparrow$}}
\newcommand{\Lalpha}{\mbox{Lyman-$\alpha$}}

\newcommand{\deriv}[2]{\frac{d#1}{d#2}}
\newcommand{\derivc}[3]{\left. \frac{d#1}{d#2}\right|_{#3}}
\newcommand{\pd}[2]{\frac{\partial #1}{\partial #2}}
\newcommand{\pdc}[3]{\left. \frac{\partial #1}{\partial #2}\right|_{#3}}

\newcommand{\bra}[1]{\left\langle #1\right|}
\newcommand{\ket}[1]{\left|#1\right\rangle}
\newcommand{\braket}[2]{\left\langle #1 \left|#2\right.\right\rangle}
\newcommand{\braOket}[3]{\left\langle #1\left|#2\right|#3\right\rangle}

\def\cal#1{\mathcal{#1}}

\def\avg#1{\left< #1 \right>}
\def\abs#1{\left| #1 \right|}
\def\recip#1{\frac{1}{#1}}
\def\vhat#1{\hat{{\bf #1}}}
\def\smallfrac#1#2{{\textstyle\frac{#1}{#2}}}
\def\smallrecip#1{\smallfrac{1}{#1}}

\def\be{\begin{equation}}
\def\ee{\end{equation}}
\def\bea{\begin{eqnarray}}
\def\eea{\end{eqnarray}}
\def\bean{\begin{mathletters}\begin{eqnarray}}
\def\eean{\end{eqnarray}\end{mathletters}}

\newcommand{\DEL}{\mbox{\boldmath $\nabla$}}

\newcommand{\sint}{\oint \! d\mbf{s} \,} 
\def\gint{\oint_\Gamma \!\! d\mbf{s} \,} 
\newcommand{\lint}{\oint \! ds \,}	
\def\infint{\int_{-\infty}^{\infty} \!\!}	

\def\ie{{\it i.e.\ }}
\def\eg{{\it e.g.\ }}
\newcommand{\ibid}{{\it ibid.\ }}

\def\gap{\hspace{0.2in}}

%

\newcounter{eqletter}
\def\mathletters{%
\setcounter{eqletter}{0}%
\addtocounter{equation}{1}
\edef\curreqno{\arabic{equation}}
\edef\@currentlabel{\theequation}
\def\theequation{%
\addtocounter{eqletter}{1}\thechapter.\curreqno\alph{eqletter}%
}%
}
\def\endmathletters{\setcounter{equation}{\curreqno}}

%
%


\hsp



\title{Studies with Ultracold Metastable Hydrogen}
\author{David Paul Landhuis}
\degreemonth{May} 
\degreeyear{2002}
\degree{Doctor of Philosophy}
\field{Physics}
\department{Physics}
\advisor{Daniel Kleppner (MIT)\\Gerald Gabrielse (Harvard)} 

\maketitle
\copyrightpage

\begin{abstract}

This thesis describes the first detailed studies of trapped metastable
($2S$-state) H.  Recent apparatus enhancements in the MIT Ultracold
Hydrogen group have enabled the production of clouds of at least
$5\times10^7$ magnetically trapped $2S$ atoms at densities exceeding
$4\times10^{10}$~\percc\ and temperatures below 100~$\mu$K.  At these
densities and temperatures, two-body inelastic collisions between $2S$
atoms are evident.  From decay measurements of the $2S$ clouds,
experimental values for the total two-body loss rate $K_2$ are
derived: $K_2=1.8^{+1.8}_{-0.7}\times10^{-9}$~cm$^3$/s at 87~$\mu$K, and
$K_2=1.0^{+0.9}_{-0.5}\times10^{-9}$~cm$^3$/s at 230~$\mu$K.  These values are
in range of recent theoretical calculations for the total $2S$-$2S$
inelastic rate constant.  Experimental upper limits for $K_{12}$, the
rate constant for loss due to inelastic $1S$-$2S$ collisions, are
also determined.  As part of the discussion and analysis, results from
numerical simulations to elucidate $2S$ spatial distributions,
evolution of the $2S$ cloud shape, and fluorescence behavior in the
magnetic trap are presented.  This work serves as a bridge to future
spectroscopy of trapped metastable H with the potential to test
quantum electrodynamics (QED) and improve fundamental constants.

\end{abstract}

\newpage
\addcontentsline{toc}{section}{Table of Contents}
\tableofcontents


\begin{acknowledgments}
This thesis is joyfully dedicated to my parents, Jesse and Kim
Landhuis.  From an early age, they instilled in me a curiosity about
the natural world and a yearning for truth.  They spent much time with
me as a child and opened every door they could for my intellectual
growth.  I am grateful for the many good things which I have
inherited from my parents, including the desire to pursue excellence
in whatever task is at hand, great or small.  My hope is that they can
be proud of the work represented in these pages.

I am deeply grateful to my research advisors, the MIT professors Dan
Kleppner and Tom Greytak.  Because they welcomed me, a Harvard
student, into the MIT Ultracold Hydrogen Group more than six years
ago, I have been able to experience the best of both the MIT and
Harvard physics worlds.  In their research group, I have come of age
as an experimental scientist.  I learned a tremendous amount in these
years, not only about atomic and low-temperature physics but also
about tenacity, perseverance, and thinking for oneself.  I very
much appreciated Dan's and Tom's generosity and hospitality as well,
including the regular invitations to their homes both in Boston and
among the lakes and mountains to the north.

The experimental results described here would not have been possible
without Stephen Moss and Lorenz Willmann, my two closest colleagues
over the last few years.  I have appreciated their friendship,
camaraderie, and tremendous work ethic.  Stephen was tireless in
ensuring that the cryogenic apparatus was working well, improving
our computer systems, and writing essential code to run the
experiments for this thesis.  His stamina, commitment, and sense of
humor through grueling weeks of debugging and taking data did much for
my morale.  At the same time, I was lucky to share an office with
Lorenz, an excellent experimentalist.  He was the source of many good
ideas, and I learned a lot from our spirited discussions at the
whiteboard.  The numerical simulation of \oneStwoS\ excitation
developed by Lorenz was crucial for the analysis of metastable H
decay.

It was also good fortune for me that Julia Steinberger, Kendra Vant,
and Lia Matos joined our research group.  In addition to the talents
they brought, I am grateful for the positive attitudes they maintained
while confronting numerous onerous tasks.  Each of them stayed late on
many nights to help take data and also performed many of the regular
chores necessary to keep the experiment running.  With these able
graduate students in charge, the prospects for groundbreaking new
measurements are bright.

I also thank Walter Joffrain, a visting graduate student in
our group during the metastable H studies.  He cheerfully helped
with data acquisition and sometimes dispelled the late-night monotony
by breaking spontaneously into song.

A major reason I joined the hydrogen experiment was the outstanding
students who preceded me and were to become my mentors.  These are
Dale Fried and Tom Killian, who are not only excellent scientists but
also highly personable.  It was a pleasure to learn from them.

Several theoretical atomic physicists were my educators in the
relevant theory for metastable H collisions.  In particular, I
thank Alex Dalgarno, Piotr Froelich, and Bob Forrey for extensive and
helpful conversations.  Several years ago, they and their
collaborators began calculating the collision cross-sections for
metastables.  Their work provides a basis for interpretation of
experimental results in this thesis.

A number of administrative staff at both MIT and Harvard helped smooth
my path through graduate school.  I only name two here, for whom I am
particularly grateful: Sheila Ferguson of the Harvard Physics
Department, and Carol Costa of the Center for Ultracold Atoms.

I am fortunate to have a cadre of solid friends outside the lab, both
in Boston and far away, who have sharpened me and shared in the trials
and joys of my Harvard years.  In this respect, I especially want to
thank Stefan Haney, David Nancekivell, Paul Ashby, Lou Soiles, and
Timothy Landhuis.

Graduate school has been a tremendous time of learning and of
developing friendships with remarkable people.  The best thing that
happened to me during this era, though, is that I married Esther, my
wonderful wife.  Her generous acts of love, sympathetic ear, and
optimistic outlook have carried me through these last intense years of
studenthood.  I hope I will be as supportive and encouraging for her
as she finishes her own thesis research.

Finally, I want to express my gratitude to the Creator, whom I believe
is also Jesus Christ: Thank you for the opportunity to live, work, and
learn alongside outstanding colleagues and among caring friends.  Your
physical universe is intricate and beautiful, and I am grateful to
have been able to study a small corner of it.
 
\end{acknowledgments}

\dedication

\begin{quote}
\hsp
\em
\raggedleft

To my mom and dad

\end{quote}

\newpage

\startarabicpagination


\chapter{Introduction}
\label{ch:intro}
The $2S$ state of hydrogen is metastable because it decays by
two-photon spontaneous emission.  It has a natural lifetime of 122~ms.
This is not the only sense in which this species is long-lived,
however.  It has enjoyed a long and fruitful life in atomic physics
research.  From the explanation of the Balmer series of spectral
lines, to the groundbreaking first calculation of a two-photon decay
rate, on through numerous rf and optical measurements of the Lamb
shift, metastable hydrogen has played a key role.  The studies
described in this thesis are part of a new chapter in this unfolding
history, one which will likely lead to further insight into
fundamental physics.

Traditionally, metastable H has been studied in discharge cells or in
atomic H beams which have been excited by a laser or by electron
bombardment.  In the MIT Ultracold Hydrogen Group, the application of
a high-power, narrow-linewidth UV laser system to magnetically trapped
$1S$-state hydrogen has produced clouds of metastables at temperatures
ranging from 300~mK down to 20~$\mu$K.  Recent improvements to the
apparatus at MIT have allowed generation of clouds of more than $10^7$
$2S$ atoms at densities greater than $10^{10}$~cm$^{-3}$ and lifetimes
of nearly 100~ms.  These trapped metastable clouds are promising new
samples for collisional physics and precision spectroscopy.

In the first section of this introduction, the impetus for studying
cold, trapped metastable H is futher discussed.  This is followed by a
brief overview of concepts relevant to magnetic trapping and
\oneStwoS\ excitation.

\section{Motivations}
\label{sec:motivations}
\subsection{Metastable H Collisions}
A cold gas of metastable H is an interesting laboratory for
collisional physics.  One reason for this is that a $2S$ atom carries
more than 10~eV of internal energy, enough to cause ionization in many
targets and to drive chemical reactions.  When metastables collide
with one another, the possible outcomes include Penning ionization,
the formation of molecular ions, excitation transfer to short-lived
$2P$ states, and transitions between hyperfine states.  If the
$2S$ atoms are excited from a background $1S$ gas, then several
\oneStwoS\ collision processes can occur as well.  Various aspects of
cold, elastic \oneStwoS\ interactions have already been investigated
by our group \cite{kil99,kfw98} and by several theory collaborations \cite{jdd96,osw99,pes01}.

Collisions between metastables have previously been studied at high
effective temperatures ($T > 48$~K) \cite{ucj92}, where the total
inelastic rate constant is several orders of magnitude larger than at
$T=0$ \cite{fjf01,lmm02}.  At high temperatures, the rate constant for
excitation transfer or ``collisional quenching'' has been predicted to
limit the lifetime of a metastable gas to about 40~$\mu$s at a density
of 10$^{10}$~\percc \cite{fcd00}.  With cold, magnetically trapped
clouds of $2S$ atoms, the lifetime at this density is found to be tens
of milliseconds.  By observing the decay behavior of a metastable
cloud, we can learn about \twoStwoS\ and \oneStwoS\ inelastic
collisions.  This thesis presents the first measurements of the total
\twoStwoS\ inelastic collision rate at low temperatures.  These
measurements extend into the theoretically predicted ``ultracold''
regime, where both elastic and inelastic collisions are parametrized
by a single constant, the complex scattering length \cite{fjf01,bfd98}.

\subsection{Spectroscopy of Metastable H}
Although the collisional physics of metastables is worth studying for
its own sake, a deeper understanding of these interactions also helps
set the stage for ultraprecise spectroscopy of metastable H.  The
absolute frequencies of transitions from the $2S$ to higher-lying
states of H can be combined with the \oneStwoS\ interval, currently
the most accurately known of all optical transition frequencies, to
simultaneously determine the $1S$ Lamb shift and the Rydberg constant
\cite{uhg97,sjd99}.  The former quantity is predominantly comprised of
quantum electrodynamics (QED) corrections to the Dirac solution for
the hydrogen atom, and it has been called the most sensitive test of
QED in an atom.  The Rydberg constant, $R_\infty=m_e c \alpha^2/2h$,
is already the most accurately known of all fundamental constants.  It
provides a link between the electron mass, the speed of light, the
fine structure constant, and Planck's constant.  Thus, spectroscopy of
metastable H provides a way to not only test the state-of-the-art of
QED theory but also improve the accuracy of fundamental constants
\cite{sjd99,nhr00,pac01}.

Several $2S$-$nS/D$ frequency intervals, where $n>3$, have been
measured with impressive accuracy by F.~Biraben and collaborators
using Doppler-free two-photon spectroscopy
\cite{whs95,bbn96,bnj97,sjd99}.  Currently, the best experimental
values for both the Lamb shift and Rydberg constant are limited by
uncertainty in $2S$-$nS/D$ frequencies
\cite{sjd99}.  The Biraben experiments were conducted with atomic
beams in which the interaction time of metastables with a laser was
limited to a few hundred microseconds.  To achieve an efficient
excitation rate required laser intensities of several kilowatts per
square centimeter.  The resulting AC Stark shifts were the primary
source of uncertainty in the measurement.  With a cold, trapped sample
of metastable H, the interaction times between the excitation laser
and the atoms can be much larger, and laser intensities can be much
lower.  Another source of systematic error in the beam experiments was
the quadratic DC Stark shift of higher-lying $D$ states.  Since
excitation rates can be much higher in a trapped sample, one can
afford to concentrate on weaker $2S$-$nS$ transitions, which are less
sensitive to stray electric fields.

Once an excitation laser is tuned to a two-photon $2S$-$nS$ resonance,
its absolute frequency can be calibrated directly against the primary
cesium standard or against the \oneStwoS\ frequency, which is now a
{\em de facto} optical frequency standard \cite{nhr00}.  The frequency
metrology can be accomplished with a mode-locked femtosecond laser
which produces a broad, coherent frequency comb
\cite{urh99,djy00,udv01}.  Frequency measurements made in this way
using ultracold hydrogen will potentially allow an order of magnitude
improvement in both the Lamb shift and Rydberg constant \cite{wk00}.

Since many transitions originating in the $2S$ state are accessible
with diodes and other readily available lasers, trapped metastable H
as a spectroscopic sample opens the door to other interesting studies
as well.  These include photoassociation experiments and measurements
of cold collision parameters involving higher-lying states of H.

\subsection{Quantum Atom Optics}
Due to its large internal energy, cold metastable H is interesting for
``quantum atom optics'' experiments involving single-atom detection.
The possibilities include measurements to probe atom-atom correlations
\cite{ys96}, atom interferometry, and atom holography \cite{myk96}.  A
metastable H atom is readily detected in an ionizing collision on an
electron-multiplying detector.  Alternatively, one can detect the
Lyman-$\alpha$ photon emitted after quenching at a surface or by a
localized electric field.  To detect individual metastables generated
from ultracold hydrogen, they must be coupled out of the magnetic
trap.  This can be accomplished by inducing an appropriate rf or
optical transition to an untrapped atomic state or by simply turning
off the magnetic trap.  Several single-atom detection experiments
involving cold, metastable noble gases have already been demonstrated
\cite{ys96,myk96} or are in progress \cite{rsb01}.  It may also be
possible to generate a highly directional metastable H beam from an
ultracold sample using photon recoil in a Raman transition
\cite{kdh99} or in the initial \oneStwoS\ excitation (see 
Sec.~\ref{sec:twophotonspec}).  A bright beam of cold metastables may
be ideal for atom lithography \cite{npm96,jtd98}.

\begin{figure}
\centerline{\epsfig{file=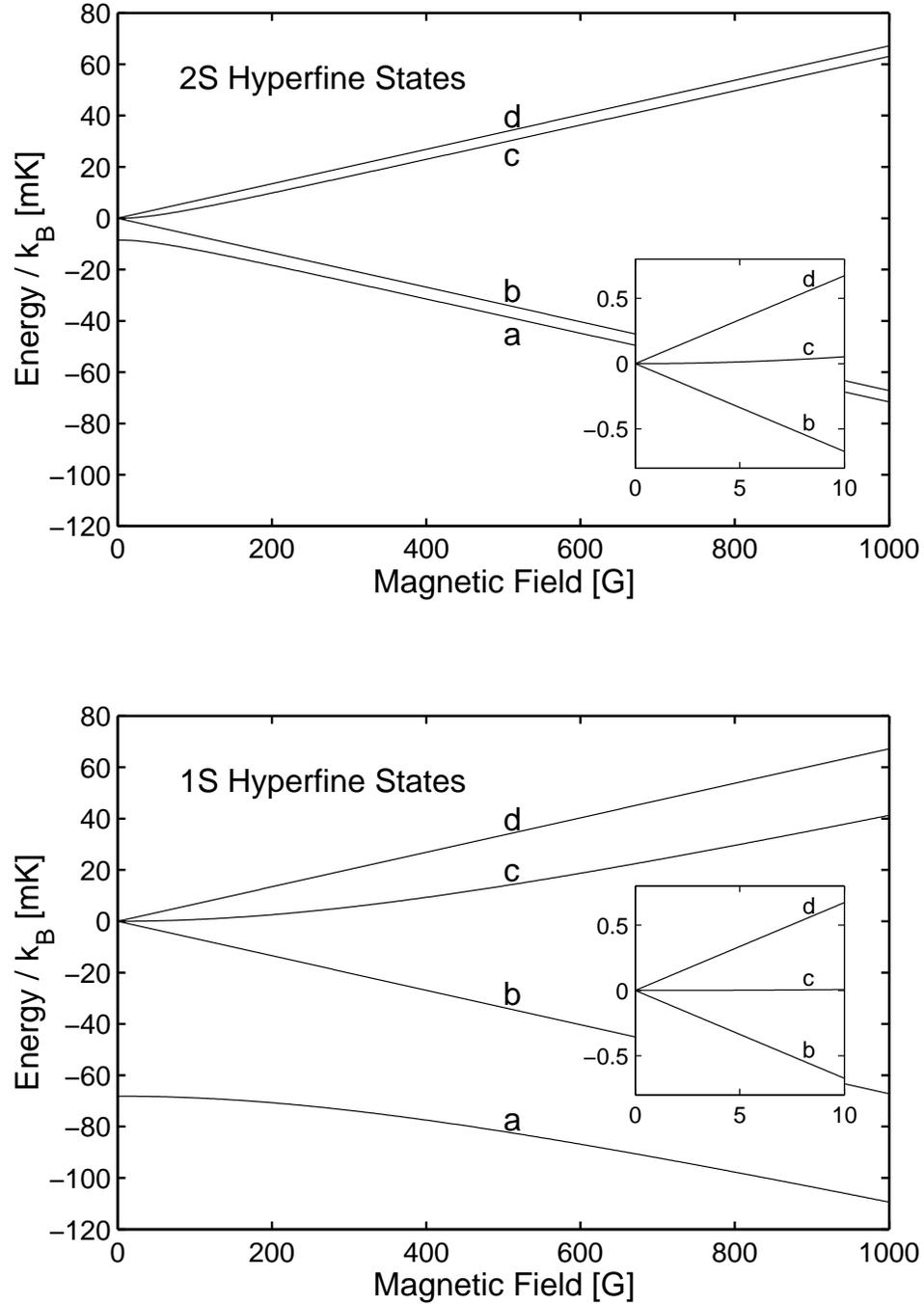,width=5in}}
\caption{Comparison of hyperfine-Zeeman diagrams for the $1S$ and $2S$
manifolds of H.  For each case, energy is referenced to the $F=1$
vertex, where the $b$, $c$, and $d$ states are degenerate.  The larger
plots show all four hyperfine states over the magnetic field range
relevant during forced evaporation of magnetically trapped H.  In the
inset plots, the $b$, $c$, and $d$ states are shown for the 0-10~G
range of magnetic fields typically experienced by samples at 100~$\mu$K. }
\label{fig:hyperfine}
\end{figure}
  
\begin{figure}
\centerline{\epsfig{file=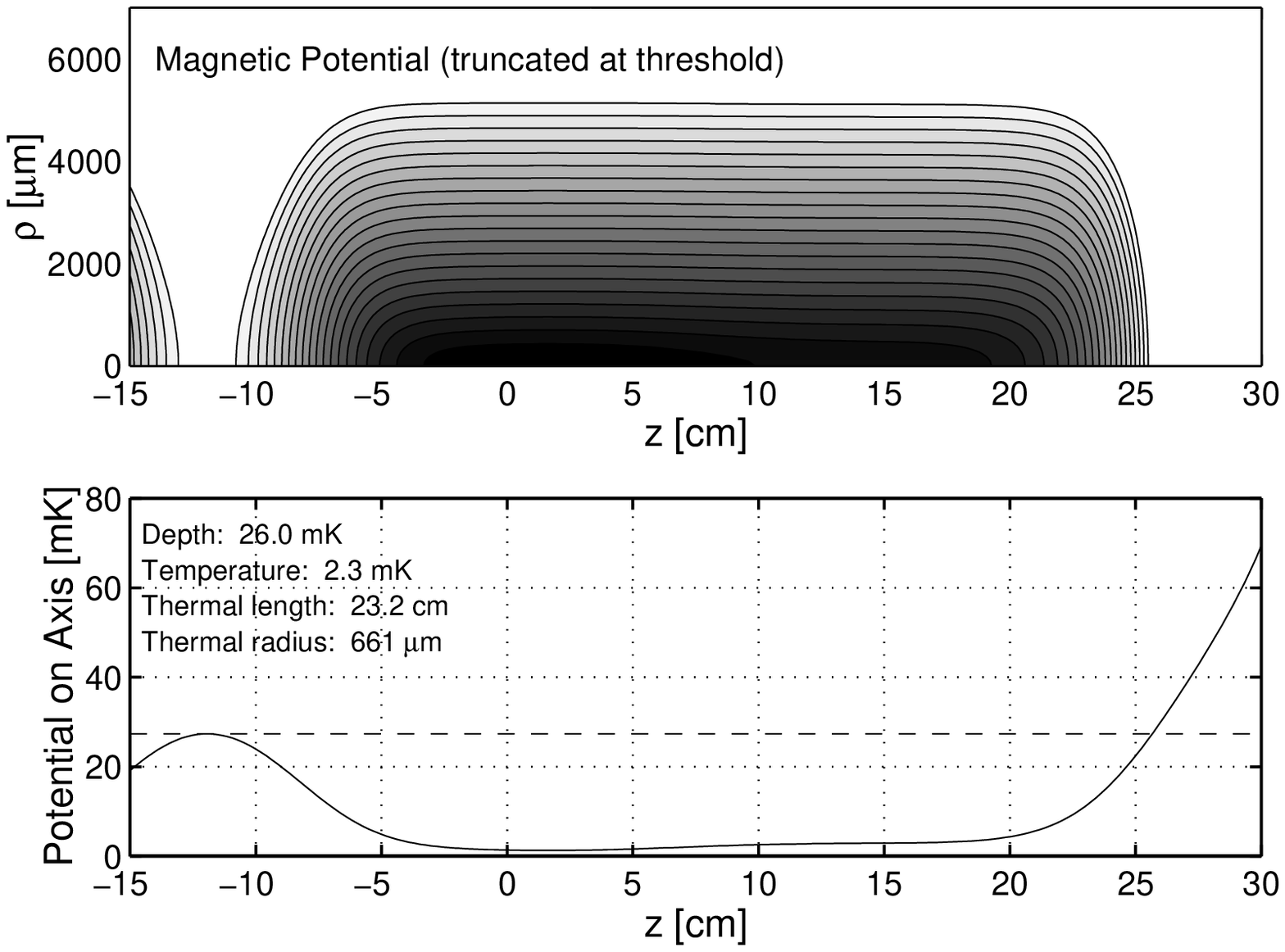,width=5in}}
\caption{Magnetic potential as a function of cylindrical coordinates
$\rho$ and $z$ for a standard trap configuration (Trap Z of
Ch.~\ref{ch:decay}).  The potential has been calculated numerically
from the known magnet geometry and currents.  In the upper half of the
figure, the region of magnetic potential below the trap threshold is
depicted by a shaded contour plot; the shading is darkest at the trap
minimum.  The lower half of the figure shows the magnetic profile
along the trap axis.  The magnetic saddlepoint, which defines the trap
threshold (dashed line), is located near $z=-12$~cm.  Dimensions are
given for the $1S$ cloud at the temperature experimentally achieved in
this trap.  Similar figures will be used in this thesis to describe
the shape and characteristics of other samples.  The coordinate system
origin follows a convention of the MIT Ultracold Hydrogen Group.}
\label{fig:trapstatsexample}
\end{figure}

\section{Hyperfine States and Magnetic Trapping}
\label{sec:hyperfine}
The hyperfine state energies for both the $1S$ and $2S$ levels of H
are plotted as a function of magnetic field in Fig.~\ref{fig:hyperfine}.
The hyperfine interaction energy, which scales as $1/n^3$ for $S$
states, is only $1/8$ as large for metastable hydrogen as for the
ground state.  Except for this hyperfine energy scaling, the $2S$
manifold is basically identical to the $1S$ manifold.

In both levels, the hyperfine states are conventionally labeled
$a$, $b$, $c$, and $d$, in order of increasing energy.  At low
magnetic field, these states correspond to the total angular momentum
states $(F=0,m_F=0)$, $(F=1,m_F=-1)$, $(F=1,m_F=0)$, and
$(F=1,m_F=1)$, respectively.  In an inhomogeneous magnetic field, the
``low-field seeking'' $c$ and $d$ states can be trapped around a
minimum of the magnetic field \cite{hkd87,doy91}.  Since $c$-state
atoms rapidly undergo spin-exchange collisions and transition to other
hyperfine states, the sample which collects in a magnetic trap
ultimately contains only $d$-state atoms.  The trap loading process
for $1S$ atoms will be discussed extensively in Sec.~\ref{sec:loading}.

Trapped metastable hydrogen is produced by two-photon excitation of
trapped $(1S,d)$ atoms \cite{cfk96}.  The two-photon selection rules are
$\Delta F=0$ and $\Delta m_F=0$; thus, only trapped $(2S,d)$ atoms are excited.
Furthermore, since the magnetic potential energy of both $(1S,d)$ and
$(2S,d)$ atoms is to an excellent approximation\footnote{The magnetic
moments of $2S$ and $1S$ electrons differ by less than 5 parts in
$10^4$ due to a relativistic correction \cite{bs77p214}.  Also, the
nuclear Zeeman contribution is more than three orders of magnitude
smaller than the electron contribution responsible for $\mu_B$.}
\be
U = \mu_B B,
\ee
where $\mu_B$ is the Bohr magneton and $B$ the local magnetic field
strength, the trapping potentials experienced by both species is the
same.

The trap in our apparatus is of the Ioffe-Pritchard variety \cite{pri83}.
To a good approximation, the magnetic field has cylindrical symmetry
about an axis known as the ``trap axis.''  With an aspect ratio typically
between 150:1 and 400:1, the axial dimension of the trap is much
longer than the radial dimension.  A typical field geometry is
depicted in Fig.~\ref{fig:trapstatsexample}.  The depth of the trap is
determined by a saddlepoint in the magnetic field located on the axis
at one end of the trap.  Our methods for evaporatively cooling the
atoms are described in Ch.~\ref{ch:apparatus}.
\begin{figure}
\centerline{\epsfig{file=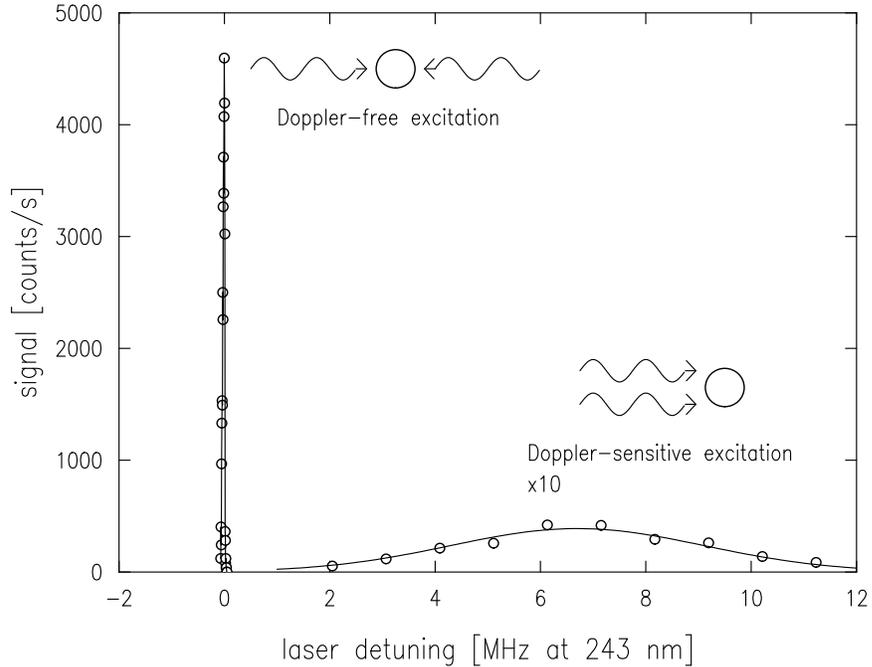,width=4.5in}}
\caption{Composite \oneStwoS\ spectrum of a thermal trapped hydrogen
sample, showing both Doppler-free and Doppler-sensitive contributions.
In the Doppler-free excitation mode, a hydrogen atom is promoted to
the $2S$ state by absorbing counter-propagating 243~nm photons.
Since there is no first-order Doppler-broadening, the corresponding
spectral feature is narrow and intense.  In the Doppler-sensitive
excitation mode, where two co-propagating photons are absorbed, the
spectrum is both recoil-shifted and Doppler-broadened.}
\label{fig:panoramicspectrum}
\end{figure}

\boldmath
\section{Two-Photon \oneStwoS\ Excitation}
\unboldmath

\label{sec:twophotonspec}

Excitation of the \oneStwoS\ transition is accomplished using two
243~nm photons from a UV laser.  The hydrogen \oneStwoS\ spectrum has
both Doppler-free and Doppler-sensitive resonances, separated by a
recoil shift (see Fig.~\ref{fig:panoramicspectrum}).  Although the
natural linewidth of the \oneStwoS\ transition is only 0.65~Hz at
243~nm, the experimental Doppler-free linewidth is at least a few
kilohertz due to a combination of transit-time broadening
\cite{cfk96}, finite laser linewidth, and the cold-collision frequency
shift \cite{kfw98}.  By comparison, the Doppler-sensitive line has a
width of $\sim4$~MHz at the coldest experimental temperatures, and the
width grows with increasing temperature.  To excite large numbers of
metastables, the laser is tuned to the narrow, intense Doppler-free
spectral feature.

\begin{figure}
\centerline{\epsfig{file=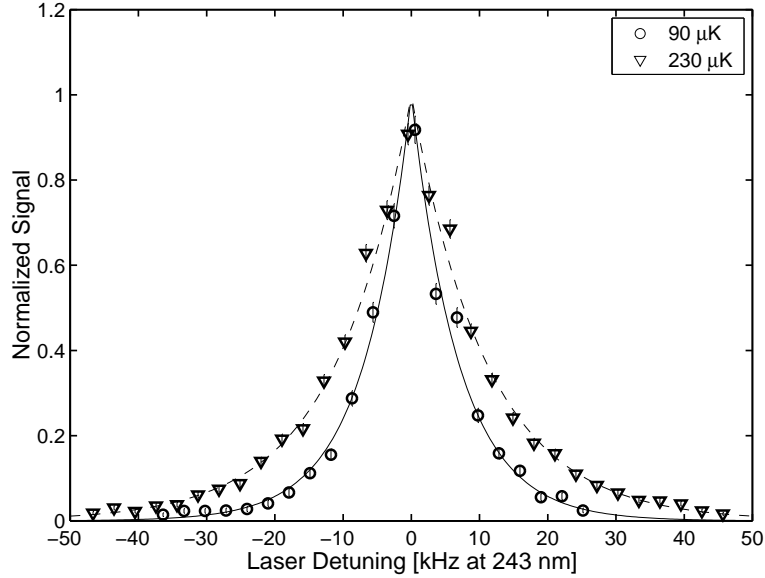,width=4in}}
\caption{Examples of Doppler-free \oneStwoS\ lineshapes for
low-density $(\sim10^{13}$cm$^{-3})$ $1S$ samples at two different
temperatures.  At these densities, the width of the spectrum is
dominated by transit-time broadening, and the lineshape is
double-exponential, characteristic of Doppler-free two-photon
excitation by a Gaussian laser beam.}
\label{fig:DFexamples}
\end{figure}

The hallmark of the Doppler-free resonance is the cusp-like lineshape
which arises in a thermal gas due to transit-time broadening.  For the
case of a perfectly collimated, Gaussian laser beam and a homogeneous
gas density, it can be shown \cite{bbc79} that the Doppler-free line
of the gas has a spectrum proportional to $e^{-|\Delta
\nu_L|/\gamma_o(T)}$ where $\Delta \nu_L$ is the laser detuning from
resonance and $\gamma_o(T) \propto \sqrt{T}$ is the $1/e$ half-width
of the line.  This lineshape is called ``double-exponential'' because
it consists of two exponential functions meeting at a cusp.  For
temperatures at which the above criteria for gas density and laser
collimation are approximately satisfied, the relative width of the
Doppler line is a useful measure of relative sample temperature
(Fig.~\ref{fig:DFexamples})
\cite{cfk96,kil99}.  Low
sample densities must be employed to resolve these double-exponential
lineshapes.  At high $1S$ densities ($> 10^{13}$ \percc), the
\oneStwoS\ cold-collision shift flattens the cusp and significantly
broadens the line \cite{kfw98}.

\begin{figure}
\centerline{\epsfig{file=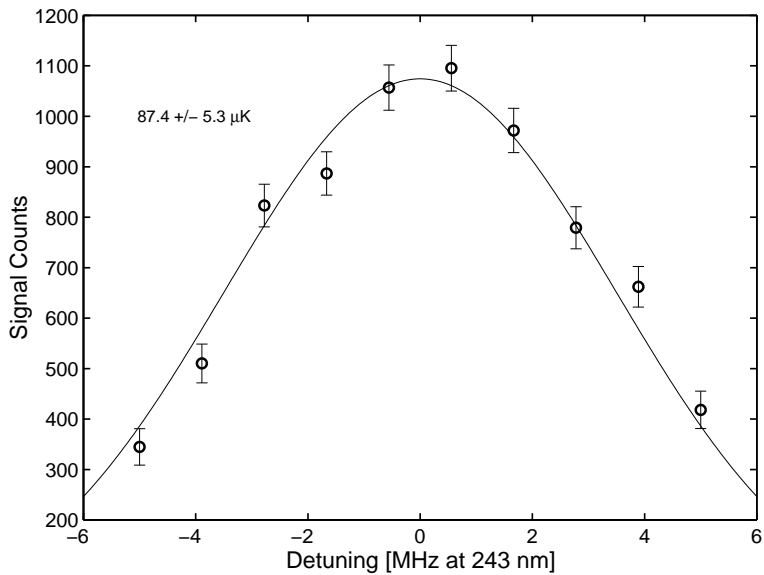,width=4in}}
\caption{Temperature determination from a Gaussian fit to
the Doppler-sensitive component of the H \oneStwoS\ spectrum.  In this
example, spectroscopic data from six consecutive trap cycles are
summed together.}
\label{fig:DStemperature}
\end{figure}

The Doppler-sensitive spectrum (Fig. \ref{fig:DStemperature}) is
proportional to the momentum distribution of the sample along the
direction of the excitation laser
\cite{fkw98}.  As long as the sample is not close to quantum
degeneracy, the distribution is Maxwell-Boltzmann, and the lineshape
is Gaussian; this is always the case for samples in this thesis.  Like
the Doppler-free double-exponential, the Doppler-sensitive width
scales as $\sqrt{T}$.  Unlike the Doppler-free case, the width is
simply related to the absolute temperature and is independent of the
laser and trap geometry.  A Gaussian fit to the Doppler-sensitive
spectrum provides the most direct way of calibrating the temperature
of a trapped hydrogen sample.  Unfortunately, at temperatures above
$\sim100$~$\mu$K the line is so broad and weak that, in our current
apparatus, many trap cycles are required to accumulate enough
statistics for a temperature measurement.

For more thorough and formal descriptions of \oneStwoS\
excitation in hydrogen, see Refs.~\cite{bea86,san93,ck99,kil99}.

\section{Overview of the Thesis}
\label{sec:overview}
The next chapter will begin by describing the experimental sequence
applied to produce a cloud of trapped $2S$ atoms.  The later sections
of Ch.~\ref{ch:apparatus} focus on apparatus and techniques which have
not yet been detailed in other theses and publications; these include
the method for metastable decay measurements.  Chapter~\ref{ch:theory}
delves into several aspects of the physics of metastable H in a
magnetic trap, mostly from a theoretical standpoint.
Chapter~\ref{ch:decay} describes the magnetic trap configurations used
in decay measurements and introduces data suggesting the presence of
two-body loss in our metastable clouds.  The analysis of decay data to
calculate a \twoStwoS\ loss rate constant and place limits on
inelastic \oneStwoS\ collision rates is presented in
Ch.~\ref{ch:results}.  In the concluding chapter, some suggestions are
made for further experiments using the current apparatus and also for
a new apparatus optimized for spectroscopy of metastable H.

Throughout this thesis, the term ``$2S$'' is used interchangeably with the
term ``metastable''; the same applies for ``$1S$'' and ``ground state.'' 


%

\chapter{Apparatus and Experimental Techniques}
\label{ch:apparatus}

The apparatus used for experiments described in this thesis was
developed with two principal goals in mind: (1) observation of
Bose-Einstein condensation in atomic hydrogen and (2) high resolution
spectroscopy of ultracold hydrogen on the \oneStwoS\ transition.
These goals required the parallel development of a sophisticated
cryogenic apparatus and a highly frequency-stable UV laser system.
Many of the important developments have been described in other PhD
theses from the MIT Ultracold Hydrogen Group
\cite{doy91,san93,ces95,fri99,kil99}.  This chapter will recapitulate
the methods used to generate and probe ultracold hydrogen with
emphasis on recent refinements.  The first section provides an
overview of the trap cycle used for measurements in this thesis, while
the following sections discuss details of our techniques
which, for the most part, have not been recorded elsewhere.

\section{A Typical Trap Cycle}
\begin{figure}
\centerline{\epsfig{file=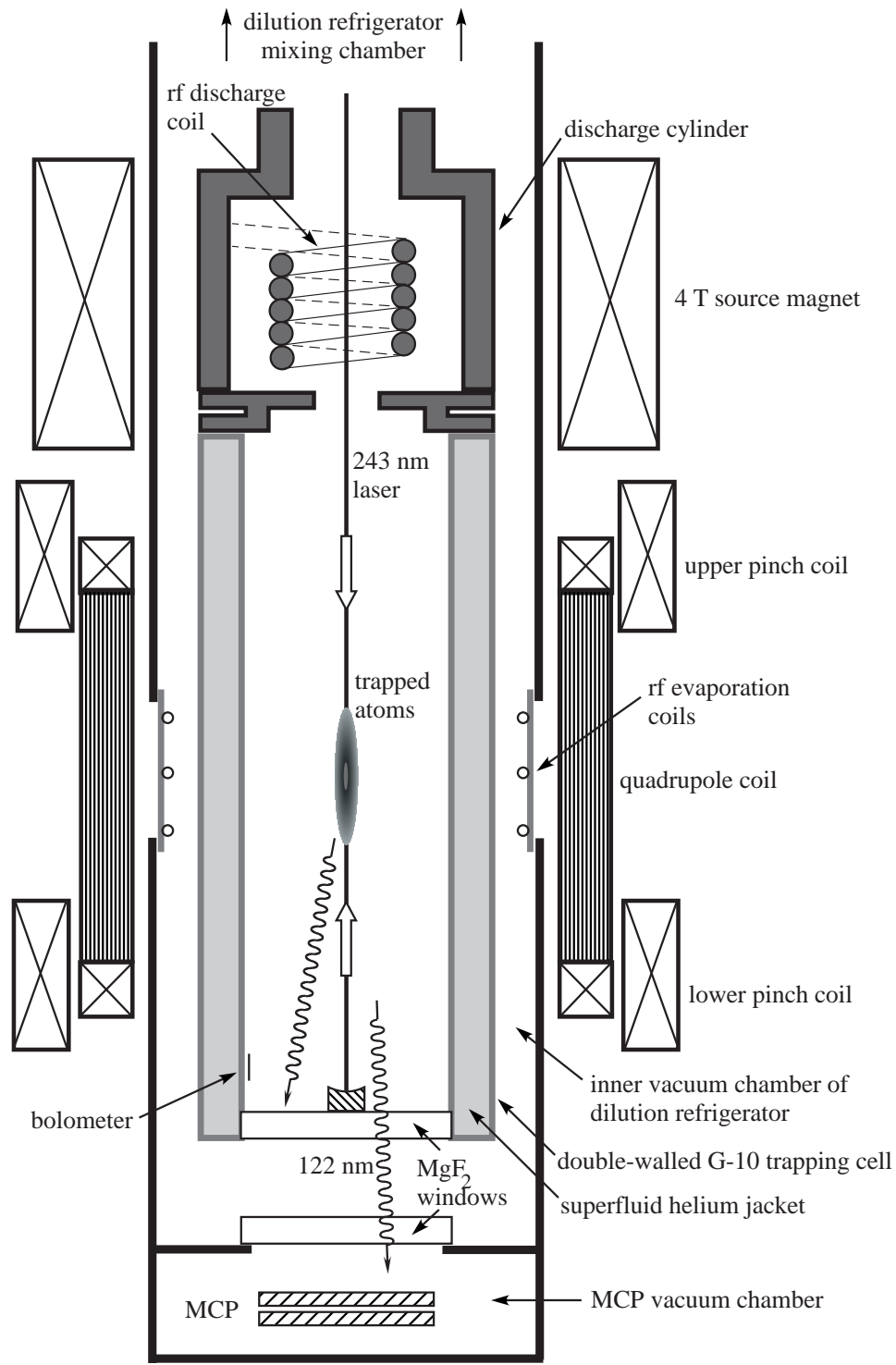,width=5in}}
\caption{Cartoon of cryogenic apparatus, not to
scale: the ratio of the length of the trapping cell ($\sim65$~cm) to its
inner diameter (4~cm) is much larger than depicted here.  During the
loading phase, atoms accumulate in a Ioffe-Pritchard trap with radial
confinement provided by an elongated set of quadrupole coils and axial
confinement by two ``pinch'' coils.  As the sample is evaporatively
cooled, the fields of other smaller, fine-control magnets come into
play.  All the magnets are immersed in liquid helium and are superconducting.}
\label{fig:apparatus}
\end{figure}

Samples of atomic hydrogen are generated, trapped, cooled, and probed
in a cycle lasting typically 5--15 minutes depending on the desired
final sample temperature and type of measurement.  At the beginning of
each cycle, pulses of rf power are supplied to a copper coil in a
cylindrical resonator \cite{kil99} which is thermally anchored
(Fig.~\ref{fig:apparatus}) to the mixing chamber of a dilution
refrigerator.  The resulting rf discharge vaporizes and dissociates
frozen molecular hydrogen from surfaces in the resonator.  Atoms in
all four $1S$ hyperfine states are injected into a trapping cell.  The
two high field seeking states ($a$ and $b$) are drawn back to the 4 T
magnetic field in the discharge region, while the low field seekers
($c$ and $d$) are attracted to the 0.6~K deep Ioffe-Pritchard magnetic
trap \cite{pri83} in the trapping cell (see
Figs. \ref{fig:loadingtrap} and \ref{fig:bigtrapandCAD}).

In order for trapping to be successful, a sufficiently thick film of
superfluid He must coat the walls of the cell.  The low field seekers
are cooled initially by interaction with the cell walls.  While the
discharge operates, the cell is held at a temperature of 320~mK, which
is warm enough for the residence time of H atoms on the He film
to be shorter than the time it takes for atoms to spin-flip on
the walls, and cold enough for a sizeable fraction of H atoms
to settle into the trap after exchanging energy in elastic collisions
\cite{doy91}.  The $c$-state atoms disappear in a few seconds due to
spin-exchange collisions, and a pure $d$-state ($F=1$, $m_F=1$) gas
remains in the trap.  After the loading period is finished, the cell
is rapidly cooled to below 100~mK by the dilution refrigerator.  At
this temperature, atoms with enough energy to reach the walls are
permanently lost from the trapped sample.  Evaporative cooling begins,
and the sample becomes thermally disconnected from the cell.  Once
thermal disconnect is complete, the atom cloud equilibrates at a
temperature of typically 40~mK.

The threshold energy of the magnetic trap is determined by a 
saddlepoint of the magnetic field at one end of the trap.  Atoms with
energy above threshold can escape over the saddlepoint, leading to
evaporative cooling of the sample.  By progressively lowering the
saddlepoint field, evaporation is forced.  In this way, the sample can
be cooled to temperatures as low as $\sim200$~$\mu$K in 4-5~minutes.

In order to reach sample temperatures down to $\sim20$~$\mu$K, rf
evaporation is employed.  For a given rf frequency $\nu_{\rm rf}$,
there is a three-dimensional ``surface of death'' consisting of all
points having a magnetic field such that the $d$-$c$ and $c$-$b$
hyperfine transitions of the atoms are resonant.  Atoms with energy
greater than $h\nu_{\rm rf}$ can cross the surface of death several
times per millisecond and quickly transition to an untrapped state.
Thus, $\nu_{\rm rf}$ sets an efficient threshold for the trap.  The
coldest temperatures in the MIT hydrogen experiment are achieved by
ramping $\nu_{\rm rf}$ from 35~MHz (significantly above the magnetic
trap threshold) to the desired final threshold value, usually between 5
and 20~MHz (240 and 960~$\mu$K).

The dramatic cooling of the sample by several orders of magnitude in
temperature is accomplished at the expense of atom number.  The
magnetic trap, initially loaded with $\sim10^{14}$ atoms, contains
only $\sim10^{11}$ atoms at the 200~$\mu$K limit of magnetic saddlepoint
evaporation and fewer than $10^{10}$ atoms at the lowest temperatures
accessible by rf evaporation.  

Atoms are lost during the trap cycle not only to evaporation but also
to two-body ``dipolar'' decay collisions \cite{skv88}.  These are
$d+d$ collisions in which the magnetic dipole interaction causes at
least one of the two atoms to change spin state.  They occur
preferentially in the highest density region of the sample, where the
average energy of an atom is less than $kT$.  Thus dipolar decay leads
to heating of the sample.  To maintain a favorable ratio of elastic to
inelastic collisions, the radially confining fields are reduced
simultaneously with the saddlepoint field to prevent the atom density
from growing significantly as the temperature drops.  During rf
evaporation, however, the peak density grows rapidly, and dipolar
losses set a limit on the temperature which can be reached.

For experiments with metastable hydrogen, the last phase of the trap
cycle involves repeated two-photon excitation of the ground state
sample with a 243~nm UV laser nearly resonant with the Doppler-free
\oneStwoS\ transition.  A typical excitation pulse lasts
$1\sim3$~ms.  The resulting $2S$ atoms are quenched by an
electric field pulse after a wait time of up to 100~ms; these quenched
atoms are mostly lost from the trap \cite{san93}.  However, the quenching
process also results in Lyman-$\alpha$ fluorescence photons, some of which
are detected on a microchannel plate (MCP) detector.  The number of
signal counts recorded is proportional to the number of metastables
present at the time of quenching.  Typically, $\sim10^8$ or fewer
$2S$ atoms are excited per laser pulse.  Thus, metastable clouds can be
generated hundreds of times with different wait times and laser
frequencies before depleting the ground state sample.

For measurements on the ground state atoms, two methods of probing are
available.  First, there is conventional \oneStwoS\ excitation
spectroscopy \cite{cfk96,kil99}, already introduced in
Ch.~\ref{ch:intro}.  In this mode of taking data, the laser is
typically chopped at 100~Hz with a 10\% duty cycle, and a quenching
electric field pulse is applied at a fixed time of $\sim6$~ms after
excitation.  The laser frequency is stepped back and forth across a
specified range, each sweep of the range lasting less than 1~s.  This
provides a series of snapshots in time of some portion of the
\oneStwoS\ spectrum.  The spectroscopic data provides crucial information
about the sample (see Ch.~\ref{ch:intro}).  At higher sample
densities, the broadening of the Doppler-free line due to the
cold-collision shift provides information about the density
distribution of the sample.  At low densities, the Doppler-free width
is proportional to the square root of the temperature.  At low
temperatures, the width of the Doppler-sensitive line is an excellent
measure of absolute temperature.

The second and older method of probing the ground state sample
involves ``dumping'' the atoms from the trap and detecting the
recombination heat on a bolometer \cite{doy91}.  This method involves
rapidly lowering the saddlepoint magnetic barrier, allowing the atoms
to escape to a zero field region and/or forming a zero field region in
the trap.  The atoms quickly spin-flip and recombine.  The integral of
the resulting bolometer signal is proportional to the number of atoms
in the trap.  An important application of this technique is the
determination of initial sample density from the decay in $1S$ number
due to dipolar loss (Sec.~\ref{subsec:densitymeasurements}).

Throughout the trap cycle a dedicated computer controls the
necessary power supplies, frequency sources, relays, heaters, and data
acquisition electronics.

\section{Enhancements to the Cryogenic Apparatus}
\label{sec:enhancements}
In 1998, our group successfully demonstrated rf evaporation in a
cryogenic trapping apparatus.  The enabling technology was a new
nonmetallic trapping cell consisting of two concentric G-10 tubes
\cite{fri99}.  By eliminating metal parts from the cell, sufficient rf
power for evaporation could be delivered to coils wrapped directly on
the inner G-10 tube without placing an excessive heat load on the
dilution refrigerator.  In the nonmetallic design, heat transport to
the mixing chamber was provided by filling the volume between the tubes
with superfluid helium.

The cryogenic trapping cell used for experiments described in this
thesis represents a second-generation nonmetallic design.  A major
problem of the first-generation G-10 cell was the presence of
relatively large stray electric fields ($\sim500$~mV/cm) which limited
metastable lifetimes to between a few hundred microseconds and a few
milliseconds.  Worse yet, the stray fields sometimes changed
unpredictably on a time scale of minutes.  Many spectroscopy runs were
rendered almost useless by these fields.  Furthermore, since short
$2S$ lifetimes did not afford the opportunity to wait for
laser-induced background fluorescence to die away after excitation,
the stray fields limited the signal-to-noise ratio in \oneStwoS\
spectroscopy.  The problem of stray fields was solved in the
second-generation G-10 cell by adding a thin copper film to the inside
cell wall.  This film was thin enough to avoid significant absorption
of rf power, yet thick enough to suppress stray fields to the 40~mV/cm
level.  With the copper film, $2S$ lifetimes in excess of 80~ms have
been observed, and the stability of stray fields is much improved.
Since the film is divided into several externally controllable
electrodes, partial compensation of the residual stray fields is
possible (Sec.~\ref{subsec:strayfieldcomp}).

Another change in the second-generation design was the relocation of
rf evaporation coils to the inner vacuum chamber (IVC) ``tailpiece,''
which separates the surrounding magnets and liquid helium bath at 4~K
from the vacuum space containing the trapping cell.  In the
first-generation G-10 cell, the heat load on the dilution refrigerator due to
the evaporation coil leads placed a limit on the allowable rf drive
power.  Furthermore, intolerable heating of the cell occurred at
absorptive resonances above 25~MHz.  By thermally anchoring the coils
to the 4~K bath instead of the dilution refrigerator, heating of the
rf leads was no longer a limitation.  The absorptive resonances were
also shifted such that rf evaporation could occur in a continuous
downward sweep from 35~MHz.  This permits more efficient rf
evaporation and may also enable better shielding of the sample from a
high-energy atom ``Oort cloud'' which forms during magnetic
saddlepoint evaporation \cite{cew99,die01}.\footnote{In colder samples we
have observed evidence for large numbers of atoms far above the
magnetic threshold.  The effect of these energetic atoms on the cold
thermal cloud has not yet been studied systematically.}  To accomplish
relocation of the coils, the middle portion of the brass IVC
tailpiece was replaced by a G-10 section of slightly smaller diameter.
The nonmetallic section was necessary to allow propagation of the rf
from the coils to the atoms.

In the second-generation design, minor changes were made to the
geometry of the detection end of the trapping cell, slightly
increasing the solid angle of detection.  Further details on the
construction of the current cryogenic apparatus are given by Moss
\cite{mos01}.

The introduction of a nonmetallic trapping cell allowing rf
evaporation in a cryogenic environment was not only necessary for the
achievement of BEC, but also enabled us to generate higher densities
of metastables than ever before.  These higher densities, together
with the addition of electrical shielding to the ``plastic'' G-10
tube, were crucial prerequisites for the observation of two-body
metastable loss described in Chapters~\ref{ch:decay} and \ref{ch:results}.

\section{The Bolometer}
\label{sec:bolometer}

\subsection{Construction}

The current bolometer design is a refinement of those described in the
theses of Yu \cite{Yu93} and Doyle \cite{doy91}.  Following a
suggestion of D.~G.~Fried, the area of the electrodes was reduced to
ameliorate rf pickup.  A new construction method was developed after
thermal cycling tests with different glues and wire-bonding methods,
and it has proven to be highly robust against thermal stress and
mechanical shock.  For the benefit of those wishing to construct such
a bolometer themselves, a detailed description of this
important diagnostic device is presented here.

\begin{figure}
\centerline{\epsfig{file=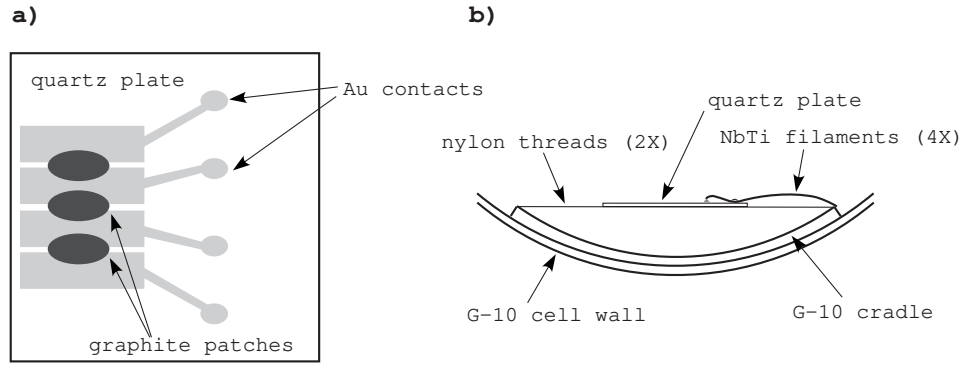,width=5in}}
\caption{(a) Face view of quartz bolometer.  The NbTi filament leads,
which are bonded to the circular ends of the Au contact pads, are
omitted.  (b) Edge view showing bolometer suspended from its cradle
and a filament lead.}
\label{fig:bolometer}
\end{figure}

The ideal bolometer has a very small heat capacity, good internal
thermal conductivity, and good thermal and electrical isolation from
its surroundings.  Quartz was chosen as the substrate because its heat
capacity at millikelvin temperatures is tiny, it has sufficient
thermal conductivity, and it is commercially available in very thin
plates.  The current bolometer substrate is a chemically polished
plate (1~cm $\times$ 1~cm $\times$ 50~$\mu$m) \cite{valpeyfisher}.
The heat capacity of the substrate is probably only a few percent of
the total bolometer heat capacity (roughly $\sim10^{-8}$~J/K at
350~mK), which is likely dominated by amorphous graphite patches
(Fig.~\ref{fig:bolometer}(a)) and the superfluid film which is
ubiquitous in the trapping cell.  The time resolution of the bolometer
is limited ultimately by the time it takes for heat deposited at one
point on its surface to spread throughout its bulk.  An upper limit on
the time constant for this spreading, $\tau_s$, is obtained by
dividing the heat capacity by the thermal conductivity between two
edges of the quartz plate.  At a typical operating temperature of
350~mK, the result is $\tau_s<40$~$\mu$s.  A detection bandwidth of
several kilohertz has been demonstrated with the bolometer described
here.  For most measurements, however, a low-pass filter is used to
suppress 60~Hz and other noise pickup in the sensitive bolometer
circuit.

A physical mask was made for the electrode pattern by cutting an
aperture in a thin metal plate and soldering three $\sim38$~$\mu$m
diameter wires across the aperture.  An evaporator was used to deposit
5~nm of chromium followed by 100~nm of gold onto the quartz.  The
resulting pattern is shown in Fig.~\ref{fig:bolometer}(a).  Electrical
leads were bonded to the gold pads using silver epoxy to establish
good electrical contact, while droplets of Stycast 1266 epoxy
\cite{1266} near the edge of the plate provided a strong mechanical
bond between the leads and the quartz (see
Fig.~\ref{fig:bolometer}(b)).  To avoid adding unnecessary heat
capacity to the bolometer in the form of excess epoxy, the droplets
were applied with the end of a fine wire.  To minimize the thermal
link between bolometer and cell, 30~$\mu$m diameter superconducting
NbTi filaments were used as electrical leads.  These filaments were
obtained by dissolving the copper matrix of a multi-filament magnet
wire in nitric acid.

Across the gaps between gold pads, amorphous graphite resistors were deposited by repeated applications of Aerodag \cite{aerodag}.  The
graphite patches have resistances of $1$-$2$~k$\Omega$ at room
temperature, but increase to several tens of k$\Omega$ below 300~mK.  In
practice, only two electrical connections and one resistor is used at
a time; the others provide redundancy in case of a failure.
As explained below, the resistor serves as both temperature sensor and
heater.

The quartz plate was laid across two nylon threads, about 10~$\mu$m in
diameter, which were stretched taut across a G-10 cradle piece having
the same radius of curvature as the inside of the cell.  Small drops
of Stycast 1266 were allowed to wick along the threads, providing a
large-surface-area bond between the threads and quartz plate.  The
G-10 cradle was then mounted by epoxy on the cell wall so that the
bolometer sits vertically below the magnetic trap.

\subsection{Operation}

In typical operation, the bolometer is held at a temperature between
200~mK and 350~mK, considerably above the cell wall temperature in the
later stages of the trap cycle.\footnote{The bolometer resistance as a
function of temperature was calibrated by introducing $^3$He exchange
gas into the cell.}  The bolometer acts as a ``thermistor'' in a
resistance bridge feedback circuit; its temperature setpoint is
determined by a variable resistor outside of the cryostat.  A signal
is derived by amplifying the changes in feedback voltage required to
hold the bolometer at constant temperature.  This feedback mode of
operation allows for much higher linearity and bandwidth than simply
monitoring changes in the bolometer resistance.  Different temperature
setpoints are selected depending on the type of bolometric
measurement; the bolometer is more sensitive at lower temperatures
because heat capacity is lower, and the temperature coefficient of its
resistance is larger.

The primary purpose of the bolometer in our apparatus is to detect the
recombination heat released when atoms from the trap are allowed to
reach the cell wall or spin-flip at a zero of the magnetic
field.  The resulting H$_2$ molecules retain much of the 4.6~eV
recombination energy in their rovibrational degrees of freedom
\cite{gtb81}.  Since the molecules are insensitive to the magnetic
field, they bounce freely around the cell, transferring some of their
energy in collisions with surfaces.  In this way, a small fraction of
the recombination energy reaches the bolometer.  Doyle estimated this
fraction to be $4\times10^{-4}$ for a similar bolometer \cite{doy91}.
Operating at a temperature of $\sim200\mu$K, the current bolometer is
sensitive enough to detect a flux of $\sim10^9$~atoms/s.  If Doyle's
fraction is still appropriate, this flux corresponds to a power of
$\sim10^{-13}$~W at the bolometer.

\subsection{Density Measurements}
\label{subsec:densitymeasurements}
As mentioned in Sec.~\ref{sec:enhancements}, the relative number of
atoms in the trap can be determined by dumping the atoms from the trap
and integrating the bolometer signal.  Since the atom number after
loading and evaporation processes is reproducible to a few percent in
successive trap cycles, it is possible to map out the decay curve of
an equilibrium sample by waiting different lengths of time before
dumping.  From the local equation for dipolar density decay,
\be
\frac{dn_{1S}}{dt}=-gn_{1S}^2,
\ee
it is possible to derive the following equation involving atom number $N_{1S}$
as a function of time: 
\be
\frac{N_{1S}(0)}{N_{1S}(t)} = 1 + (1+\xi
f)\frac{Q_{1S}(T)}{V_{1S}(T)}gn_{1S,o}(0)t.
\label{eq:densitydecay} 
\ee
Here, $n_{1S,o}$ is the peak density in the trap, and
$g=1.2\times10^{-15}$~cm${^2}$/s is the theoretical rate coefficient
for dipolar decay of $d$-state atoms\cite{skv88}.  The quantity
$V_{1S}$ is known as the effective volume and is defined as the ratio
$N_{1S}/n_{1S,o}$.  It can also be expressed in terms of the
normalized spatial distribution function $f_{1S}({\bf r})=n_{1S}({\bf
r})/n_{1S,o}$: $V_{1S}=\int f_{1S}({\bf r}) d^3{\bf r}$.  The quantity
$Q_{1S}=\int f^2_{1S}({\bf r}) d^3{\bf r}$ can be considered the
effective volume for two-body loss.  Both $V_{1S}$ and $Q_{1S}$ depend
on the temperature $T$.  The factor $(1 + \xi f)$ is a correction factor
accounting for the fact that some evaporation must occur simultaneously
with dipolar decay if the sample is in equilibrium.  More
specifically, $\xi$ represents the ratio of elastic collisions to
inelastic collisions, and $f$ is the fraction of elastic collisions
which result in an atom leaving the trap.

\begin{figure}
\centerline{\epsfig{file=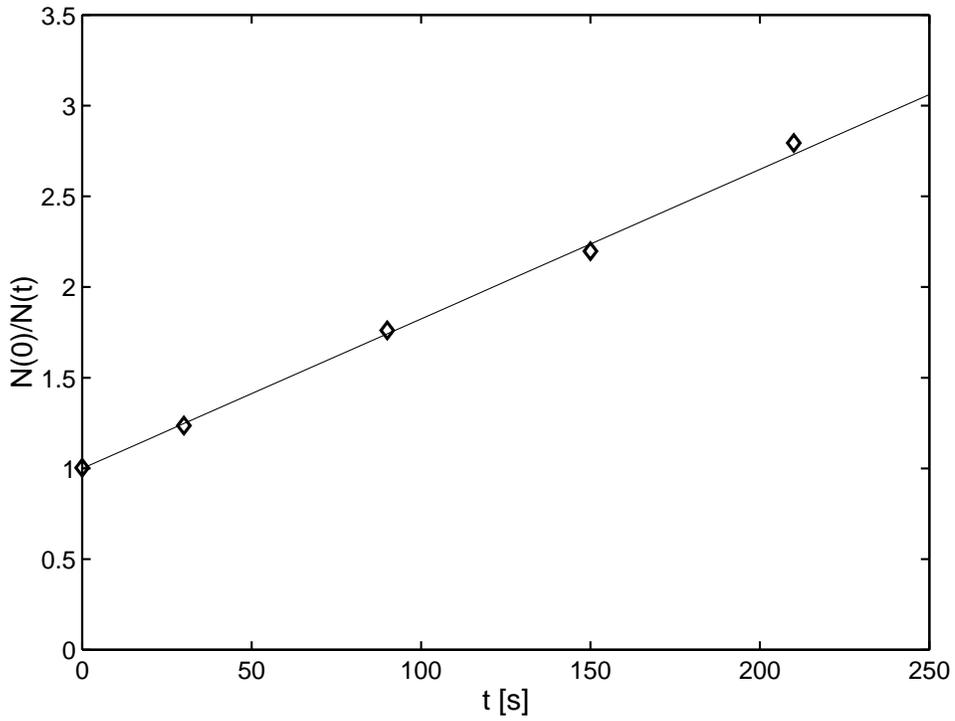,width=5in}}
\caption{Inverse of integrated bolometer signal, normalized to its
value at $t=0$, when dumping atoms after different wait times $t$.
The example shown here is for a sample in Trap~Y of
Ch.~\ref{ch:decay}.  The initial peak density derived from the linear fit
is $2.4\times10^{13}$~\percc.}
\label{fig:density}
\end{figure}

Equation \ref{eq:densitydecay} describes the time-dependence of the
inverse normalized atom number, $N_{1S}(0)/N_{1S}(t)$.  This can be
determined from integrated bolometer signals without knowing the
bolometer detection efficiency.  As shown in Fig.~\ref{fig:density},
the inverse atom number has a linear time dependence, characteristic
of two-body decay.  The slope of $N_{1S}(0)/N_{1S}(t)$ can be used to
determine the initial density $n_{1S,o}$ assuming the other quantities
are well known.  The theoretical value of $g$ is believed accurate to
better than 10\% \cite{sto01}.  Furthermore, the ratio $Q_{1S}/V_{1S}$ can
be calculated numerically from the known geometry of the magnets
to about 10\%.  The uncertainty stems from both imperfect knowledge of
the trapping field and uncertainty in the sample temperature.  

From thermodynamic considerations, it can be shown that the
evaporation correction factor is given by \cite{gre01}
\be
(1 + \xi f) = \frac{\eta+\beta-3/2-\delta/2}{\eta+\beta-3/2-\delta}.
\label{eq:corrfactorexact}
\ee 
In this equation, $\eta$ is the ratio of trap depth to temperature ($8<\eta<11$ for samples analyzed in this thesis);\ $\beta$ is the
average fraction of thermal energy $kT$ carried away by an evaporated
atom in excess of the the threshold energy $\eta kT$.  Numerical
calculations by the MIT group as well as theoretical calculations by
the Amsterdam group \cite{lrw96} have shown that $\beta\simeq0.8$ for
the range of $\eta$ found in our experiment, assuming a cylindrical
quadrupole trap.  The constant $\delta$, defined by $dV_{1S}\propto
U^{\delta-1}dU$, where $U$ is potential energy, is equal to 2 for a
cylindrical quadrupole trap, 1 for a cylindrical harmonic trap, and 3
for a spherical quadrupole trap.  For the traps relevant here, it is a
good approximation to take $\delta=2$, and
Eq.~\ref{eq:corrfactorexact} simplifies to
\be
(1 + \xi f) \simeq \frac{\eta-1.7}{\eta-2.7}.
\label{eq:corrfactorapprox}
\ee 
This expression is believed accurate to a few percent for all but the
coldest samples in the MIT hydrogen experiment.  (An alternative
approach to calculating the correction factor is outlined in
\cite{lrw96,fri99}).  Depending on $\eta$, which generally decreases
as the sample temperature decreases, the evaporation correction
amounts to a 10-20\% reduction in the apparent density.

The uncertainties in $g$, the evaporation correction factor, and the
effective volumes are the dominant contributions to the error in
ground state density measurements.  By adding these contributions in
quadrature, the total uncertainty is estimated to be about 20\%.

\section{Loading the Magnetic Trap}
\label{sec:loading}
Over the past several years, a number of studies have been undertaken
to better understand what is arguably the most physically complex
stage of the trapping cycle: the loading phase.  The goal of these
studies has been to understand what limits the number of atoms which
can be accumulated in our trap.  This section summarizes the current
state of knowledge about the physics of hydrogen trap loading,
highlighting our recent investigations.

\begin{figure}
\centerline{\epsfig{file=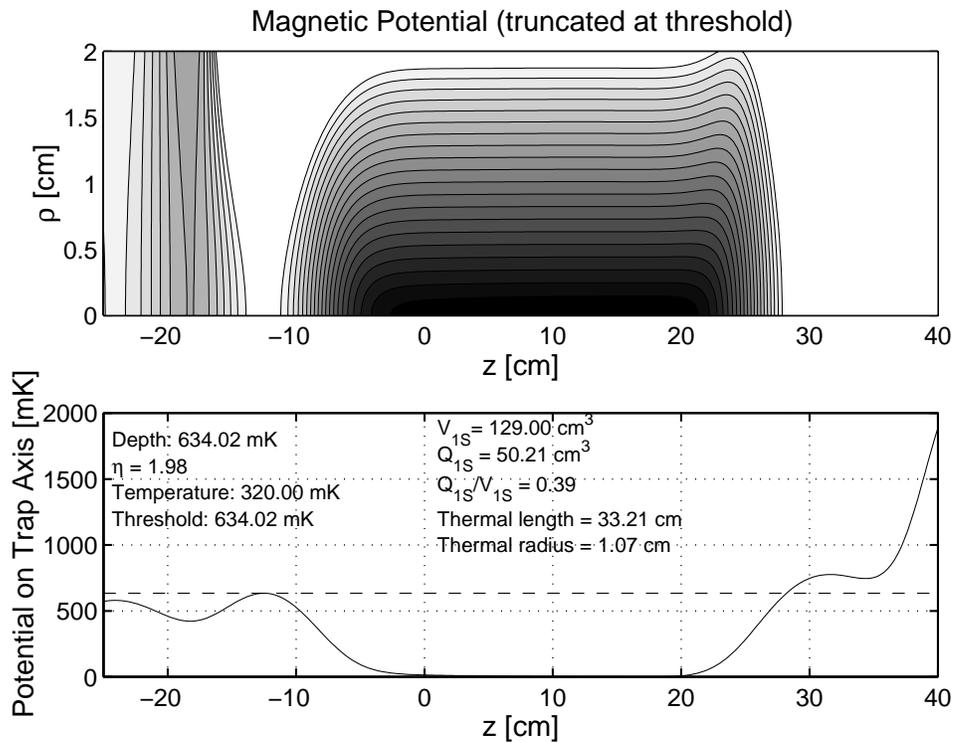,width=5in}}
\caption{Trap shape during accumulation; the sample temperature is
determined by the 320~mK cell wall temperature.  The trap volumes are
calculated assuming a truncated Maxwell-Boltzmann distribution.  The leftmost edge of the plots
correspond to the MgF$_2$ window at the bottom of the trapping cell
(see Fig.~\ref{fig:apparatus}), and the cell wall is at $\rho=1.9$~cm.
Local maxima of the field on axis near $z=-24, -12, +32$~cm result from
contributions by the loading, lower pinch, and upper pinch magnets,
respectively.  See also Fig.~\ref{fig:bigtrapandCAD}.}
\label{fig:loadingtrap}
\end{figure}

\begin{figure}
\centerline{\epsfig{file=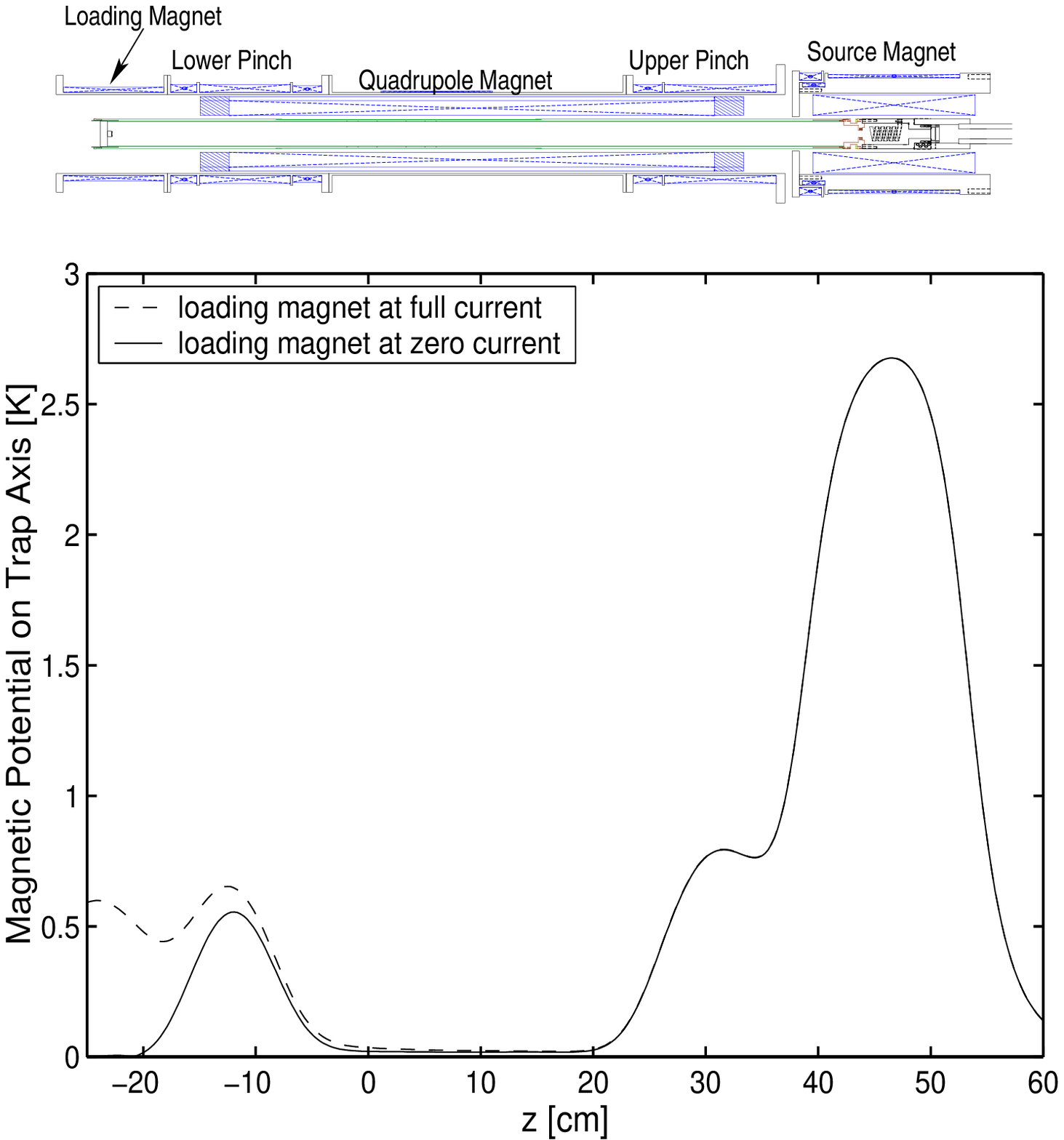,width=5in}}
\caption{Magnetic field profile along the trap axis showing the 4~T
(2.7~K) source field in the discharge resonator.  Atoms in all four
$1S$ hyperfine states are generated in the high source field and
stream into the lower field regions.  After thermalizing with the cell
walls, $d$-state atoms settle into the trap between $z=-10$ and
$z=25$~cm.  Once accumulation is complete, the loading magnet current
is ramped to zero.  At top and aligned with the magnetic field profile
is a CAD drawing of the principal magnets and trapping cell.}
\label{fig:bigtrapandCAD}
\end{figure}

The loading phase can be divided into two parts: accumulation and
thermal disconnect.  In preparation for accumulation, the largest of
the computer-controlled magnets are ramped to maximum current,
generating a 0.6~K deep Ioffe-Pritchard trap.  At this time, a
``loading magnet'' is energized to prevent the existence of a zero
field region near the lower end of the trapping cell
(Figs. \ref{fig:loadingtrap} and
\ref{fig:bigtrapandCAD}).  In the next few seconds, the cell is
heated above 300~mK by heaters inside the superfluid helium jacket,
making the cell walls less ``sticky.''  At this temperature, atoms are
adsorbed on the helium film for less than 1~$\mu$s before
returning to the gas \cite{doy91}.  (The binding energy for hydrogen
atoms on superfluid helium is about 1~K
\cite{bhk86}).  Once the cell is sufficiently warm, discharge pulses
begin puffing atoms into the trapping cell, at a typical pulse rate of
50~Hz.  The atoms thermalize with the walls and explore the entire
cell volume.  The population of $d$-state atoms in the magnetic trap
region grows to a saturation value.  While the discharge pulses feed
this population, a number of loss processes can occur, including
dipolar decay and other two-body inelastic collisions in the gas,
one-body spin-flips on the cell walls, and two-body recombination on
the walls.  As will be explained below, however, most of these
processes can be ruled out as the limiting factor in determining how
many atoms are trapped.

When accumulation is complete, the cell heater is switched off, and
the cell temperature drops below 150~mK in about 2~s.  After about
10~s, the temperature reaches 100~mK.  During the rapid cooldown, the
flux of atoms to the wall from the magnetic trap minimum decreases.
However, the residence time of atoms on the walls dramatically
increases, and the relatively slow wall loss processes begin to remove
atoms with enough energy to reach the wall.  When the cell passes
through a temperature of roughly 150~mK, the magnetically trapped
sample thermally disconnects from the walls; the probability of atoms
returning from the walls goes to zero.  Meanwhile, the sample
continues to cool by evaporation of atoms to the walls, and its
temperature drops below the wall temperature.  In a typical trap
cycle, the loading magnet is ramped down simultaneously with thermal
disconnect, and evaporation over the magnetic saddlepoint begins.
About $1\times10^{14}$ atoms equilibrate at a temperature of 40~mK and
a peak density of $2\times10^{13}$~\percc.

The primary means for studying the loading process is to introduce
variations in the sequence and observe the effect on atom number.  The
atoms can be dumped directly from the initial deep magnetic trap and
detected on the bolometer.

\subsection{Pulsed Discharge}
The 300~MHz discharge resonator currently in use is described by Killian
\cite{kil99}.  Several studies were performed to optimize the duty
cycle, pulse length, and rf power parameters of the discharge.  These
studies show that all of these parameters can be varied over a wide
range without changing the saturation atom number by more than a few
percent.  The time constant for reaching saturation, however, depends on the
average power supplied to the discharge.  

For constant rf power, it was found that larger duty cycles result in
more atoms trapped per unit energy supplied to the discharge.  An
energy efficient discharge is desirable since it requires less heat
to be removed from the copper resonator by the dilution refrigerator.  As
will be explained in Sec.~\ref{subsec:disconnect}, maximum cooling
power is desired during thermal disconnect.  Very short
$(<100$~$\mu$s) discharge pulses are inefficient because it takes
15-25~$\mu$s for the discharge to start firing after the beginning of
each rf pulse.  The energy deposited in the resonator during this time
heats the fridge but does not produce atoms.  The reason for this
``dead time'' may be that sufficient helium vapor pressure must
develop in the resonator before discharge activity can begin.

Some effort was expended to optimize the shape of the rf pulses.  An
rf switching scheme was used to apply two different power levels, a
lower level during the dead time (25~$\mu$s) and a 5-9~dB higher level
during the firing time (100~$\mu$s-10~ms) of each pulse.  Typical peak
powers at the amplifier output driving the discharge were 50-100~W,
applied at duty cycles of 0.5-5\%.\footnote{These values imply average
powers far greater than can be tolerated by the dilution refrigerator.
Since less than 10\% of the power is reflected from the resonator,
most of the rf power must be dissipated in parts of the apparatus not
thermally anchored to the dilution refrigerator.}  The bolometer was
used to record atom pulses as the discharge was firing.  Also, studies
were made of the trapped atom number as a function of accumulation
time.  A number of different pulse lengths and power levels were
tried.  Although the size of the atom pulses scaled linearly with the
product of RF power and firing time, it was found that almost
regardless of pulse characteristics the trapped atom number after
thermal disconnect saturated at nearly the same value.  A couple
conclusions were drawn from these studies.  First, the trapped atom
number is not limited by the number of atoms produced in a single
discharge pulse.  Second, since we were unable to trap more atoms
using attenuated rf power during the dead time of each pulse, it seems
that overheating of the resonator is not the number-limiting factor.
The power switching scheme provided no clear advantage.

Even though the saturation atom number is not better, a high average
flux from the discharge is desirable because it shortens the
accumulation time.  This means less total heat is required to hold the
cell at its relatively high accumulation temperature; after the
loading phase the cell can more quickly reach a cold temperature
advantageous for sensitive bolometric detection.  For the experiments
described in the remainder of this thesis, discharge parameters were
chosen which loaded the trap 3-4 times faster than in the initial
nonmetallic cell experiments.  Typically, the discharge operates with
1~ms pulses repeated at 50~Hz for a total of 8~s.  The peak power of
the (square) rf pulses at the amplifier output is $\sim90$~W.

A sufficient helium film thickness is important for efficient loading
of the trap from the discharge.  If the the film is too thin, bare
spots can form in the cell and resonator while the discharge fires.
The bare spots serve as sticky sites where atoms spin-flip and
recombine.  The discharge itself can become erratic with a very thin
film, probably due to low vapor pressure in the resonator.  

\subsection{The Accumulating Sample}
\label{subsec:accumulating}

The discharge injects a flux of both high field and low field seekers
into the trapping cell.  The average flux (counting only $d$ states)
is more than 10$^{13}$ atoms/s, and a single 1~ms pulse may inject
nearly 10$^{12}$ atoms.  The high field seeking $a$ and $b$ states are
eventually pulled back toward the 2.7~K source field
(Fig.~\ref{fig:bigtrapandCAD}), though this can take some time if
they reach the trap region and equilibrate with the wall.  In the trap
region, the gradient in $z$ is low, and high field seekers bounce
randomly on the cell wall until reaching the high gradient in $z$.
From the ratio of the cell wall area in this region to the cell
cross-section, we can estimate that the the atoms must bounce 25 times
before leaving the low-gradient region.  At 300~mK, the average time
between bounces is about 1~ms, implying a time constant of $\tau_{\rm
hf}\sim25$~ms for high field seekers to return to the source.  High
field seekers reaching the relatively high field at the opposite end
of the trap may linger there and take even longer to return.  In the
meantime, the high field seekers can undergo hyperfine-changing
collisions with $d$ atoms, potentially causing trap loss.

The other low field seekers, $c$-state atoms, disappear due to spin
exchange collisions on a time scale of seconds.  As a result, the
sample becomes purely $d$-state within seconds after the end of
accumulation.

Some insight was gained from studies in which the bolometer was
deliberately made bare of helium film for a time during or after
accumulation.  This was accomplished by applying several hundred
microamps to the bolometer.  When the bolometer was bare throughout
accumulation, the number of atoms loaded into the trap was reduced by
nearly two orders of magnitude.  What is more, the number remaining
depended on the magnetic potential $E_{\rm bolo}$ at the bolometer,
which sits at approximately $z=-23$~cm in the coordinate system of the
Figures.  We varied $E_{\rm bolo}$ using the loading magnet current.
Under the assumption that the bare bolometer surface was the dominant
loss mechanism during loading, the dependence of the atom number on
$E_{\rm bolo}$ was found roughly consistent with a thermally
distributed sample at the cell temperature.  Furthermore, it was found
that the sample could be depleted by baring the bolometer at a time
long after the discharge stopped firing as long as the cell
temperature was maintained.  If the bolometer was not bared, then the
sample could be held at the cell temperature for 30~s or more after
accumulation without significant depletion.  Finally, if the cell was
allowed to cool after accumulation, a subsequent baring of the
bolometer had no effect on the atom number.  This was interpreted as a
signature of thermal disconnect.


These observations provided evidence that the assumption of the
accumulating sample being in equilibrium at the cell temperature is a
good one.  In addition, we found evidence that the peak trap density
reaches a saturation value during accumulation which is too low for
significant loss due to dipolar decay.  Since the evaporation process
during thermal disconnect is believed to be well understood
(Sec.~\ref{sec:thermaldisconnect}), measurements of atom number
following thermal disconnect allow an order of magnitude estimate of
the sample density before thermal disconnect.  These measurements led
to the conclusion that the peak loading density is
$(\sim10^{12}$~\percc$)$, at which the decay time due to dipolar decay
is many minutes.  The fact that the atom number after thermal
disconnect does not change appreciably when the atoms are held for
many seconds at the loading temperature following accumulation also
suggests a low density at the end of accumulation.  This would mean
that the key limiting process during loading occurs in the
accumulation phase.  However, in the absence of more direct
measurements of the density before thermal disconnect, we admit the
possibility that an unknown process during thermal disconnect may cap
the atom number at a definite maximum.

If indeed the sample can be held in thermal contact with the walls for
many seconds without loss, an implication is that surface loss
processes at the cell loading temperature are not important for a
$d$-state sample.  In other words, one-body spin flip and surface
recombination are not limiting processes during accumulation.

Since the bolometer is the most poorly thermally anchored part of the
trapping cell, there was some concern that the bolometer was heating
enough to become bare while the discharge was firing even under normal
circumstances.  This possibility was ruled out by the bare bolometer
studies.

One unexpected observation was that the time to reach saturation
number in the trap was unaffected by the bare bolometer.  This implies
that there is another loss mechanism other than the bare bolometer
which was determining the saturation atom number in the above studies.

\subsection{Phenomenological Model}
We will assume for the moment that the most important loss mechanism
occurs during accumulation.  The fact that the saturation atom number
is independent of discharge pulse parameters suggests a phenomenological
model for trap loading which involves a loss rate proportional to the
discharge flux.  Consider the simple model
\be
\frac{dN_d}{dt}=F_d - \beta F_d N_d,
\label{eq:loadingmodel}
\ee
where $N_d$ is the number of $d$ atoms in the trap, $F_d$ is the average
flux of $d$ atoms from the discharge, including $d$ atoms resulting
from hyperfine-changing collisions, and $\beta$ is a proportionality
constant for the dominant loss mechanism.  The solution to
Eq.~\ref{eq:loadingmodel} is
\be
N_d(t)=\frac{1}{\beta}(1-e^{-\beta F_d t}).
\label{eq:loadmodsolution}
\ee  
The saturation time constant $\tau_{\rm sat}=(\beta F_d)^{-1}$ depends inversely on the
average flux, consistent with data from discharge parameter studies.
In addition, the saturation value depends only on $\beta$ and not the
flux.  The question remains: what physical process gives rise to
$\beta$?

One candidate is the inelastic collisions between $d$ atoms and
high field seekers in the thermal gas during the time $\tau_{\rm hf}$
after a discharge pulse (Sec.~\ref{subsec:accumulating}).  According
to theoretical calculations \cite{skv88}, there are several $b+d$
inelastic processes with large cross sections.  The channels $b+d
\rightarrow a+a$ and $b+d \rightarrow c+c$, for example, both have
rate coefficients of approximately $4\times10^{-13}$~cm$^3$/s at
300~mK and zero field.  Rate coefficients for $a+d$ processes have not
been calculated at finite temperature, but in the zero temperature
limit, the rate coefficients of the most likely $a+d$ channels are
several orders of magnitude below the $b+d$ channels at all magnetic
fields.  Taking $b+d$ collisions to be the dominant source of loss,
the local rate of $d$-state density change due to collisions with high
field seekers can be expressed
\be
\left. \frac{dn_d}{dt}\right|_{\rm loss} = -g_{bd} n_b n_d,
\ee
where $g_{bd}$ is the local total loss rate constant for $b+d$
collisions, and $n_d$ and $n_b$ are the $d$ and $b$ atom densities,
respectively.  The rate constant is a function of position because it
depends on the magnetic field.  Using equilibrium distributions for
the atoms, the rate of change of $d$-state number during accumulation
is
\be
\frac{dN_d}{dt}=F_d - \frac{N_b}{V_d V_b} \left( \int g_{bd}({\bf r})
d^3{\bf r} \right) N_d,
\label{eq:dbmodel}
\ee
where $N_b$ is the total number of $b$ atoms in the trapping cell which
are not already accelerating in a high gradient region back to the
source; $V_d=\int {\rm exp}(-U({\bf r})/kT)d^3{\bf r}$ and 
$V_b=\int {\rm exp}(+U({\bf r})/kT)d^3{\bf r}$ are effective volumes for low field and high field seekers,
respectively.  The integration is to be taken over the cell volume
occupied by the quasi-static $b$-state population $N_b$.  From comparison
of Eq.~\ref{eq:dbmodel} with Eq.~\ref{eq:loadingmodel}, we have
\be
\tau_{\rm sat} = (\beta F_d)^{-1} = \frac{V_d V_b}{N_b (\int g_{bd}({\bf r}) d^3{\bf r})}.
\label{eq:betaFd}
\ee 

To test the plausibility of this physical model, we plug some numbers
into Eq.~\ref{eq:betaFd} and compare it with an observed time constant
for saturation of $N_d$.  Since this is an order of magnitude
calculation, we arbitrarily take the integration volume to be the
entire cell volume excluding the region of large gradient at
$z>36$~cm.  At 320~mK, numerical integration yields
$V_d\simeq150$~cm$^3$ and $V_b\simeq 4000$~cm$^3$.  For the currently
preferred discharge parameters, $\tau_{\rm sat}$ is observed to be
$\sim4$~s.  Since there are about $1\times10^{14}$ atoms accumulated
in the trap, $\beta \simeq 1\times10^{-14}$, implying
$F_d\simeq3\times10^{13}$~s$^{-1}$.  The discharge fires at 50~Hz, so
each pulse brings about $6\times10^{11}$ $d$ atoms into the cell.  The
discharge pulses are separated by a time comparable to $\tau_{\rm hf}$.
If the number of $b$ atoms per pulse is not too different from
the number of $d$ atoms, then $N_b$ is $\sim10^{11}$.  If we further
assume that $g_{bd}$ averaged over the cell is $~10^{-12}$~cm$^3$/s,
the result is $(\beta F_d)^{-1}\sim10^4$~s, which is 3-4 orders of
magnitude too long.  It seems unlikely that either $g_{bd}$ or $N_b$
have been grossly underestimated here.  In other words, $b+d$
collisions in the gas probably do not cause the observed saturation of
$N_d$.

Inelastic $c+d$ collisions are another possibility.  However, the
rate coefficients for $c+d$ channels are probably far smaller
than the $c+c$ channels which remove $c$ atoms from the trap.  At zero
temperature, the sum of theoretical $c+d$ rate coefficients is
comparable to the sum of rate coefficients for $d+d$ processes, which
are known not to cause significant loss during loading.

Not only the gas densities but also the surface densities of the four
hyperfine states should be proportional to the discharge flux.  For
state $i$, the equilibrium surface density $\sigma_i$ is
given by
\be
\sigma_i = n_{i,{\rm w}} \lambda_{\rm dB} e^{E_b/kT},
\ee
where $n_{i,{\rm w}}$ is the gas density of species $i$ near the
wall, $\lambda_{\rm dB}$ is the thermal deBroglie wavelength, and
$E_b$ is the binding energy on the film.  Can inelastic processes on
the helium surface give rise to the loss term in the phenomenological
model?  Loading studies with different film thicknesses have shown
that as long as the helium film is not too thin, the saturation atom
number is about the same.  Since changing the film thickness changes
the binding energy of the hydrogen atoms on the surface, the rate of
surface collisions should be affected by changing the film thickness.
Thus, it also seems unlikely that inelastic surface collisions limit
the atom number.

Another idea for a loss mechanism proportional to discharge flux is
that high energy particles from the discharge knock $d$
atoms out of the trap.  The discharge likely produces many other
species besides the $1S$ hyperfine states of H.  These could include
metastable H, excited states of He, molecular states of H, and various
ions.  As these other species stream through the trapping region, they
could cause loss of $d$ states.  

To see if direct flux from the discharge causes loss, in one
experimental run a series of baffles was installed in the opening
between the discharge resonator and cell volume.  The baffles were
made of copper and were thermally anchored to the copper resonator.
They were constructed such that any particles leaving the
discharge would have to bounce several times on a cold, helium-covered
surface before entering the cell.  Most of the species mentioned above
would not be able to bounce because of their relatively large surface
binding energies.  Furthermore, any particles streaming into the cell
would have energies determined by the cold baffles rather than the
discharge.  The trap loading behavior was found, however, to be very
similar to loading without the baffles.  The maximum number of atoms
loaded was again the same.  We concluded that energetic particles from
the discharge were not limiting the atom number.

Since the baffles were located in a high field gradient, the baffle
studies also provided experimental evidence that high field seekers do not
cause significant loss during loading.  The baffles should have
prevented most high field seekers from reaching the trap, and yet the
saturation behavior of loading was unchanged.

\subsection{Thermal Disconnect}
\label{sec:thermaldisconnect}
\begin{figure}
\centerline{\epsfig{file=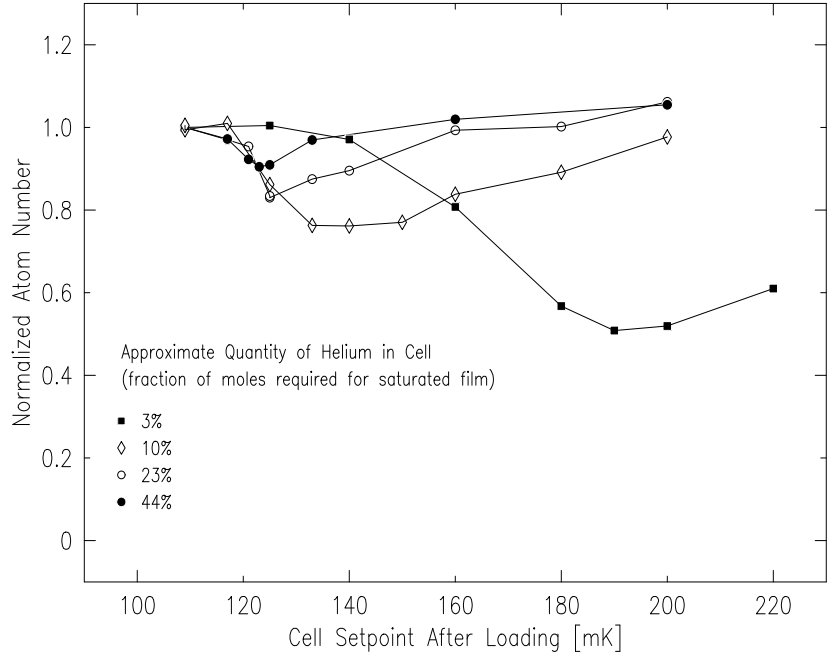,width=5in}}
\caption{Relative number of atoms trapped as a function of the cell setpoint
temperature following accumulation with different helium film
thicknesses.  For the warmer points, the setpoint was reached within a
few seconds and maintained until 30~s after the end of accumulation.
For the coldest points, the setpoint was not reached until the latter
part of the 30~s post-accumulation period.  Since the bolometer
sensitivity was different for each film thickness, each curve has been
normalized to the lowest temperature point.}
\label{fig:disconnect}
\end{figure}

Some light was shed on the physics of thermal disconnect by holding
the cell at various temperatures following accumulation and observing
the effect on atom number.  The typical hold time was 30~s.  A dip in
the atom number was consistently observed at hold temperatures between 100
and 200~mK (See Fig.~\ref{fig:disconnect}).  This dip was interpreted
as the crossover between the residence time on the surface, which
varies as $e^{E_b/kT_{\rm wall}}$ and the flux of atoms to the wall,
which varies as $n_{1S,o}T^{1/2}e^{-U_{\rm wall}/kT_{\rm atoms}}$.  Since the
temperature of the atoms tracks with the temperature of the cell
before thermal disconnect, there is a temperature range where
significant numbers of atoms can be depleted by surface loss processes since the wall is
sticky and the flux to the wall from the sample is
significant.  At higher temperatures, the residence times on the wall
are shorter, and the wall is less lossy.  At lower temperatures, the wall is so sticky that
efficient evaporation is initiated, and $T_{\rm atoms}$ quickly falls
below $T_{\rm wall}$, resulting in thermal disconnect.  Thus, the dips in
Fig.~\ref{fig:disconnect} are believed to be closely associated with
the temperature at which thermal disconnect occurs.

The binding energy $E_b$ of H on a helium film is a function of the
film thickness.  For a very thin film, $E_b$ can be much larger than
1~K because H atoms on the surface can interact with the substrate.
For a thin film and larger $E_b$, the wall becomes sticky at a higher
temperature, and the dip should occur at a higher temperature.
Furthermore, for a given hold time, it is reasonable to expect the
fractional depletion of the sample to be larger since a larger
fraction of atoms have energies large enough to reach the wall.
These qualitative trends can be observed in Fig.~\ref{fig:disconnect}.
In principle, it should be possible to derive quantitative information
about $E_b$ from such data, but this has not been attempted.

One conclusion drawn from the temperature setpoint studies is that to
maximize atom number, it is crucial to cool the cell as quickly as
possible through the temperature range where significant depletion
occurs.  This underscores the importance of reducing the heat
deposited during accumulation to a manageable level.  For this reason
also, it is important to have a dilution refrigerator with ample cooling power.

For the sake of completeness, it should also be mentioned that a
number of loading studies were performed with about 1 part $^3$He to 3
parts $^4$He in the cell.  The presence of $^3$He reduces the binding
energy of the film \cite{svk01}, and, for a given temperature,
increases the vapor pressure in the cell.  Accumulation was optimized
at a lower cell temperature than before, presumably because of the
higher vapor pressure.  Bolometric detection was impaired, however,
and it was only possible to determine that the maximum number of atoms
loaded was the same as without $^3$He to within a factor of 2.
 
\label{subsec:disconnect}

\subsection{Trap Loading: Conclusion}
The constancy of the number of atoms loaded is probably the most
remarkable feature of the trap loading process.  The saturation number
does not change for wide variations in the film thickness, discharge
parameters, cell temperature, and the nature of the conduit between
discharge and magnetic trap.  Our studies have enabled us to rule out
a number of possible loss mechanisms as the limiting factor in
loading.  A phenomenological model involving a loss term proportional
to discharge flux is consistent with our observations, but a physical
model remains elusive.  Although several aspects of loading are now
better understood, we still do not know what process causes the
atom number to saturate.  To trap more atoms, it may be necessary to
increase the effective trap volume, which will require a new magnet
system design.

\section{UV Laser System}
Excitation on the \oneStwoS\ transition in H requires a powerful,
frequency-stable source of 243~nm radiation.  In order to achieve the
high excitation rates necessary to observe two-body metastable H
effects, UV power on the order of 10~mW is required.  To sufficiently
resolve the Doppler-free spectrum at typical sample temperatures and
densities, the UV laser linewidth must be no more than a few
kilohertz.  With currently available laser technology, these
requirements are best satisfied by frequency doubling the output of a
486~nm dye laser which is locked to a stable frequency reference.
Following the design of H\"{a}nsch and co-workers \cite{kzm89},
Sandberg \cite{san93} and Cesar \cite{ces95} developed such a system
at MIT; further improvements were introduced by Killian \cite{kil99}.
The remainder of this section describes the UV laser system of the
Ultracold Hydrogen Group as used for measurements in this thesis.

\subsection{System Overview}
\begin{figure}
\centerline{\epsfig{file=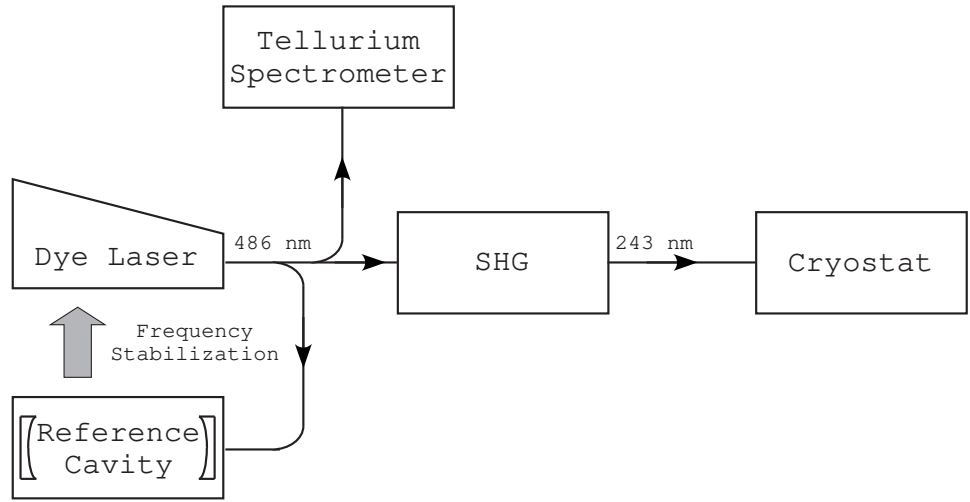,width=5in}}
\caption{Block diagram of UV laser system.}
\label{fig:lasersystem}
\end{figure}

Stable, narrowband light originates in a 486~nm dye laser
(Fig.~\ref{fig:lasersystem}), which is a modified Coherent 699 system
\cite{coherent}.  The dye solution is circulated at a pressure of
8~bar by a high pressure circulation system purchased from Radiant Dyes
\cite{radiantdyes}.  A dye jet of 200~$\mu$m thickness is produced
using a polished stainless steel nozzle from the same manufacturer.
An intracavity EOM allows for high-bandwidth locking of the laser
frequency.  When fresh dye solution is pumped by 4~W of violet from a
Krypton ion laser, around 650~mW of single-frequency output power is
obtained.  This power is more than 3 times the specification for an
unmodified commercial system.

Frequency stability is achieved by locking the dye laser to a Fabry-Perot
reference cavity by the Pound-Drever-Hall phase modulation technique
\cite{dhk83}.  The cavity, which has a spacer made of zerodur glass,
is enclosed in a temperature-stabilized vacuum chamber.  It has a free
spectral range of 598~MHz and a linewidth of about 600 kHz.  When
locked to the reference cavity, the laser has a linewidth of about
1~kHz at 486~nm.  The reference cavity has a long term drift rate of
less than 1~MHz per week.

For finding the cavity resonance closest to the H \oneStwoS\ frequency,
the absorption spectrum of a temperature-stablized Te$_2$ gas cell serves
as an absolute frequency reference.  Precise tuning of the laser
frequency is achieved by locking the first order diffraction beam from
an AOM to the cavity.  The AOM drive frequency, which is the offset
frequency between laser and cavity, is determined by a high resolution
frequency synthesizer.  By saturated absorption spectroscopy of a specific
Te$_2$ line \cite{mci87,san93,kil99}, the dye laser can be tuned to within several hundred kilohertz of the \oneStwoS\ frequency.

Most of the 486~nm laser power is coupled into a bow-tie
enhancement cavity where the blue light is frequency doubled in a
10~mm long Brewster-cut BBO crystal \cite{san93}.  The
H\"{a}nsch-Couillaud \cite{hac80} scheme is used to lock the cavity in
resonance with the incoming blue light.  An enhancement factor of
$~100$ is achieved, resulting in circulating powers as high as 50~W.  More
than 40~mW of 243~nm light can be generated by this doubling cavity.
Due to the large walk-off angle of BBO, the UV is generated in a
highly astigmatic transverse mode.

\subsection{UV Beam Transport and Alignment}
\begin{figure}
\centerline{\epsfig{file=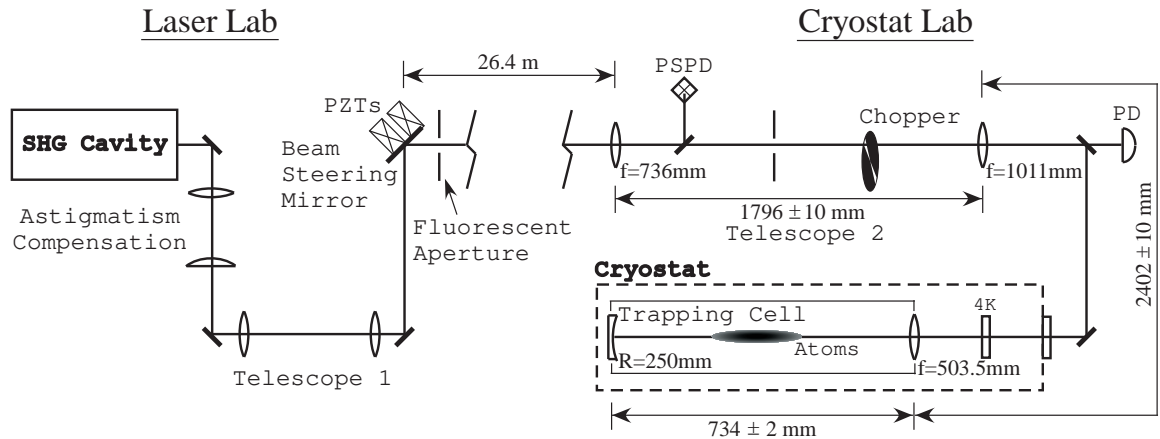,width=6in}}
\caption{Layout of UV optics in the laser lab, where the UV is
generated, and in the cryostat lab, where the atoms are trapped.
Several folding mirrors have been omitted, and the drawing is not to
scale.  The dimensions in the cryostat lab indicate optical path
lengths; estimates of thermal contraction have been included in the
cryogenic path lengths.  A pair of servo circuits tilts the
PZT-driven beam steering mirror in the laser lab to fix the beam on
a position sensing photodiode (PSPD) in the cryostat lab.  A
photodiode (PD) monitors the weak beam transmitted from a folding
mirror between Telescope 2 and the cryostat.  Amplified pulses from
this photodiode trigger a timing generator in the data acquisition
electronics.  In the cryostat, a window thermally anchored at 4~K
blocks most of the blackbody radiation from the room-temperature
entrance window.}
\label{fig:uvlayout}
\end{figure}

After passing through astigmatism compensation optics \cite{san93}, the beam is
widened in a telescope for low-divergence propagation to the cryostat,
which is 30~m away in a different laboratory.  As shown in
Fig.~\ref{fig:uvlayout}, the beam passes through another telescope
near the cryostat before propagating into the dilution refrigerator.
A lens on top of the discharge resonator brings the beam to a focus
near the magnetic trap minimum.  A concave retromirror glued to
the MgF$_2$ window at the bottom of the trapping cell reflects the
beam onto itself, producing the standing wave configuration necessary
for Doppler-free two-photon excitation of the \oneStwoS\ transition.

Since the cryostat is located in a different lab, an active beam
steering system is required to maintain a steady pointing of the beam
into Telescope 2 (see Fig.~\ref{fig:uvlayout}).  The beam steering
system, which consists of a position sensing photodiode, a servo
circuit, and a mirror mounted on piezoelectric transducers (PZT's), is
necessary to compensate for relative motion of the laser table and
cryostat and for UV pointing variations due to dye laser pointing
instability.  The PZT-mounted mirror is 12.7~mm in diameter and 6.4~mm
thick.  It was mounted using a 5-minute epoxy on a fulcrum and two low
voltage PZT stacks \cite{thorlabspzt}; the PZT's cause deflection in
orthogonal directions.  We have found this construction to be
mechanically robust.  In previous incarnations, much thinner (1-2~mm)
and lighter mirrors were used to maximize PZT bandwidth.  However, the
stress on these thin mirrors caused significant distortion of the UV
beam at the cryostat.  Furthermore, the distortion changed constantly
as the PZT voltages required to maintain lock drifted on a time scale
of minutes.  These problems were eliminated by going to a thicker
mirror.  The low voltage PZT's currently in use have very high
resonance frequencies.  In spite of the relatively large mass of the
mirror, a servo bandwidth of several kilohertz is achieved.

To ensure a good overlap of incoming and return beams at the
atoms, the position of the return beam is monitored on a fluorescent
aperture near the beam steering mirror.  The aperture is just large
enough to allow the incoming beam to pass through without diffraction
effects.  While spectroscopy is in progress, the return beam alignment
on the aperture is monitored from the cryostat lab via video camera.
If the return beam is not centered on the aperture, corrections are
made by micromotor translation of one of the Telescope 2 lenses.

The UV beam immediately following the astigmatism compensation optics
is nearly square in shape.  After propagating nearly 30 m to Telescope
2, the transverse mode consists of a bright
central lobe flanked by a series of weaker lobes.  The outer lobes
are cut off by an iris aperture in Telescope 2.  The remaining
transverse mode structure is a slightly astigmatic Gaussian beam.  By
adjusting the length of Telescope 1, more than 60\% of the power after
Telescope 1 remains after the aperture in Telescope 2.  The UV power
reaching the atoms is typically 10-20~mW.

In previous spectroscopic experiments with our apparatus, the beam
waist near the atoms had a radius of $\sim50$~$\mu$m.  For BEC
experiments, it was desirable to employ a tighter focus, since the
excitation rate for nearly motionless condensate atoms should scale
with the square of laser intensity.  It was also desirable to change
the size of the waist to study the effects of laser geometry on the
\oneStwoS\ lineshape.  For these reasons, the length of Telescope 2 was reduced,
resulting in a waist radius of about 21~$\mu$m and a Rayleigh range of
6~mm.  These values are calculated by using previously determined
Gaussian beam parameters for the input to Telescope 2 and adjusting
the length of the telescope for perfect overlap of incoming and return
beam modes.  In the case of perfect overlap, the position of the waist
in the cell is fixed by the radius of curvature of the retromirror.
If there is a small mismatch between incoming and return modes, then
the incoming and return waists are slightly separated on the trap
axis, with the midpoint between waists located at the waist position
of ideal overlap.  To optimize overlap, the spot sizes of the incoming
and return beams are viewed on a folding mirror in Telescope 2, and
the length of the telescope is adjusted until they are the same.

\begin{figure}
\centerline{\epsfig{file=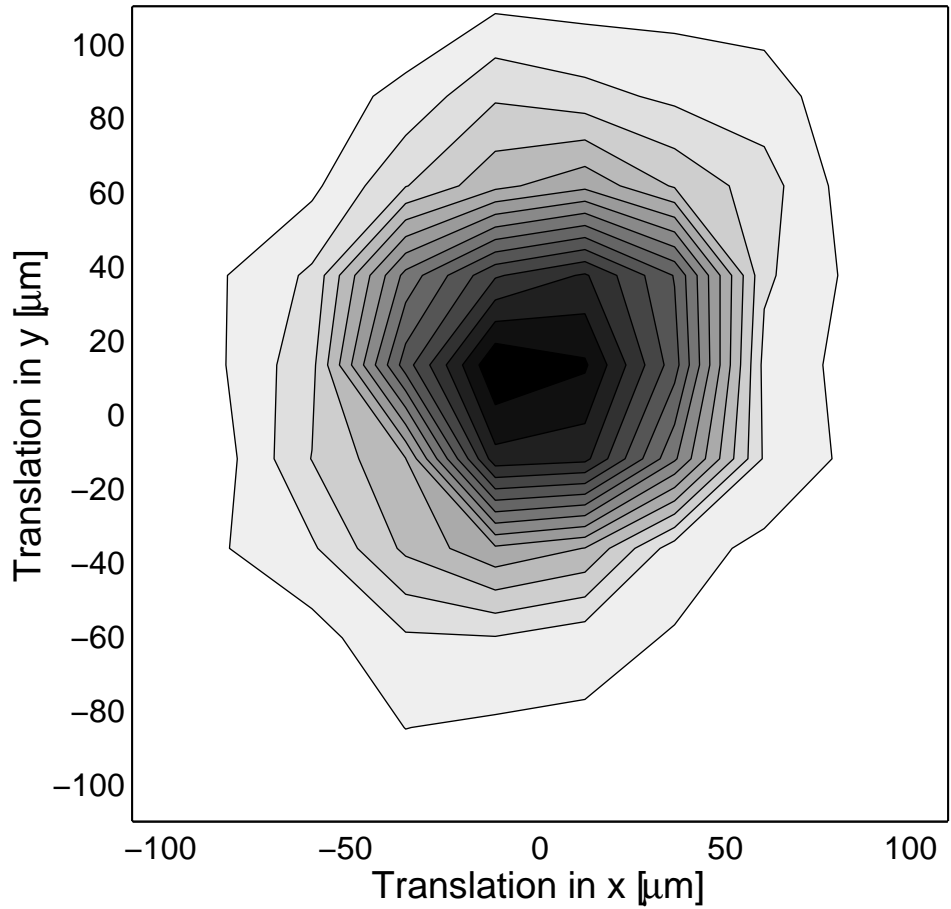,width=5in}}
\caption{``Lateral scan'' image of a trapped hydrogen gas.  The number
of fluorescence counts is recorded while translating the atoms in a
raster pattern with respect to the laser beam.  The laser frequency is
detuned to the red such that atoms from the central, densest part of
the sample (where the cold-collision shift is largest) are excited
preferentially.  In this example, the sample is slightly offset from
the center of the scan range.}
\label{fig:lateralscan}
\end{figure}

Good alignment of the laser beam with the center of the atom cloud is
essential for high signal rates and reproducible results.  An
effective method was developed for achieving this alignment.  In our
magnet system, there are four ``racetrack'' quadrupole coils which
give rise to the radial confining field of the trap.  By adding a
small trim current to one coil in each pair of opposing quadrupole
coils, it is possible to precisely displace the trap minimum in both
directions perpendicular to the trap axis.  By placing the trim current in
both coils under computer control, it is possible to move the atom
cloud rapidly in a lateral raster pattern while exciting the
atoms at a constant laser frequency.  The resulting plot of excitation
signal as a function of trim currents is a kind of spatial image of
the atom cloud (Fig.~\ref{fig:lateralscan}).  From such a ``lateral
scan'' image, the trim currents can be chosen which center the atom
cloud on the laser beam.  In addition, if the laser beam axis
intersects the trap axis at a significant angle, a highly elliptical
distribution results; the lateral scan is a useful diagnostic for
overlapping the laser and trap axes.
    
\section{Data Acquisition}

\subsection{Lyman-$\alpha$ Detection}
The metastables excited in our trap are detected by applying an
electric field of about 10 V/cm, which quenches them within a few
microseconds.  The detection efficiency for the resulting
Lyman-$\alpha$ photons is small, however.  Due to the closed geometry
of the magnet system, the MCP detector sits beyond the lower end of
the trapping cell, about 30 cm from the center of the atom cloud.  The
detection solid angle is only $1.1\times10^{-2}$~sr.  Lyman-$\alpha$
photons must pass through not only the MgF$_2$ window at the end of
the cell but also a second window on the MCP vacuum chamber.  Each
window transmits at most 40\% at 122~nm.  The best case quantum
efficiency for the MCP itself is about 25\%.  An upper limit for
metastable detection efficiency $\epsilon$ is thus $5\times10^{-5}$.
From measurements described in Ch.~\ref{ch:results}, $\epsilon$
appears to be about $2\times10^{-6}$.  Most of the additional losses
are likely due to absorption on window surfaces and sub-optimal
MCP quantum efficiency.  Futhermore, at typical $1S$ densities up to
30\% of the Lyman-$\alpha$ fluorescence in the detection solid angle
is scattered into other directions before it can escape the sample.  A
discussion of this radiation trapping effect will be presented in
Sec.~\ref{sec:radtrap}.

\subsection{Microchannel Plate Detector}
\begin{figure}
\centerline{\epsfig{file=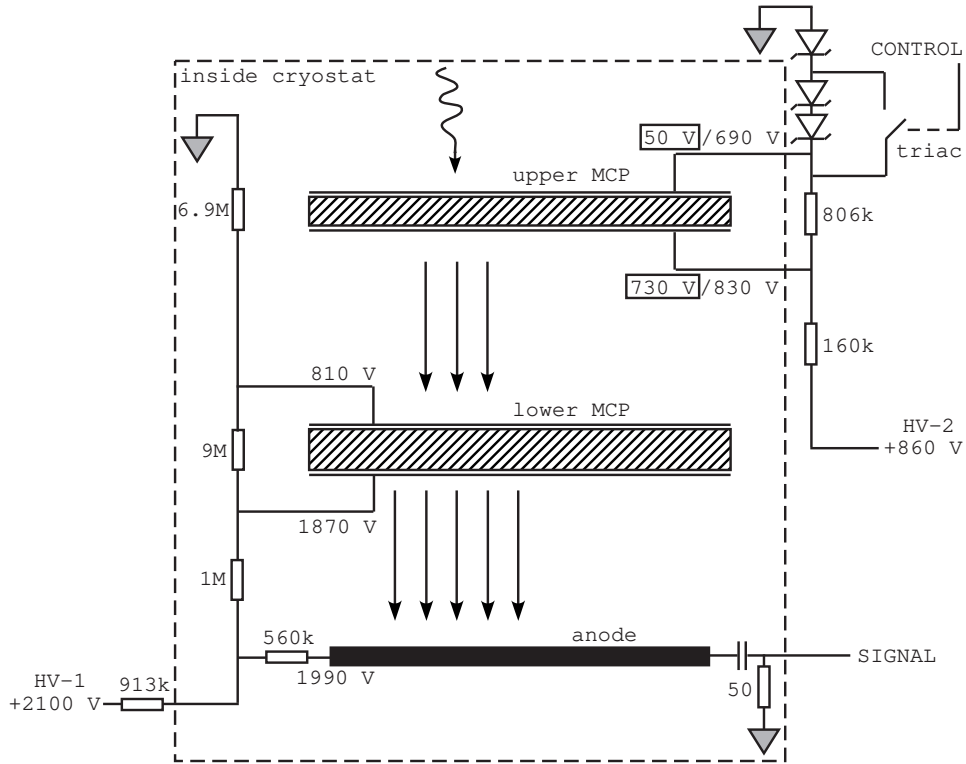,width=5in}}
\caption{Schematic of MCP switching and biasing circuits.  A
Lyman-$\alpha$ photon incident on the upper MCP gives rise to a shower
of electrons which is further amplified by the lower MCP.  The boxed
(unboxed) voltage values on the upper MCP correspond to the on (off)
state of the detector, controlled by a high voltage triac switch.
Room temperature values are given for the resistances and voltages;
these may differ by a few percent at low temperature.  With HV-1 at
2100~V, significant heating of the MCP occurs on a time scale of
minutes.  To minimize blackbody heating of the trapping cell, HV-1
is turned on just before the laser excitation phase of the trap cycle
and turned off afterwards. }
\label{fig:mcp}
\end{figure}

The microchannel plates in the current detector are similar to those
described by Killian \cite{kil99}.  In previous work, a Lyman-$\alpha$
filter was used to prevent the detector from being saturated by scatter
from the 243~nm laser.  The transmission of the filter at 122~nm was only 10\%,
however.  To increase the signal detection efficiency by a factor of 4, the
Lyman-$\alpha$ filter was replaced by a MgF$_2$ window, and a
switching scheme was developed which turns on the gain of the upper MCP
plate only long enough to detect the Lyman-$\alpha$ fluorescence at
each quench pulse.  

The relevant circuit diagrams are given in Fig.~\ref{fig:mcp}.  A triac
switch, whose state is governed by a digital signal, allows the
voltage across the upper plate to be switched between 140~V and 680~V
in a few microseconds.  At 140~V the detector is blind to incoming
radiation.  At 680~V on the upper plate and with the detector biased
as shown in Fig.~\ref{fig:mcp}, the detector gain is $\sim10^6$.  In
the present scheme, the channels of the upper plate have a 40:1 length to
diameter ratio and the lower plate a 60:1 ratio.

\subsection{Data Acquisition Electronics}
\begin{figure}
\centerline{\epsfig{file=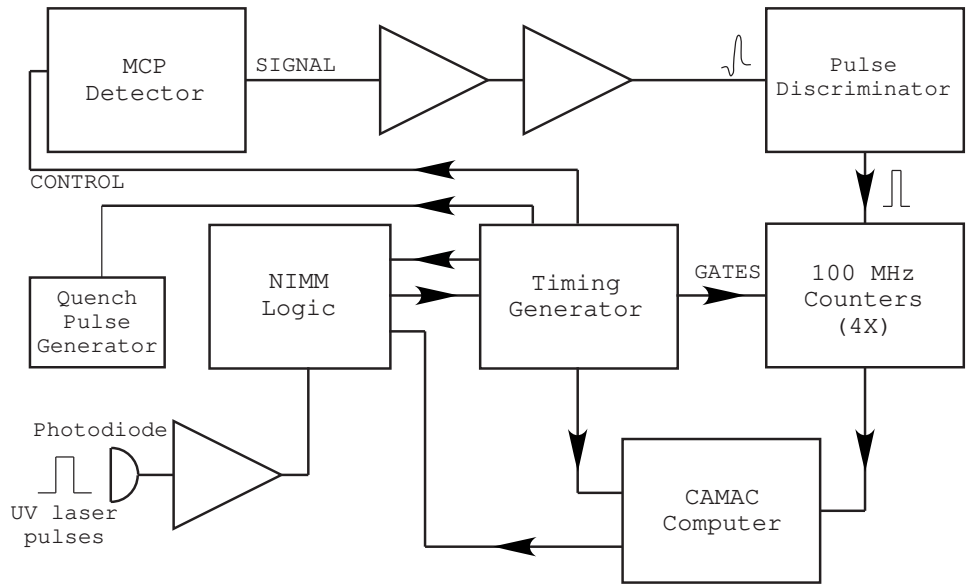,width=5in}}
\caption{Block diagram of data acquisition electronics used during the
laser excitation phase of the trap cycle.  Not shown are the rf
frequency synthesizers used to tune the laser, drive the rf
evaporation coils, and drive the discharge.  The synthesizers are
controlled by the CAMAC computer via GPIB.  The same CAMAC computer
also controls an array of A/D and D/A converters for control and
monitoring of magnet currents, rf switches, the MCP HV-1 supply, cell
temperature, and other diagnostics. }
\label{fig:daq}
\end{figure}

An overview of the data acquisition system is given in
Fig.~\ref{fig:daq}.  The sequence of events during a trap cycle is
determined by a program running on a CAMAC crate computer.  When the
laser excitation phase begins, a mechanical shutter near the cryostat
entrance window opens, and laser pulses are allowed to trigger a
timing generator with 8 output channels and sub-microsecond
resolution.  The timing generator allows precise timing control of the
MCP switch and quench (electric field) pulses with respect to the
excitation pulses.  It also gates four counters to count MCP pulses
when signal is present.  When the timing generator sequence is
finished, the CAMAC computer reads out the counters, the counters are
reset, and the timing generator waits to be triggered by the next
laser pulse.

\subsection{Metastable Decay Measurements}
\label{subsec:decaymeasurements}
\begin{figure}
\centerline{\epsfig{file=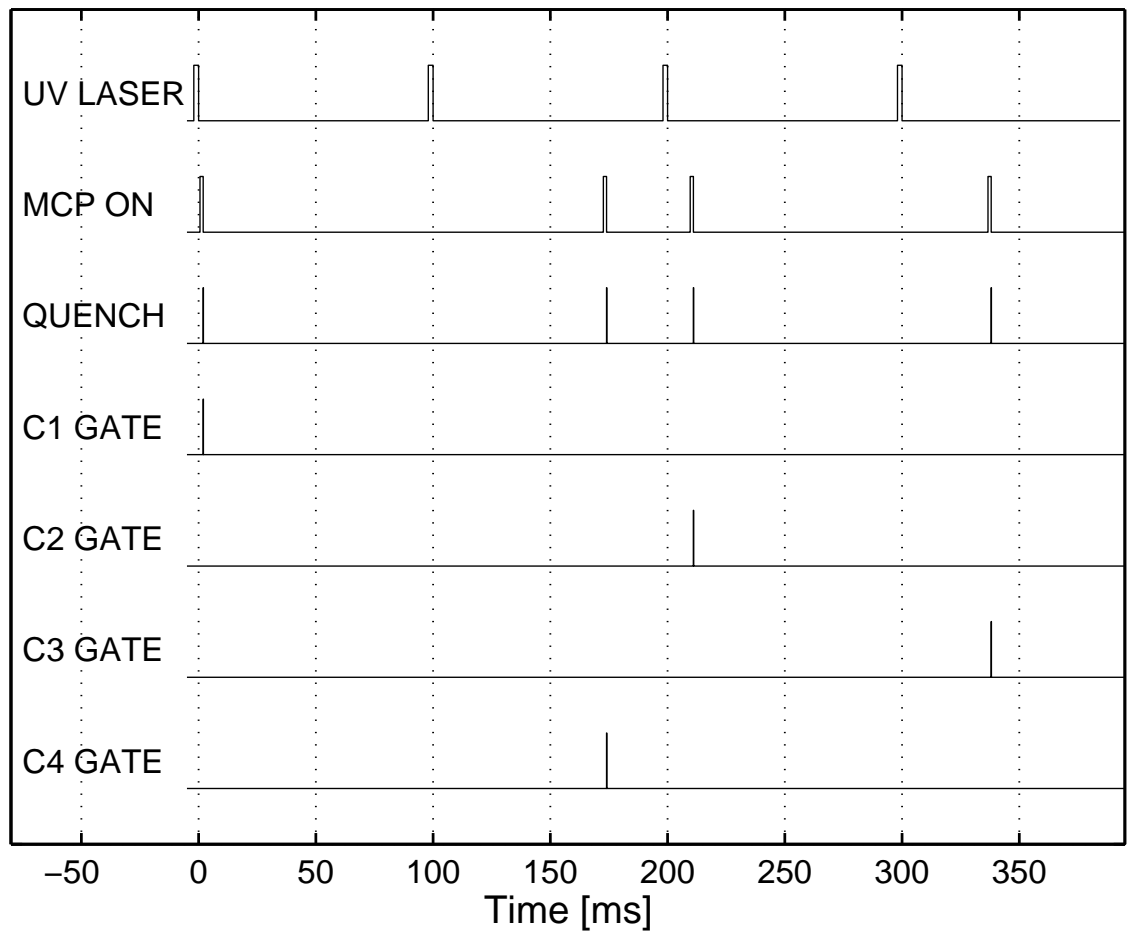,width=6in}}
\caption{Example timing sequence for a single trigger of the timing
generator.  C1-C4 indicate counters 1-4.  The time axis is referenced
to the end of the laser pulse.  In this example, triggering occurs
with the rise of the first laser pulse at -2~ms.  As explained in the
text, an additional delay is introduced before every other trigger,
resulting in eight different wait times for eight consecutive laser
pulses.}
\label{fig:timingdiagram}
\end{figure}

Figure~\ref{fig:timingdiagram} shows the timing sequence used for the
decay measurements described in Ch.~\ref{ch:decay}.  The sequence
determines wait times for quench pulses following four consecutive
laser pulses.  Each quench pulse lasts for 90~$\mu$s, and a counter is
gated on for the first 40~$\mu$s of each.  The MCP is switched on
1.4~ms before each quench in order to fully charge the plates.  To
obtain more time points in a single decay measurement, a delay
generator is enabled before every other trigger, introducing an
additional delay between the laser pulse and the start of the timing
sequence.  For example, the sequence shown in
Fig.~\ref{fig:timingdiagram} contains wait times of 2, 11, 38, and
74~ms (not in this order) after the end of a laser pulse; with an
additional delay of 18 ms before every other trigger, eight
consecutive laser pulses are associated with wait times of 2, 11, 20,
29, 38, 56, 74, and 92~ms.  The accuracy of these delay times is
limited by phase jitter in the mechanical chopper used to pulse the
laser.  In the worst case, this jitter leads to a 0.4~ms wait time
inaccuracy.

\begin{figure}
\centerline{\epsfig{file=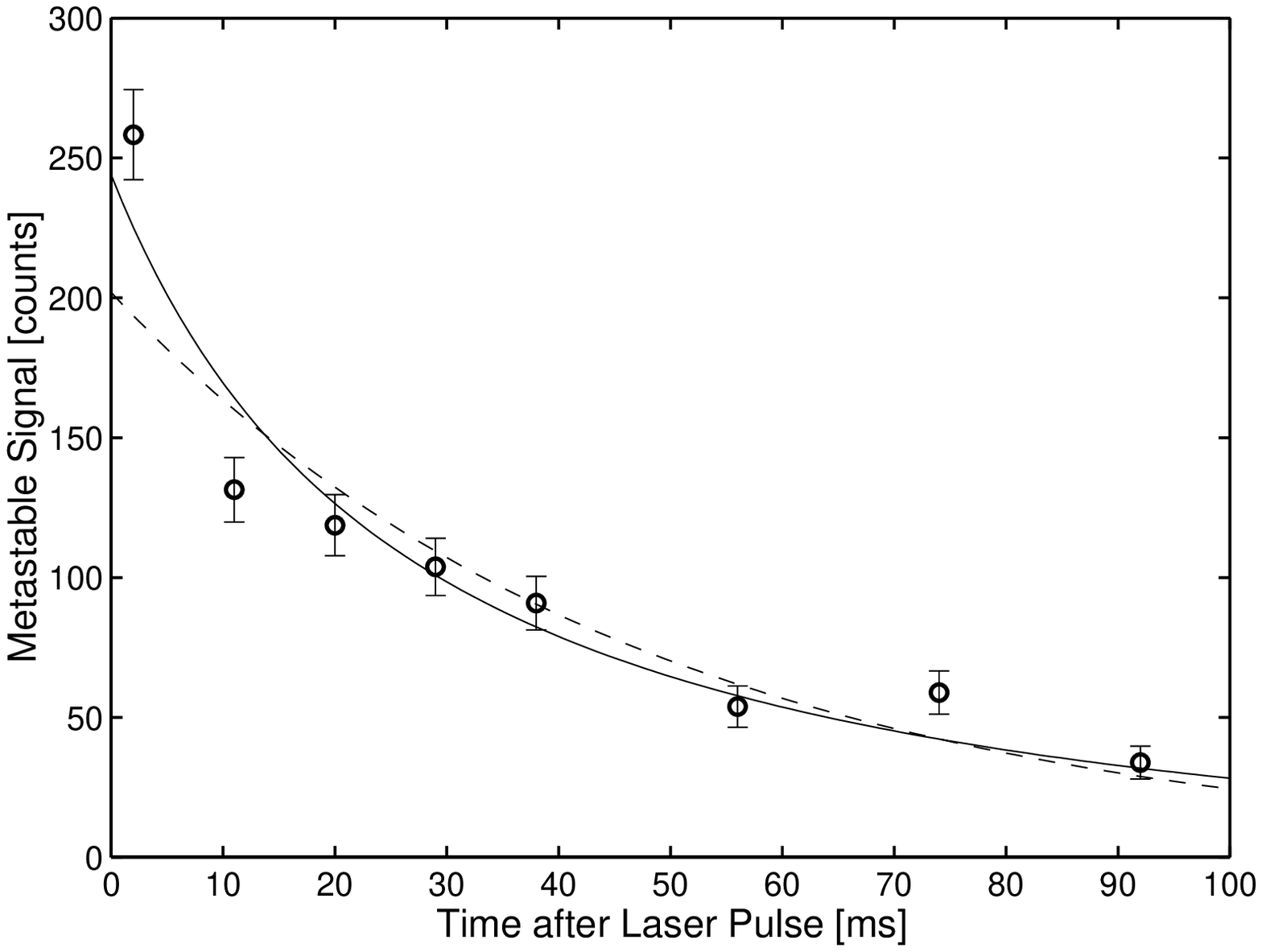,width=5in}}
\caption{A single metastable decay measurement from a $\sim100$~$\mu$K
sample, recorded with eight laser pulses over 800~ms.  The solid line
results from fitting with a model including two-body decay, while the
dashed line is the best fitting simple exponential.}
\label{fig:singledecaynoavg}
\end{figure}

Results from a single eight-point decay measurement are shown in
Fig.~\ref{fig:singledecaynoavg}.  The laser detuning is constant to
assure the same metastable density at the beginning of each decay, and
the ground state density does not change appreciably during the 800~ms
required to make the measurement.  The nonstatistical scatter, which
arises from fluctuations in laser alignment and power, is typical for
a single measurement.  To reduce this scatter, many decay curves made
at similar laser detuning and $1S$ density are averaged together.  As
will be explained further in Chapters~\ref{ch:decay} and
\ref{ch:results}, the decay curves are analyzed by fitting them with
either a simple exponential or a model including both one-body and
two-body decay terms.

\begin{figure}
\centerline{\epsfig{file=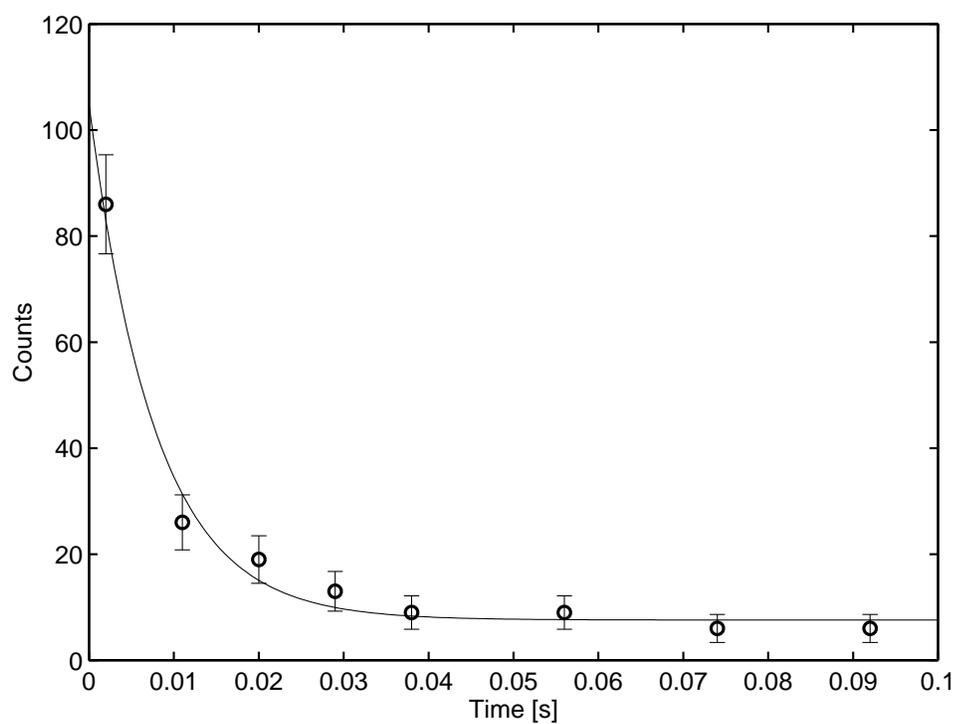,width=5in}}
\caption{Decay of background fluoresence after a laser pulse.  The
data here is the sum of 50 decay curves taken with the laser off
resonance.  The decay is adequately described by the sum of an exponential decay and a constant.}
\label{fig:backcorr}
\end{figure}

In the example of Fig.~\ref{fig:singledecaynoavg}, the raw data has
been corrected for laser-induced background fluorescence, which has
its own characteristic decay behavior (Fig.~\ref{fig:backcorr}).
Although the MCP is $10^4$ times less sensitive to 243~nm than to
Lyman-$\alpha$, fluorescence photons at 243~nm and longer wavelengths
are so numerous after each laser pulse that a few give rise to MCP
pulses.  The background fluorescence decay is measured at the end of
each trap cycle by detuning the laser far off resonance and recording
decay curves in the manner described above.  The fluorescence decay
can be fit to the sum of a fast exponential component and a very
long-lived component well approximated by a constant over 100~ms.
This fit is used to subtract the background contribution from
individual decay curves.  For most data, however, the background
correction results in only small changes to the metastable decay
curves.

\subsection{Stray Field Compensation}
\label{sec:strayfield}
\begin{figure}
\centerline{\epsfig{file=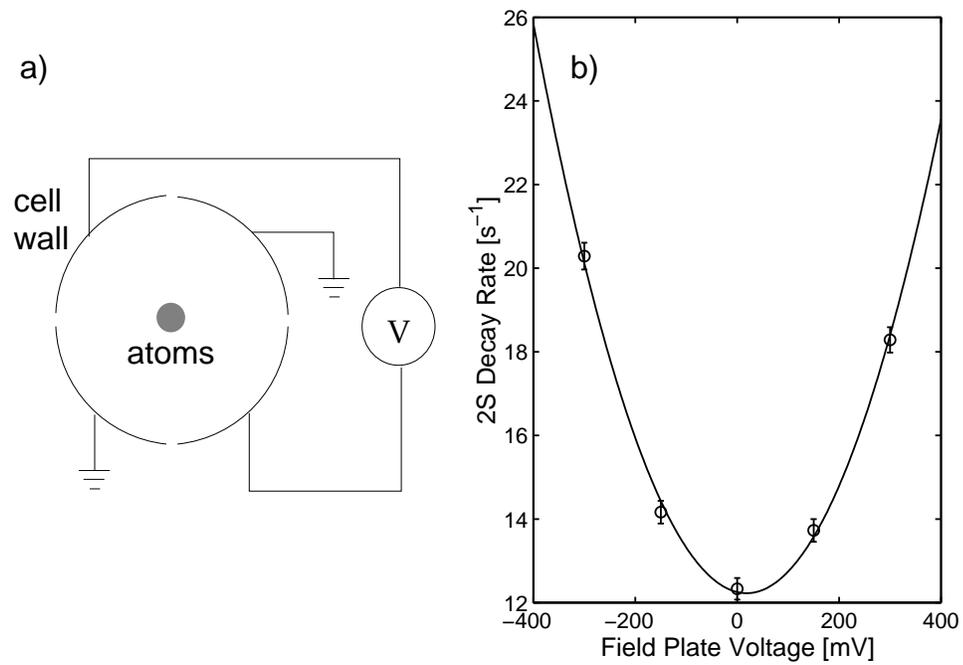,width=5in}}
\caption{(a) Schematic of trapping cell in cross section, showing
electrical connections for the copper film electrodes used for both
quench pulses and stray field compensation.  (b) Decay rate measurements
for different dc electric fields in a single trapped sample.  The
expected parabolic dependence is observed.}
\label{fig:strayfieldcomp}
\end{figure}

In the presence of an electric field of strength $E$, metastable H
quenches at a rate 
\be
\gamma_s=2800E^2 ({\rm cm}^2/{\rm V}^2){\rm s}^{-1}
\label{eq:gammas}
\ee
due to Stark-mixing with the $2P$ state \cite{bs77p287}.  To minimize the
stray field in our apparatus, we apply a compensation dc offset voltage
across the copper film electrodes used for detection quench pulses
(Fig.~\ref{fig:strayfieldcomp}(a)).  The total electric field 
${\bf E}_{\rm tot}$ experienced by the atoms is the sum of applied and stray
fields:
\be
{\bf E}_{\rm tot} = {\bf E}_a + {\bf E}_s.
\ee
At any given point in space, the rate $\gamma_s$ is proportional to
\be
E_{\rm tot}^2 = (E_a + E_{\|})^2 + E_{\bot}^2,
\label{eq:quaddeponEa}
\ee
where $E_{\|}$ and $E_{\bot}$ are, respectively, the stray field
components perpendicular and parallel to the applied field.  If both
the applied field and the stray field are nearly uniform over the atom
cloud, then the optimal compensation voltage is the one which gives
$E_a=-E_{\|}$.  In a situation where only one-body loss mechanisms are
important, the total decay rate is the sum of $\gamma_s$, the natural
decay rate of 8.2~s$^{-1}$, and any other one-body rates which may be
present.  One-body loss dominates in metastable clouds excited from warm, low-density samples.
In this case, Eqs.~\ref{eq:gammas} and \ref{eq:quaddeponEa} imply a
parabolic dependence of the total decay rate on $E_a$, as shown in
Fig.~\ref{fig:strayfieldcomp}(b).  The best compensation voltage is
determined by fitting a parabola to the one-body decay
rates measured at several dc voltages.  The minimum of the parabola
has been found to be stable over several hours.

\label{subsec:strayfieldcomp}

In the current apparatus, two of the electrodes are hard-wired to the
electrical ground of the cryostat.  This means that stray field
compensation is only possible along one direction, and there is a
residual stray field with magnitude $E_{\bot}$.  An upper limit for
the decay rate due to $E_{\bot}$ is 4~s$^{-1}$, obtained by subtracting
the natural decay rate from the rate at the parabola minimum.  This
implies that the residual stray field is less than 40 mV/cm.

In future experiments, it should be possible to float all four
electrodes, allowing compensation in two dimensions.  The stray field
component in the third dimension, along the cell axis, is probably
very small.


\chapter{Trapped Metastable Hydrogen}
\label{ch:theory}
The physics of metastable hydrogen excited in a magnetic trap is
potentially very rich.  Metastable H can participate in several types
of inelastic collisions, including ones that result in energetic ions.
Diffusion is also important.  Due to the presence of the $1S$
background gas, a metastable cloud does not immediately fill the
volume of the ground state sample, but instead diffuses slowly outward
from a region defined by the focus of the excitation laser.  In
addition, the intense UV radiation can subsequently photoionize many
of the metastables.

This chapter summarizes some of the microscopic and macroscopic
physics relevant to metastable H in a trap.  An understanding of this
physics is important for both the interpretation of metastable decay
measurements described in Ch.~\ref{ch:results} as well as the
preparation for future high resolution spectroscopy measurements.

\section{Collisions between Metastables}
Stimulated by our experimental work, theoretical investigations of
\twoStwoS\ collisions were begun recently.  Jonsell \etal\ have
calculated the complex interaction potentials between two metastable
hydrogen atoms \cite{jsp02}. Forrey \etal\ used the potentials to
compute elastic, double excitation transfer, and total ionization
cross sections
\cite{fcd00}.  The initial calculations neglected the Lamb shift and
fine structure, and the results were considered valid for thermal
energies large compared to the splittings between $2S_{1/2}$,
$2P_{1/2}$, and $2P_{3/2}$ states (that is, for $T\gg1$~K).  In the
past year, Forrey, Jonsell, and collaborators have included the Lamb
shift and fine structure in their potentials, allowing the first
calculations of elastic and inelastic \twoStwoS\ rate constants down
to $T=0$ \cite{fjf01}.  The effect of the hyperfine interaction has
not yet been included, but the rate calculations which neglect hyperfine
structure are expected to be accurate to within an order of magnitude
\cite{for01}.

As described in Ch.~\ref{ch:intro}, the trapped metastables are in the
$d$ state $(F=1, m_F=1)$ of the $2S$ manifold.  Just as with the $1S$ $d$-state
atoms, dipolar decay collisions can change the hyperfine state of the
metastables, causing them to be lost from the trap.  The rate
constants for the inelastic $d+d$ channels of the $2S$ manifold have
not been calculated.  However, the approach of Stoof, \etal\ for $1S$
dipolar decay collisions \cite{skv88} is applicable to the $2S$ case
as well.  The rate for $2S$ dipolar decay is governed by weak
magnetic dipole interactions between atoms.  These interactions in the
\twoStwoS\ case are similar in strength to the \oneSoneS\ case, and
the corresponding inelastic rates are not expected to differ by more
than an order of magnitude \cite{sto01}.  At the metastable densities
achieved in the experiments described in Ch.~\ref{ch:results}, $2S$
dipolar decay should be undetectable.

Several \twoStwoS\ collision processes which may occur in the trap are
outlined below in more detail.  Unless denoted otherwise, the label
``$2S$'' refers to the magnetically trapped $d$ state of the $2S$ manifold.

\boldmath
\subsection{Summary of \twoStwoS\ Collision Processes}
\unboldmath

\label{subsec:twoStwoSsummary}

\noindent{\bf Elastic Collisions.}
\be
H(2S) + H(2S) \longrightarrow H(2S) + H(2S).
\ee
According to recent theoretical calculations \cite{fjf01}, these
collisions occur more frequently than \twoStwoS\ inelastic collisions
at all temperatures accessible in our trap.  However, due to the fact
that the metastable density is several orders of magnitude smaller
than the ground state density, elastic \twoStwoS\ collisions are much
less frequent than elastic \oneStwoS\ collisions.  They contribute
insignificantly to the dynamics of the metastable cloud, and they are
not detectable in current experiments.

\vspace{0.25in}

\noindent{\bf Excitation Transfer.}  During a collision between
metastables, either one or both of the atoms may undergo a transition
to the $2P$ state.  The processes are formally known as single
excitation transfer,
\be
H(2S) + H(2S) \longrightarrow H(2P) + H(2S),
\ee
and double excitation transfer, 
\be
H(2S) + H(2S) \longrightarrow H(2P) + H(2P).
\ee
These collisions are, in principle, directly detectable because the
$2P$ products radiate Lyman-$\alpha$ after a lifetime of 1.6~ns.  At
high temperatures, where the $2S$ and $2P$ levels can be considered
degenerate, the allowed molecular symmetries for a pair of metastables
do not permit single excitation transfer
\cite{fcd00}.  Single excitation transfer may be allowed at low
temperatures, but it would require coupling between the orbital angular
momentum of the nuclei and the electronic angular momentum.  The
calculations of Forrey, \etal\ to date have assumed that this coupling
is zero.  Forthcoming theoretical work will determine whether this
assumption is justified for spin-polarized metastables \cite{for01}.

The rate constant for excitation transfer collisions is predicted to
be larger than the rate constant for ionizing collisions at all
temperatures.   

\vspace{0.25in}

\noindent{\bf Ionization.}  Two types of ionization can occur when
metastables collide.  One type is Penning ionization,
\be
H(2S) + H(2S) \longrightarrow H(1S) + H^{+} + e^{-},
\ee
which liberates 6.8~eV ($7.9\times10^4$~K) of energy.  Most of the
energy is imparted to the electron, which leaves the trap instantly.
A fraction of the energy approximately equal to $m_e/m_p$, the
electron-to-proton mass ratio, becomes the kinetic energy associated
with the center of mass of the H ion (proton) and ground state atom.
This small fraction still amounts to more than 40~K, and the proton
and atom will generally leave the trapped sample within a couple
microseconds.  The magnetic field at the center of the trap is far too
weak to hold the proton in a cyclotron orbit; the proton will quickly
move away from the trap axis and towards the cell wall, guided by the
radially-pointing magnetic field of the quadrupole magnets.

The second type of ionization is associative,
\be
H(2S) + H(2S) \longrightarrow H_2^{+} + e^{-}.
\ee
Between 6.8~eV and 9.45~eV of translational energy is liberated,
depending on the rovibrational state of the molecular ion.  By
arguments similar to those of the preceding paragraph, both the ion
and electron products leave the trap quickly.

The cross section for associative ionization has been measured in beam
experiments for collision energies down to 4.1~meV (48~K)
\cite{ucj92}.  At this energy, the measured cross section for
associative ionization is 100~times smaller than the theoretical total
ionization cross section \cite{fcd00}, indicating that Penning
ionization is far more prevalent than associative ionization.  The
experiments showed that the associative cross section varied as
$E^{-1}$ at high temperatures, while theory predicts the total
ionization cross section to vary as $E^{-2/3}$.  It has not yet been
established whether Penning ionization also dominates at low temperatures.

\vspace{0.25in}

\noindent{\bf Dipolar Decay.}    
Due to the electron-electron and electron-proton magnetic dipole
interactions between two $2S$ atoms, inelastic collisions can occur in
which the sum of magnetic quantum numbers $m_F$ of the two atoms is
not conserved.  The hyperfine state of one or both atoms can change
from $d$ to another hyperfine state which is weakly-trapped or
anti-trapped.  Under typical trap conditions, enough kinetic energy is
liberated in the transition to eject even the weakly-trapped products
from the trap.  Very generally, these dipolar decay transitions can be
written
\be
H(2S,d) + H(2S,d) \longrightarrow H(2S,\mu) + H(2S,\nu),
\label{eq:twoSdipdecay}
\ee
where at least one of $\mu$ and $\nu$ is not $d$.  These magnetic
interactions are weak compared to the central forces responsible for
other inelastic processes.  For ground state H, the sum of rate
constants for dipolar decay of $d$ states was determined to be
$\sim10^{-15}$~cm$^3$/s across the range of temperatures and magnetic
fields relevant to hydrogen trapping \cite{skv88,rbj88}.  The same
order of magnitude is expected for the sum of rates corresponding to
Eq.~\ref{eq:twoSdipdecay}.

\subsection{Theoretical Rate Constants}
\label{sec:theoreticalrateconstants}
\begin{figure}
\centerline{\epsfig{file=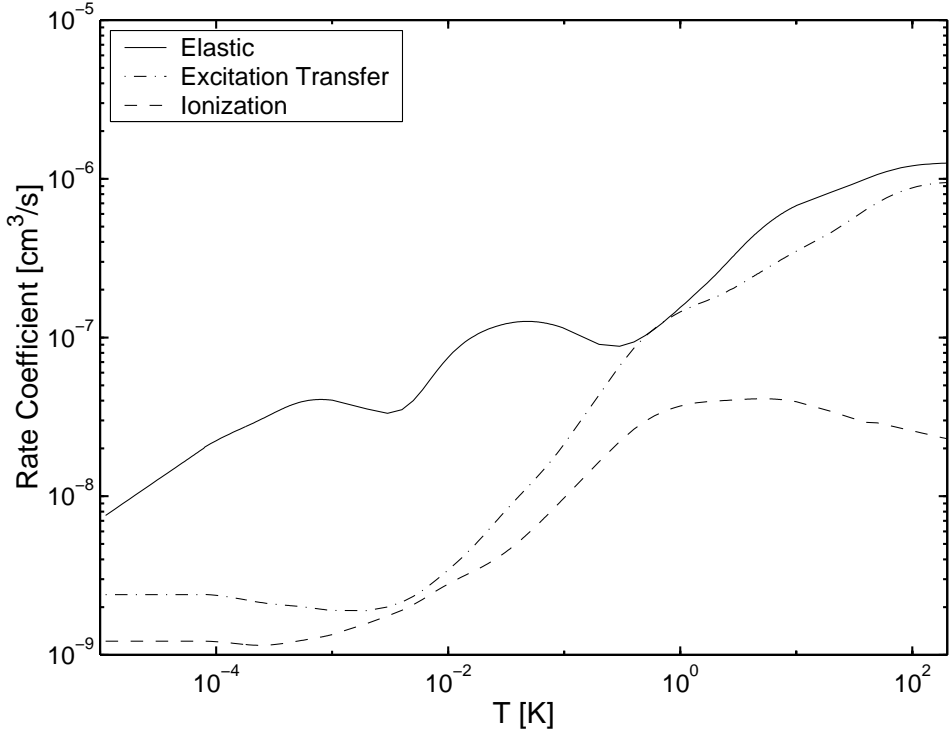,width=5in}}
\caption{Theoretical rate constants for \twoStwoS\ collisions as a
function of temperature, courtesy of Forrey, \etal\ }
\label{fig:forreyrates}
\end{figure}

The rate constants for double excitation transfer and ionizing
\twoStwoS\ collisions have recently been calculated down to $T=0$ and
are summarized in Fig.~\ref{fig:forreyrates}.  The methods of
calculation are outlined in \cite{fcd00} and a forthcoming paper
\cite{fjf01}.  For each collision channel $i$, the corresponding rate
constant $R_i$ is defined such that $R_i n_{2S}^2$, where $n_{2S}$ is the
local density of metastables, is the local rate of collision events of
type $i$.  If we assume that double excitation transfer and
ionization are much more prevalent than other inelastic
collisions, the total two-body loss constant $K_2$ can be expressed
\be
K_2 = 2(R_{\rm et}+R_{\rm ion}),
\label{eq:K2fromtheory}
\ee
where $R_{\rm et}$ and $R_{\rm ion}$ are the rates for double
excitation transfer and ionization.  The factor of 2 appears since two
metastable atoms are lost from the trap in each collision event.
Chapter~\ref{ch:results} describes how $K_2$ was determined experimentally.
 
As can be seen from Fig.~\ref{fig:forreyrates}, below $10^{-4}$~K the
elastic collision rate varies as $T^{1/2}$, and the inelastic rates
are nearly independent of temperature.  In this regime, only
$s$-wave collisions are important, and both the elastic and inelastic
collisional behavior of metastable H are characterized by a single
quantity, the complex scattering length $a_{2S-2S}$.  In general, for
a complex $s$-wave scattering length $a=\alpha-i\beta$ involving identical
bosons, the elastic cross section is given by
\be
\sigma_{\rm el} = 8\pi(\alpha^2 + \beta^2),
\label{eq:elasticxsection}
\ee  
and the inelastic cross section by
\be
\sigma_{\rm inel} = 8\pi\beta/k,
\label{eq:inelasticxsection}
\ee
where {\bf k} is the relative wave vector of the colliding atoms
\cite{bfd98}.\footnote{For distinguishable particles, the expressions
in Eqs. \ref{eq:elasticxsection} and \ref{eq:inelasticxsection} are
reduced by a factor of 2.}  The elastic cross section at collision
energy $kT$ is thus independent of $T$ while the inelastic cross
section has a $1/k\sim T^{-1/2}$ dependence.  This $1/k$ dependence of
$\sigma_{\rm inel}$ is an example of the Wigner threshold law
\cite{wig48,wbz99}.  To find the rate constants for a thermal sample,
the average over a Maxwell-Boltzmann distribution is taken:
$R_i=\langle \sigma_i v
\rangle$.  The resulting elastic rate goes as $T^{1/2}$, and the inelastic
rate is given by $R_{\rm inel}=8\pi\hbar\beta/\mu$, where $\mu$ is the
two-body reduced mass.  The results of Forrey and co-workers
\cite{fjf01} imply that in the zero temperature limit
\be
a_{2S-2S} = \alpha - i(\beta_{\rm et} + \beta_{\rm ion}),
\ee 
where $\alpha=20.4$~nm,
$\beta_{\rm et}=0.75$~nm, and 
$\beta_{\rm ion}=0.38$~nm.

Since metastable hydrogen can be excited at temperatures below
100~$\mu$K, our experiments are able to probe the theoretically
predicted Wigner threshold regime.  It should be noted, however, that
even up to temperatures as high as 10~mK, the theoretical inelastic
collision rates do not change by more than a factor of 2.

The accuracy of the theoretical rates is difficult to estimate.
Hyperfine structure has not yet been included in the calculations, nor
an allowance for coupling between electronic angular momentum and
nuclear orbital angular momentum.  In addition, the effects of the
trapping magnetic field on the collision rates have not been examined.
For the magnetic fields of a few gauss experienced by colder
samples, the rate constants are predicted to be accurate to within an
order of magnitude.

\section{\sloppy Collisions between Metastable and Ground State Atoms}
In our experiments, the peak metastable densities are typically three
or more orders of magnitude smaller than the peak ground state
density.  Although the \oneStwoS\ cross sections are smaller than the
\twoStwoS\ cross sections, the metastables still collide with ground
state atoms much more frequently than with each other.  Most of the
\oneStwoS\ collisions are elastic, and most of what has been learned
theoretically and experimentally about \oneStwoS\ interactions so far
pertains to elastic collisions.

\boldmath
\subsection{Elastic \oneStwoS\ Collisions}
\unboldmath
With the advent of {\em ab initio} calculations for the short range
\oneStwoS\ molecular potentials, it became possible in the last decade
for theorists to predict $a_{1S-2S}$, the $s$-wave elastic scattering
length.\footnote{More precisely, $a_{1S-2S}$ denotes here the
scattering length corresponding to the $e^3\Sigma_u$ molecular
potential on which a 1S $d$-state atom interacts with a 2S $d$-state
atom.  Other \oneStwoS\ potentials have been studied as well, but they
are not relevant for the present discussion.}  Since \oneStwoS\
collisions are predominantly $s$-wave at the temperatures of magnetically
trapped hydrogen, this single parameter is sufficient to describe the
elastic collisions.  The elastic cross section, $\sigma_{1S-2S}=4\pi
a_{1S-2S}^2$, is independent of collision energy.

Calculations of $a_{1S-2S}$ depend sensitively on the molecular
potential.  A first calculation of $a_{1S-2S}$ was made by Jamieson,
\etal\ \cite{jdd96} using the best available {\em ab initio} potential
of the time, which extended only to 20 bohr.  The result was
$a_{1S-2S}=-2.3$~nm.  The accuracy of this calculation was limited by
(1) the mismatch between the short range potential and the asymptotic
long range potential and (2) uncertainty in the very short range part
of the {\em ab initio} potential.  More recently, Orlikowski and
collaborators recalculated the entire short range potential, extending
it to 44 bohr and realizing a much better match to the long range
potential \cite{osw99}.  From their improved potential, they found
$a_{1S-2S}=-3.0$~nm.  Although the uncertainty is difficult to
estimate \cite{wol01}, this is believed the most accurate calculation
of $a_{1S-2S}$ to date.

As mentioned in Ch.~\ref{ch:intro}, there is a density-dependent
cold-collision shift of the \oneStwoS\ transition frequency due to the
difference in \oneStwoS\ and \oneSoneS\ interactions.  Since the
\oneSoneS\ scattering length is theoretically known to be
$a_{1S-1S}=0.0648$~nm with a relatively small uncertainty
\cite{jdk95}, the cold-collision shift parameter $\chi=\Delta E_{1S-2S}/hn_{1S}$
can be used to measure $a_{1S-2S}$ provided that the relationship
between the shift and scattering lengths is well understood.
According to the arguments given by Killian \cite{kil99}, the energy
shift of the \oneStwoS\ transition for homogeneous weak excitation in
a homogeneous ground state gas of density $n_{1S}$ is
\be
\Delta E_{1S-2S}= \frac{8 \pi \hbar^2 n_{1S}}{m}(a_{1S-2S}-a_{1S-1S}),
\label{eq:killianshift}
\ee
where $m$ is the atomic mass.  The fact that the two scattering
lengths in this formula are multiplied by the same factor follows from
assuming uniform coherent excitation of the many-body wave function of the
ground state gas.  The bosonic bunching correlations which enhance the
\oneSoneS\ interactions are maintained as each $1S$ atom in the
many-body wave function gains $2S$ character.
Eq.~\ref{eq:killianshift} has been derived by Stoof and
Pethick using a microscopic approach \cite{pes01}, and a more
general sum rule extension of Eq.~\ref{eq:killianshift} to
inhomogeneous cases has recently been posited by Oktel and
collaborators \cite{okk02}.  If instead we consider
the case where a $2S$ atom is introduced into a homogeneous $1S$ gas
by means other than coherent excitation, then Hartree-Fock theory
predicts that the energy of the $2S$ state will be shifted only half
as much; this is because $1S$ and $2S$ atoms are distinguishable
particles and, in this case, they are spatially uncorrelated.  The
energy difference between the two states would then be
\be
\Delta E_{1S-2S}= \frac{4 \pi \hbar^2 n_{1S}}{m}(a_{1S-2S}-2a_{1S-1S}).
\label{eq:HFshift}
\ee
Both Eq.~\ref{eq:killianshift} and Eq.~\ref{eq:HFshift} assume that
the contributions of inelastic processes to the energy shifts are
negligible.

The MIT Ultracold Hydrogen Group used Eq.~\ref{eq:killianshift} to determine
$a_{1S-2S}=-1.4\pm0.3$~nm from the experimental cold-collision shift
\cite{kfw98}, which is a factor of two smaller than the value of
Orlikowski, \etal\ \ Interestingly, if Eq.~\ref{eq:HFshift} is used,
$a_{1S-2S}=-2.9\pm0.6$~nm is obtained, in good agreement with the
theoretical value.  This suggests that Eq.~\ref{eq:HFshift} is closer
to the correct description of the cold-collision shift in our
experiment.  Further support for this conclusion is provided by recent
time-resolved spectroscopic measurements of hydrogen BEC's performed in our
group \cite{mos01,mlm02}.  A possible reason why Eq.~\ref{eq:HFshift}
may be favored over Eq.~\ref{eq:killianshift} is that \oneStwoS\
collisions occurring during the excitation may destroy the spatial
correlations of the ground state gas.  Furthermore, it remains an open
question how to correctly describe the cold-collision shift in terms
of scattering lengths for the experimentally realistic case of an
inhomogeneous thermal gas and an inhomogeneous laser field.

In the remainder of this thesis, the theoretical value of $a_{1S-2S}$
will be assumed correct.  This implies an elastic scattering cross
section $\sigma_{1S-2S}=1.1\times10^{-12}$~cm$^2$.  The magnitude of
$\sigma_{1S-2S}$ is important for determining the rate at which
metastables can diffuse through the ground state sample.

\boldmath
\subsection{Inelastic \oneStwoS\ Collisions}
\unboldmath
At least two types of inelastic collisions are possible between
metastables and ground state H in the trap.  First, there is the
analog of dipolar decay,
\be
H(1S,d) + H(2S,d) \longrightarrow H(1S,\mu) + H(2S,\nu),
\label{eq:1S2Shyperfinechanging}
\ee
where at least one of the products is not a $d$-state.  As mentioned
in Sec.~\ref{subsec:twoStwoSsummary}, these processes are
mediated by weak magnetic dipole forces, and their rates should be
small compared to the important \twoStwoS\ inelastic channels.
Second, excitation transfer can occur as follows:
\be
H(1S) + H(2S) \longrightarrow H(1S) + H(2P).
\label{eq:1S2Sexcitationtransfer}
\ee

To our knowledge, no theoretical work has been done to date on
inelastic \oneStwoS\ collisions.  If they occur at significant rates
in the hydrogen trap, then they should be observable as one-body
metastable decay processes; the decay component would be a simple
exponential since the ground state density does not change appreciably
during the metastable lifetime.  An upper limit for these rates based
on decay measurements is presented in Ch.~\ref{ch:results}.

\section{Photoionization}
\label{sec:photoionization}
\begin{figure}
\centerline{\epsfig{file=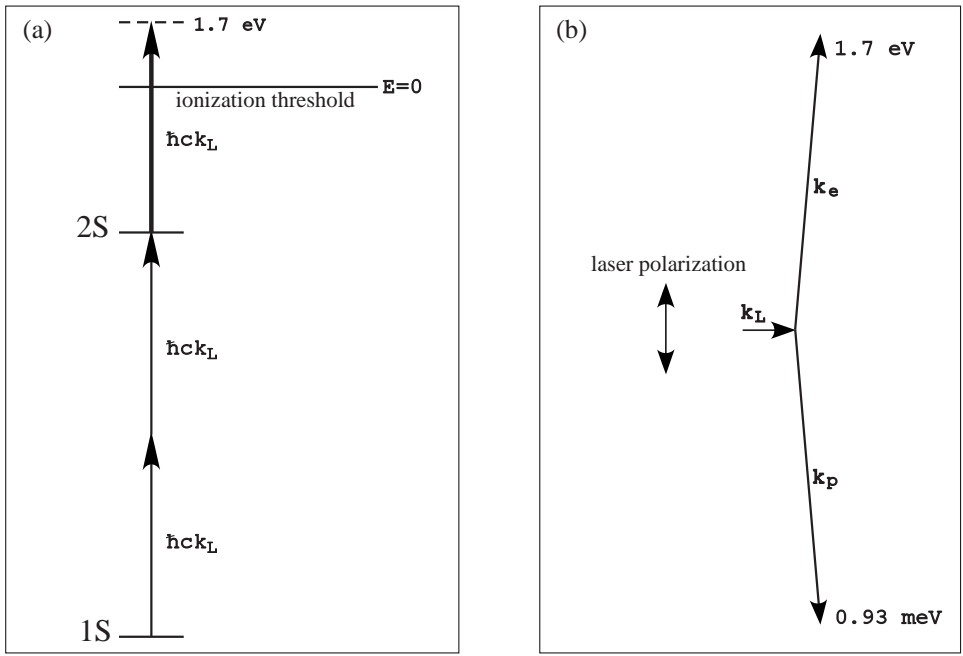,width=5in}}
\caption{(a) Energy level diagram for excitation and photoionization
of metastable hydrogen with a 243~nm laser.  The energy $\hbar c k_L$ of each
photon is 5.1~eV.  (b) Momentum diagram in the lab frame for photoionization of
a trapped metastable H atom.  In the center-of-mass frame of the atom, the most
probable direction of electron emission is along the laser
polarization.  The size of ${\bf k}_L$ in proportion to the
electron wavevector ${\bf k}_e$ and proton wavevector ${\bf k}_p$ has
been exaggerated in this diagram.}
\label{fig:photoionization}
\end{figure}

The 243~nm UV photons which create metastable H can also ionize the
metastables (Fig. \ref{fig:photoionization}).  By cubic interpolation
of the tabulated values from Burgess \cite{bur65}, the cross section for
photoionization is found to be $\sigma_{\rm
pi}=6.2\times10^{-18}$~cm$^2$ at 243~nm.\footnote{Although citing the
same reference, Killian quotes a cross section which is nearly
30\% larger \cite{kil99}.} If we express the
single-atom photoionization rate due to incident laser intensity $I$ as
$\alpha_{\rm pi}I$, the proportionality constant is $\alpha_{\rm
pi}=7.6\times10^{-3}$~(mW/cm$^2$)$^{-1}$s$^{-1}$.  

At warmer sample temperatures ($>1$~mK), the volume of the metastable
cloud which forms during the laser pulse is much larger than the laser
beam, and the fraction of atoms experiencing high laser intensity at
any given time is small.  As a consequence, only a small fraction of
the metastables are photoionized.  In colder samples, however, a
significant fraction of the metastables are in the laser field.
For samples below 100~$\mu$K, photoionization may be the most important
metastable loss process during the excitation pulse.

For a given trap shape, temperature, sample density, and laser
intensity distribution, we can numerically calculate the fraction of
metastables $\cal{F}_{\rm pi}$ which are lost to photoionization during an
excitation pulse.  This calculation is straightforward assuming that
the time required to establish a quasi-static metastable spatial
distribution $f_{2S}({\bf r})=f_{2S}(\rho,z)$ is short compared to the
length of the excitation pulse.  Since the collision time and radial
transit time are both typically 10 times or more shorter than the
pulse length at $\sim100$~$\mu$K, this is a good assumption for colder
samples (see Sec.~\ref{sec:clouddynamics}).

To find $\cal{F}_{\rm pi}$, we consider the number of atoms $N_{z,2S}(z,t)$
in a slice perpendicular to the trap axis at $z$ with thickness $dz$.
In each slice,
\be
\dot{N}_{z,2S}(z,t) = R_{12}(z) - \Phi(z)N_{z,2S}(z,t),
\label{eq:pidiffeq}
\ee
where $R_{12}(z)$ is the total rate of metastable generation due to
laser excitation, and $\Phi(z)$ is the fraction of metastables which
are photoionized per unit time.  The distribution $R_{12}(z)$ can be
calculated using the Monte Carlo simulation of excitation described in
Sec.~\ref{sec:spatialdist}.  If $t=0$ corresponds to the beginning of
the excitation pulse, the solution to Eq.~\ref{eq:pidiffeq} for times
up to the end of the pulse can be written
\be
N_{z,2S}(z,t) = R_{12}(z)C(z,t)t,
\ee
where $C(z,t)$ is a correction factor which expresses the effect of
photoionization on the otherwise linear growth of the metastable
population at $z$.  The time-dependent correction factor is given by
\be
C(z,t) = \frac{1}{\Phi(z)t} \left( 1 - e^{-\Phi(z)t} \right). 
\ee
The fraction photoionized at $z$ per unit time is found by evaluating 
\be
\Phi(z) = 
\frac{\alpha_{\rm pi} \int d\rho\ \rho f_{2S}(\rho,z) I_{\rm tot}(\rho,z)}
{\int d\rho\ \rho  f_{2S}(\rho,z)},
\ee
where $I_{\rm tot}$ is the sum of incoming and return laser beam
intensities.  The fraction of metastables photoionized in a pulse of
length $t$ is the spatial average of $C(z,t)$:
\be
\cal{F}_{\rm pi}(t) = 
\frac{\int f_{2S}({\bf r}) C(z,t)\,d^3{\bf r}}
{\int f_{2S}({\bf r})\,d^3{\bf r}}.
\ee
As an example, for perfectly overlapped UV beams in the 87~$\mu$K
sample of Trap~W in Ch.~\ref{ch:decay}, the fraction of $2S$ atoms
lost to photoionization during a 2~ms excitation pulse is found to be
$\cal{F}_{\rm pi}=0.28$ at 12~mW of UV power and $\cal{F}_{\rm pi}=0.38$ at
18~mW.  The fraction photionized has a weak dependence on the beam
radius, the laser detuning, and the location of the laser focus along
the trap axis.

At $T\sim100~\mu$K the rate of metastable production in the
trap should begin to saturate due to photoionization after a few milliseconds.  This expectation was confirmed in a preliminary experiment
which recorded signal rates in consecutive trap cycles using different
pulse lengths (see Fig.~\ref{fig:saturation}).

A natural question to ask is whether any of the charged particles
generated by photoionization remain in the trap long enough to affect
the metastable cloud.  If enough space charge lingers after a laser
pulse, the metastables could quench in the resulting electric field.
To investigate this, we consider the energy and momentum of the
resulting proton-electron pair (see Fig.~\ref{fig:photoionization}).
Almost all of the of energy above threshold, 1.7~eV
($2.0\times10^4$~K), goes to the electron, causing the electron to
leave the trap instantaneously.  The proton recoils with a fraction
$m_e/m_p$ of the electron energy, about 0.93~meV (11~K).  In the
center-of-mass frame, this recoil is opposite to the direction of
electron emission.  For an $S$-orbital electron, the angular
distribution of photoelectric emission is peaked around the laser
polarization axis (see
\cite{sak94}, for example).  The lab frame does not look much
different from the center-of-mass frame, since the initial momenta of
the photon and atom are small compared to the photoionization
momentum.  In our experiments, the excitation laser is linearly
polarized in a direction perpendicular to the trap axis; this means
that the protons will recoil in a radial direction, exiting the atom
cloud in less than 1~$\mu$s.  Similar to the case of ion collision
products described in Sec.~\ref{subsec:twoStwoSsummary}, the protons
quickly move into the quadrupole field, where the magnetic field is
parallel to the direction of motion.  In this way, the protons leave
the trap without causing much loss in the sample.

\section{Dynamics of the Metastable Cloud}
\label{sec:clouddynamics}
The dynamics of the $2S$ cloud are more complex than the
dynamics of the quasi-equilibrium $1S$ cloud.  This is because the
metastables are excited at the focus of the laser and spread along the
trap axis by diffusion.  At the densities typical of experiments
described in this thesis, the lifetime of the metastables is much
shorter than the time required to establish equilibrium along the axis.

Since Doppler-free excitation imparts no momentum to an atom, each
metastable atom initally possesses the velocity of the ground state
atom from which it was excited.  Different velocity classes in the
ground state gas have different probabilities for excitation, however,
and it is because $2S$ atoms collide frequently with $1S$ atoms that a
Maxwell-Boltzmann velocity distribution is quickly established in the
metastable cloud.  The \oneStwoS\ collision time is between 100~$\mu$s
and 1~ms for typical sample densities, and an equilibrium velocity
distribution is established within a few collision times \cite{sw89}.
Since the mean free path of the $2S$ atoms
$l=(\sqrt{2}n_{1S}\sigma_{1S-2S})^{-1}$ is typically comparable to the
thermal radius of the $1S$ cloud (0.7~mm at 2.5~mK, 0.18~mm at
100~$\mu$K), an equilibrium radial spatial distribution is achieved on
a short time scale.  In other words, the radial extent of the
metastable cloud quickly matches the radial extent of the ground state
cloud.

Much more time is required for the metastable cloud to spread out to
its equilibrium extent in the axial direction, however.  The
metastable cloud is excited in a region along the trap axis which is
defined by the depth of focus of the laser.  Initially, the length of
the $2S$ cloud between its 1/e density points is $\sim2$~cm, while the
length of the ground state cloud varies between 4~cm and 24~cm.  The
distance which $2S$ atoms must travel along $z$ to fill up the
equilibrium trap volume is more than 100 times longer than a mean free
path.  Thus, to approach equilibrium, the metastables must diffuse
along the trap axis.  The time required is typically much longer
than the lifetime of the $2S$ state.  

If we neglect for the moment \twoStwoS\ inelastic collisions, a
metastable cloud excited from a ground state sample with peak density
of at least a few times 10$^{13}$~\percc\ will have a quasi-static
spatial distribution.  This is because (1) the metastable cloud is
axially confined by \oneStwoS\ collisions and (2) radial equilibrium
is maintained over the lifetime of the metastable cloud.  To examine
the transport of the cloud more generally, we must consider the
diffusion of $2S$ atoms in a $1S$ background gas.

To accurately describe the time evolution of the metastable
distribution, the density-dependent losses due to \twoStwoS\ inelastic
collisions must also be considered.  The effects of diffusion and
two-body losses on the metastable cloud shape have been studied using
a dynamic simulation described in Sec.~\ref{sec:dynamicsimulation}.
One of the inputs to this simulation is a parameter describing the
rate of diffusion along the trap axis; the remainder of this section
summarizes an approach for determining the diffusion parameter.

\subsection{Diffusion Constant}
From the results of a hard-sphere kinetic theory, the diffusion
constant for a dilute gas of one species in a background gas of
another species with the same atomic mass is
\be
D = \frac{4}{3\sigma n} \sqrt{ \frac{kT}{\pi m}},
\label{eq:hardsphereD}
\ee
where $n$ is the background gas density, $\sigma$ is the cross section
for collisions between the two species, and $m$ is the atomic mass \cite{oha89}.\footnote{More sophisticated diffusion
theories lead to a constant which is about 10\% smaller
\cite{chc70,dal01}.  However, Eq.~\ref{eq:hardsphereD} is sufficiently
accurate for the arguments presented here.}  The dilute gas atoms are
here assumed to collide much more frequently with the background gas
atoms than with each other.  For the case of $2S$ atoms in a $1S$ gas,
Eq.~\ref{eq:hardsphereD} becomes
\be
D=6.1\times10^{15} \frac{T^{1/2}}{n_{1S}}~{\rm cm}^{-1}{\rm
s}^{-1}{\rm K}^{-1/2},
\ee
where the theoretical value of $\sigma_{1S-2S}$ has been used.

If the length scales for variations in $n_{1S}$ were much larger than
the $2S$ mean free path in all directions, then a
three-dimensional diffusion equation
\be
\frac{\partial n_{2S}}{\partial t}=D \nabla^2 n_{2S}
\ee
would describe the transport of the metastables.  However, as
mentioned above, an equilibrium distribution is rapidly established in
the radial direction.  Futhermore, the $2S$ atoms oscillate through
the radial density distribution at a frequency comparable to the \oneStwoS\
collision frequency.  A better approach is to find the effective diffusion
constant along the trap axis as a function of the peak ground
state density.  In this way, metastable transport in the trap can be
described as a one-dimensional diffusion process.

\begin{figure}
\centerline{\epsfig{file=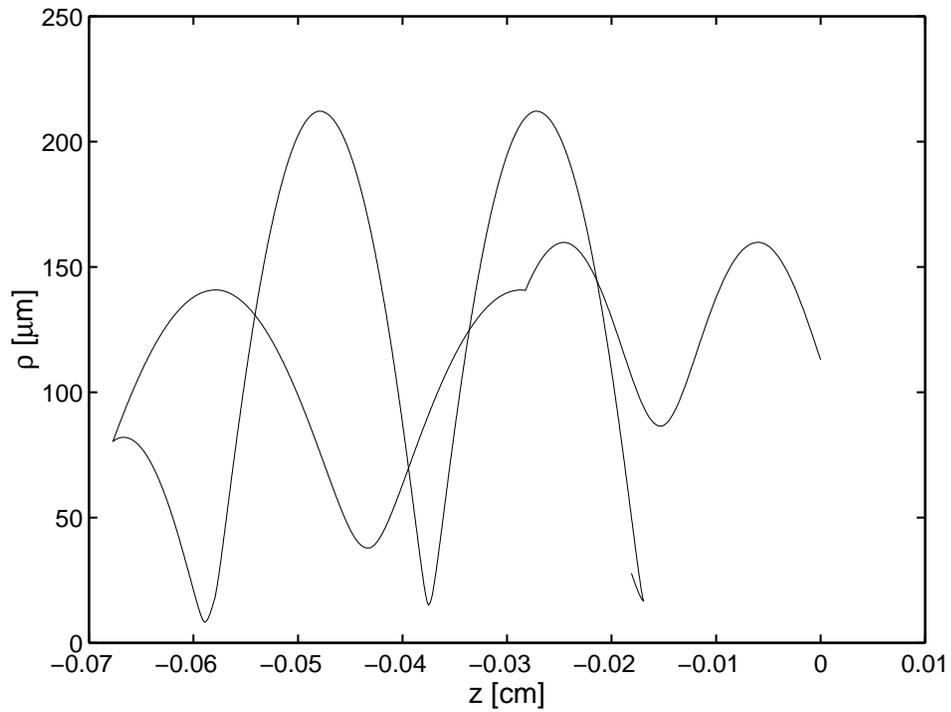,width=5in}}
\caption{Trajectory of a metastable atom in the $z$-$\rho$ plane over a
2~ms period, generated by Monte Carlo simulation.  The trajectory
begins at $z=0$, and kinks are evident at the points where
momentum-changing collisions with ground state atoms occur. The
regular oscillations in $\rho$ are due to quasi-elliptical orbits
about the trap axis occurring at a frequency of $\sim3$~kHz.}
\label{fig:trajectory}
\end{figure}

\begin{figure}
\centerline{\epsfig{file=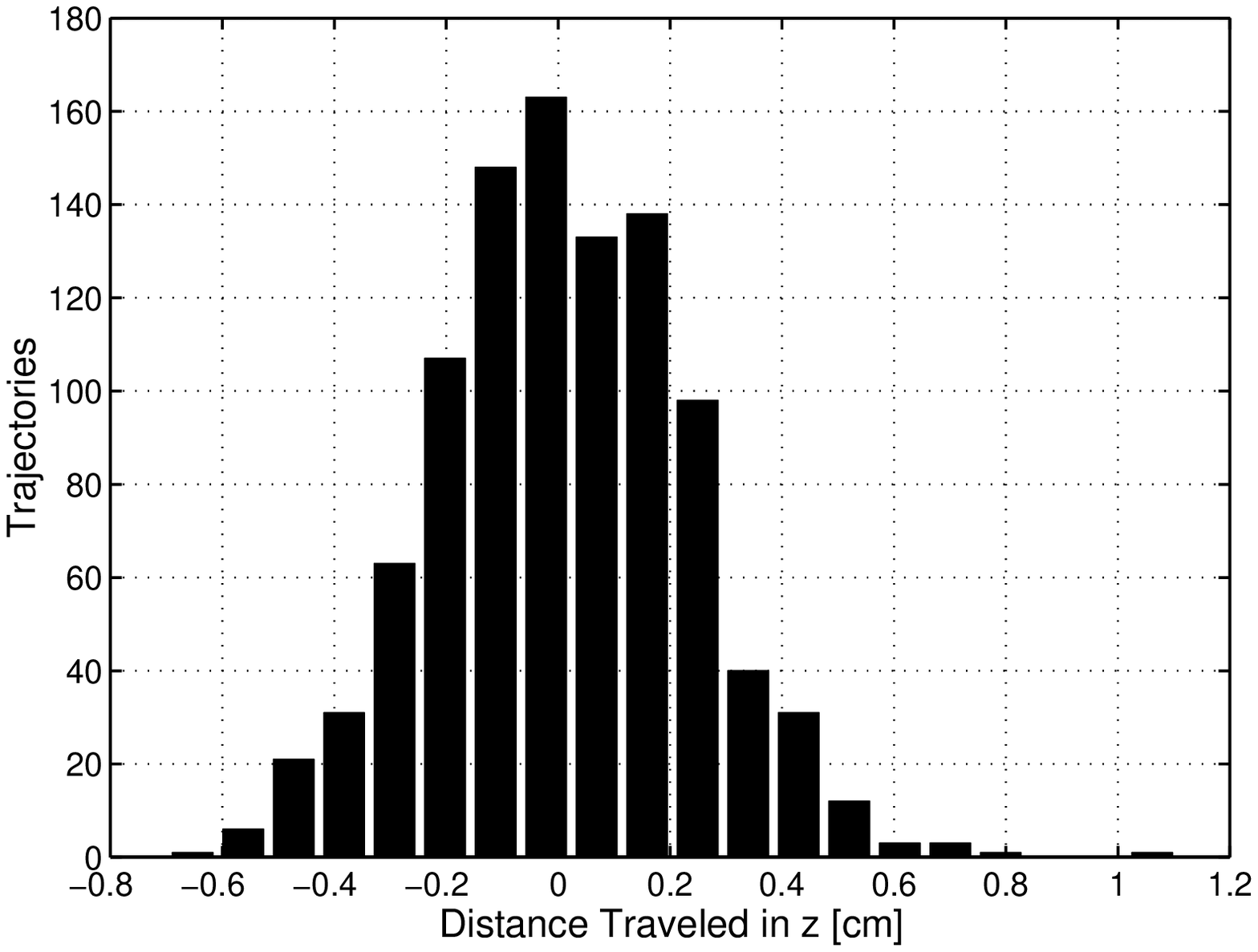,width=5in}}
\caption{Histogram of net distance traveled in $z$ for 1000
metastable atom trajectories followed over 20~ms.  The temperature is
87~$\mu$K, and peak $1S$ density is $5.7\times10^{13}$~\percc.  These
are close to the peak density conditions of Trap~W in Ch.~\ref{ch:results}.}
\label{fig:disthist}
\end{figure}

\subsection{Monte Carlo Diffusion Simulation}
\label{sec:diffsim}
The rate of diffusion in $z$ has been studied by Monte
Carlo simulation of metastable atom trajectories.  The computer
simulation uses a normally-distributed random number generator to
select the components of an initial $2S$ atom velocity ${\bf v}$ from
the Maxwell-Boltzmann distribution.  The equations of motion for the
atom are integrated in a trap potential $U$ whose shape is specified
by the sum of an axial bias field and a radial field gradient.  To
simplify the coding, the variation of the trapping field in $z$ is
neglected.  Collisions are simulated by taking time steps $dt$ much
shorter than the average collision time and using random numbers to
test for a collision in each time step.  The collision probability per
step is $n_{1S}\sigma_{1S-2S}v~dt$, where $n_{1S}=n_{1S,o}e^{-U/kT}$.
After each collision, a new random velocity is selected.

Fig.~\ref{fig:trajectory} shows an example trajectory calculated in
this way.  In a typical simulation, 1000 such trajectories are
generated, each covering a period of tens of milliseconds.  Statistics
are collected for the trajectories, including the distribution of
distances traveled in $z$ (Fig.~\ref{fig:disthist}).  The rms width of
the distribution, $\sigma_z$, serves as a measure of the
characteristic diffusion length for the given trap conditions.  At a
peak $1S$ density $n_{1S,o}=6\times10^{13}$~\percc and $T=90~\mu$K
(Trap~W of Ch.~\ref{ch:results}), $\sigma_z$ is 2.3~mm after a period
$t=20$~ms.  The \oneStwoS\ collision rate is found to be about half
that expected in a uniform sample at the peak density of the trap.

It was verified that $\sigma_z$ is proportional to $t^{1/2}$ and
inversely proportional to $n_{1S,o}^{1/2}$, as expected for a diffusion
process.  The distribution of distances traveled did not depend on the
choice of initial $\rho$.

\section{Fluorescence in a Quenching Field}
For the measurements in this thesis, metastables were detected by applying a
quenching electric field and observing the Lyman-$\alpha$
fluorescence.  The recoil energy associated with this photon is
670~$\mu$K.  Whether or not the recoiling atom leaves the trap depends
on the depth of the trap and the ground hyperfine state into which the
atom decays.  The relevant traps in Ch.~\ref{ch:results} are all
deeper than the recoil energy for the $\ket{1S_d}$ state, but not for the
weakly-trapped $\ket{1S_c}$ state.  Generally speaking, atoms returning to
the $\ket{1S_d}$ state in these relatively deep traps remain trapped,
but atoms returning to other hyperfine states will leave the sample.

To calculate the branching ratios of trapped metastable H in a
quenching electric field, it is necessary to know the eigenstate
$\ket{2S'_d}$ in the presence of simultaneous electric and magnetic
fields.  Numerical diagonalization of a Hamiltonian including
interactions with weak fields {\bf E} and {\bf B} reveals that the state is
\begin{eqnarray}
\ket{2S'_d}=\ket{2S\uparrow\upproton} - 10^{-3}E\Bigl[ 
({\rm cos}\,\gamma)\Bigl( 
1.82\ket{\uparrow\:\downarrow\upproton} 
- 0.90\ket{\rightarrow\:\uparrow\upproton}
\Bigr) + 
\nonumber\\
{} + ({\rm sin}\,\gamma)
\Bigl(
1.26\ket{\rightarrow\:\downarrow\upproton} 
- 1.51\ket{\downarrow\:\uparrow\upproton}
\Bigr)
\Bigr],
\label{eq:twoSdprime}
\end{eqnarray}
where $E$ has units of V/cm, and $\gamma$ is the angle between {\bf E}
and {\bf B} \cite{san93}.  The first label in each ket on the
right-hand-side of Eq.~\ref{eq:twoSdprime} indicates the orbital
state, while the second and third labels indicate the electron and
proton spin states, respectively.  A shorthand notation has been used
for the $2P$ components: $\uparrow$, $\rightarrow$, and $\downarrow$
as a first label denote the orbital states $\ket{2\:1\:1}$,
$\ket{2\:1\:0}$, and $\ket{2\:1\:-\!\!1}$ of the $\ket{n\:L\:m_L}$
basis, respectively.  The expression is valid provided both the Stark
shift and Zeeman shift are small compared to the Lamb shift separating
$2S$ and $2P$; the experimental situation is well within this
regime.  The relative probability of decaying to each of the ground
hyperfine states, which in the relevant magnetic field regime are
\begin{eqnarray}
\ket{1S_a} & = & \frac{1}{\sqrt{2}}\left[\ket{1S\downarrow\upproton}-\ket{1S\uparrow\downproton}\right] \\
\ket{1S_b} & = & \ket{1S\downarrow\downproton} \\
\ket{1S_c} & = & \frac{1}{\sqrt{2}}\left[\ket{1S\uparrow\downproton}+\ket{1S\downarrow\upproton}\right] \\
\ket{1S_d} & = & \ket{1S\uparrow\upproton} ,
\end{eqnarray}
is obtained from the squares of the dipole matrix elements
between these states and $\ket{2S'_d}$.  For a fixed position in the
trap, the fraction of atoms $F_d$ returning to $\ket{1S_d}$ is
\be
F_d = \frac{0.81 + 2.28\,{\rm tan}^2\gamma}{4.12 + 3.87\,{\rm tan}^2\gamma}.
\ee
This ratio ranges from 20\% when {\bf E} and {\bf B} are parallel to
59\% when the fields are perpendicular.  To find the branching ratio
for the sample as a whole, $F_d$ must be averaged over the metastable
distribution.  One way to calculate this average is by a Monte Carlo
simulation, described in Sec.~\ref{sec:radtrap}.  It turns out that
nearly 50\% of the metastables in a typical trap return to
$\ket{1S_d}$, subsequently thermalizing with the ground state cloud.

\begin{table}
\begin{center}
\begin{tabular}{||c|c|c||}
\hline
Polarization & Contributing Matrix Elements & Total Probability \\
\hline \hline
     & $\braOket{1S_d}{z}{2S'_d}$ &  \\                          
$\pi$& $\braOket{1S_c}{z}{2S'_d}$ & $(0.78\,{\rm
     sin}^2\gamma+0.81)/{\cal N}$ \\
     & $\braOket{1S_a}{z}{2S'_d}$ & \\ \hline
$\sigma^-$ & $\braOket{1S_d}{x-iy}{2S'_d}$ & $2.28\,{\rm
     sin}^2\gamma/{\cal N}$ \\ \hline
 & & \\
$\sigma^+$ & $\braOket{1S_a}{x+iy}{2S'_d}$ & $3.31\,{\rm sin}^2\gamma/{\cal
     N}$ \\
           & $\braOket{1S_c}{x+iy}{2S'_d}$ & \\ \hline
\end{tabular}
\end{center}
\caption{Probability of a Lyman-$\alpha$ fluorescence photon having each of the three possible polarizations as a
function of $\gamma$, the angle between {\bf B} and the quenching
electric field {\bf E} at the location of the emitting atom.  The
probabilities are calculated by summing the squares of the
contributing dipole matrix elements.  The normalization factor is
${\cal N}=0.25\,{\rm cos}^2\gamma + 3.87$.}
\label{tab:polarizations}
\end{table}

\begin{figure}
\centerline{\epsfig{file=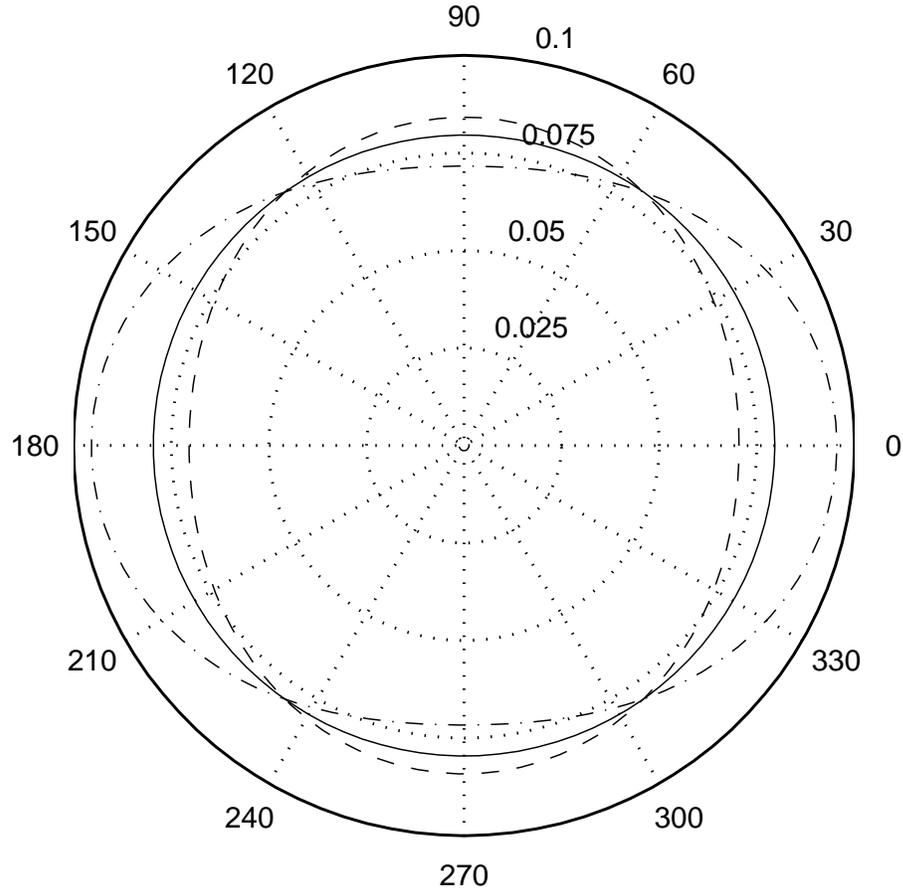,width=5in}}
\caption{Normalized angular distributions of Lyman-$\alpha$
fluorescence for the cases of ${\bf E}\perp{\bf B}$ (dashed) and
${\bf E}\parallel{\bf B}$ (dot-dashed).  The solid curve connecting
the intersection of the distributions is a circle with radius
$1/4\pi$, representing the case of isotropic fluorescence.  The angles
are in degrees with respect to the direction of the magnetic field.
All photons emitted at 0 and 180 degrees have a circular
polarization, while all photons emitted at 90 and 270 degrees have a
linear polarization.}
\label{fig:fluorescencedist}
\end{figure}

When examining the effect of sample geometry on the detection
efficiency for metastables, it is necessary to consider the
polarization of the emitted Lyman-$\alpha$ photons.  The polarization
of a photon emitted in a transition between specific $2P$ and $1S$
hyperfine states is determined by the change in $m_L$ represented in
the contributing dipole matrix element.  The relative probabilities of
obtaining $\pi$, $\sigma^+$, and $\sigma^-$ polarization as a function
of $\gamma$ are summarized in Table~\ref{tab:polarizations}.  The
dipole radiation patterns with respect to the orientation of {\bf B}
depend on the polarization (Table~\ref{tab:angulardist}) \cite{bj83}.
For a quenched metastable, taking into account the probability of
emitting into each polarization, the angular probability distribution
of fluorescence is shown in Fig.~\ref{fig:fluorescencedist} for the
limiting cases of $\gamma = \pi/2$ and $\gamma=0$.  In the lab frame,
the far-field angular distribution can be obtained by taking a spatial
average over the local emission patterns in the metastable cloud.

\begin{table}
\begin{center}
\begin{tabular}{||c|c||}
\hline
Polarization & Angular Distribution \\
\hline \hline
                      &  \\
$\pi$                 & $\frac{3}{8\pi}\,{\rm sin}^2\beta$ \\
		      &  \\ \hline
		      &  \\
$\sigma^-,\:\sigma^+$ & $\frac{3}{16\pi}(1+{\rm cos}^2\beta)$ \\

		      &  \\ \hline
\end{tabular}
\end{center}
\caption{Angular distribution of fluorescence as a function of $\beta$,
the angle between {\bf B} and the direction of photon emission, for
each of the three possible polarizations.  The distributions are
normalized so that integration over all solid angles results in unity.}
\label{tab:angulardist}
\end{table}

\section{Radiation Trapping}
\label{sec:radtrap}
The fluorescence photons which result from metastable quenching have a
frequency very nearly equal to the \oneStwoS\ transition frequency.
Since this frequency is only detuned by $1.1$~GHz (about 11 natural
linewidths of the Lyman-$\alpha$ line) from the strong $1S$-$2P_{1/2}$
transitions, the photons have a non-negligible probability of
absorption as they move through a dense $1S$ sample.  This phenomenon
is known as radiation trapping, and it has consequences for the
detection efficiency of metastable atoms when using Lyman-$\alpha$
fluorescence as the means of detection.  In particular, the detection
efficiency can potentially vary with the $1S$ density, the trap shape,
and also with the geometry of the $2S$ cloud.  This section, which
extends the work of \cite{san93}, describes calculations relevant to
predicting the effects of radiation trapping in realistic experimental
geometries.

\subsection{Cross Sections for Lyman-$\alpha$ Absorption}
In the absence of fine or hyperfine structure, the absorption
cross-section $\sigma$ for incident photons detuned by $\delta$ from
the resonant $1S$-$2P$ transition is given by
\be
\sigma=6\pi\left(\frac{\lambda}{2\pi}\right)^2{\cal L}(\delta),
\label{eq:resonancexsec}
\ee
where $\lambda$ is the transition wavelength, and ${\cal L}$ is a
Lorentzian function centered at $\delta=0$ with ${\cal L}(0)=1$ and a
FWHM of 100~MHz.\footnote{We neglect
here the Doppler broadening of the $1S$-$2P$ transition which is only
3.3~MHz at 100~$\mu$K and varies as $T^{1/2}$.}  To simplify matters, we
can neglect all transitions to the $2P_{3/2}$ level.  These transitions
are detuned 10 times further away from the fluorescence than the
$2P_{1/2}$ level, and the total cross section for $1S$-$2P_{3/2}$ absorption
is about 2 orders of magnitude smaller.

Equation~\ref{eq:resonancexsec} can be scaled to find the cross
section for a particular pair of hyperfine states and a particular
polarization.  The procedure is to multiply Eq.~\ref{eq:resonancexsec}
by the square of the appropriate matrix element of {\bf r} between hyperfine
states, normalizing by
\be
|M_{1S-2P}|^2 = |\braOket{1S}{z}{2P}|^2 = |\braOket{1S}{x\pm iy}{2P}|^2,
\ee
the square of the spinless matrix element.  We can also take into
account the variation in ${\cal L}(\delta)$.  In the low magnetic
field limit, the two possible detunings are $\delta_{F=0}=1.12$~GHz
and $\delta_{F=1}=1.06$~GHz, for transitions from $\ket{1S_d}$ into
$F=0$ and $F=1$ states, respectively, of the $2P_{1/2}$ level.  These
detunings account for the Lamb shift, the $2S$ and $2P_{1/2}$
hyperfine splittings, and the recoil shift.

At magnetic fields smaller than the hyperfine decoupling field of the
$2P_{1/2}$ level (about 40~G), the composition of the $2P_{1/2}$
hyperfine states is roughly independent of the magnetic field.  This
is a good approximation for most of the samples in this thesis.  In
terms of the $2P$ magnetic quantum number basis, the $2P_{1/2}$
hyperfine states are  
\begin{eqnarray}
\ket{2P;F=0,m_F=0} & = & 
\frac{1}{\sqrt{3}}\ket{\uparrow\:\downarrow\downproton} -
\frac{1}{\sqrt{6}}\ket{\rightarrow\:\uparrow\downproton} -
\frac{1}{\sqrt{6}}\ket{\rightarrow\:\downarrow\upproton} +
\frac{1}{\sqrt{3}}\ket{\downarrow\:\uparrow\upproton} 
\nonumber \\
\ket{2P;F=1,m_F=-1} & = & 
\sqrt{\frac{1}{3}}\ket{\rightarrow\:\downarrow\downproton} -
\sqrt{\frac{2}{3}}\ket{\downarrow\:\uparrow\downproton}
\nonumber \\
\ket{2P;F=1,m_F=0} & = &
\frac{1}{\sqrt{3}}\ket{\uparrow\:\downarrow\downproton} -
\frac{1}{\sqrt{6}}\ket{\rightarrow\:\uparrow\downproton} +
\frac{1}{\sqrt{6}}\ket{\rightarrow\:\downarrow\upproton} -
\frac{1}{\sqrt{3}}\ket{\downarrow\:\uparrow\upproton} 
\nonumber \\
\ket{2P;F=1,m_F=1} & = &
\sqrt{\frac{2}{3}}\ket{\uparrow\:\downarrow\upproton} -
\sqrt{\frac{1}{3}}\ket{\rightarrow\:\uparrow\upproton}
\end{eqnarray}
There is only one possible transition between the $1S$ and $2P_{1/2}$ manifolds
with a $\pi$-polarized photon.  The square of its matrix
element is
\be
|\braOket{1S_d}{z}{2P;F=1,m_F=1}|^2 = \frac{1}{3}|M_{1S-2P}|^2,
\ee
and the cross section $\sigma_{\pi}$ for absorption of this polarization is
\be
\sigma_{\pi} =
\frac{1}{3}\left[6\pi\left(\frac{\lambda}{2\pi}\right)^2\right]{\cal
L}(\delta_{F=1}) = 4.7\times10^{-14}\ {\rm cm}^2.
\ee
Two matrix elements contribute to $\sigma^-$ absorption:
\begin{eqnarray}
|\braOket{1S_d}{x-iy}{2P;F=0,m_F=0}|^2 & = & \frac{1}{3}|M_{1S-2P}|^2 \\
|\braOket{1S_d}{x-iy}{2P;F=1,m_F=0}|^2 & = & \frac{1}{3}|M_{1S-2P}|^2
\end{eqnarray}
It follows that the cross section for absorption of fluorescence
photons with $\sigma^-$ polarization is 
\be
\sigma_{\sigma^-} =
\frac{1}{3}\left[6\pi\left(\frac{\lambda}{2\pi}\right)^2\right]\left[{\cal
L}(\delta_{F=0})+{\cal L}(\delta_{F=1})\right] = 9.9\times10^{-14}\ 
{\rm cm}^2.
\ee
Due to the absence of $\ket{\uparrow\:\uparrow\upproton}$ components
in the $2P_{1/2}$ level, there are no matrix elements corresponding to
absorption of $\sigma^+$ photons.  This polarization can only excite
transitions to the far-detuned $2P_{3/2}$ level, and thus we neglect
$\sigma_{\sigma^+}$ in comparison to $\sigma_{\pi}$ and
$\sigma_{\sigma^-}$.

\subsection{Monte Carlo Simulation of Radiation Trapping}
\label{sec:radtrapsim}
In order to accurately predict the effects of radiation trapping under
various experimental conditions, a Monte Carlo simulation of the
process was developed.  The simulation takes as inputs the shape of
the magnetic trap, calculated numerically from magnet currents, and
the distribution of metastables along $z$.  The $2S$ distribution is
calculated using a simulation of the \oneStwoS\ excitation process
described in Sec.~\ref{sec:spatialdist}.  It has a sensitive
dependence on the geometry of the excitation laser field.  The
radiation trapping simulation assumes that the $2S$ cloud is
distributed radially according to the Maxwell-Boltzmann distribution
at a specified temperature $T$.  Meanwhile, the $1S$ cloud is assumed
to have a Maxwell-Boltzmann density distribution throughout the trap.

One Lyman-$\alpha$ photon is simulated at a time.  For each photon,
its trajectory is computed by selecting the location of the emitting
atom at random from the $2S$ cloud and then randomly choosing a
propagation direction.  Typically, the propagation direction is
restricted to be within the detection solid angle.  The quenching
electric field is assumed to be uniform and perpendicular to the trap
axis, and the photon polarization is randomly selected according to
the expressions in Table~\ref{tab:polarizations}.  To account for the
different angular distributions of circularly and linearly polarized
fluorescence, each photon is assigned a statistical weight in the
simulation according to Table~\ref{tab:angulardist}.  Next, the photon
is propagated through the $1S$ cloud in steps which are small compared
to the local absorption length $1/n_{1S}\sigma$, where $\sigma$ is the
absorption cross section corresponding to the photon polarization.
Random numbers are used to determine at each step whether the photon
has been absorbed.  Once a photon is absorbed, the simulation assumes
that, since the detection solid angle is only $\sim1\times10^{-2}$~sr,
the photon scattered into a new direction will not be
detected.

To a good approximation, the weighted fraction of photons
absorbed in the detection solid angle equals the attenuation of
metastable signal due to radiation trapping.  At low $1S$ densities
($<1\times10^{13}$~\percc), the fraction absorbed is proportional to
the density.  For a sample with peak density $6.5\times10^{13}$~\percc\
and $T=87~\mu$K (Trap~W of Ch.~\ref{ch:results}), simulations predict
that radiation trapping reduces the signal by 24-28\%, with the
uncertainty coming predominantly from the shape of the
metastable cloud. 

Most photons are not initially emitted into the detection solid angle.
One might imagine that enough of these are scattered into the
detection solid angle to significantly increase the detection
efficiency.  However, due to the fact that the detector is located on
the long axis of the trap, the scattering effect is small.  Most
photons are initially emitted in directions where the sample is
optically thin, and they have little chance to be scattered.  A
modified version of the radiation trapping simulation, in which
photons are allowed to radiate into all $4\pi$ steradians, predicts
that less than 3\% of the total fluorescence is scattered in the coldest and
densest of the relevant samples.  Furthermore, the average probability
of a scattered photon reaching the detector is less than that of an
unscattered photon, since scattering happens preferentially in the
densest part of the sample, where a second absorption event is more
probable.  For these reasons, we neglect the effects of Lyman-$\alpha$
scattering.

The simulation of fluorescence into all directions uses the matrix
elements of Table~\ref{tab:polarizations} to additionally calculate $F_d$,
the fraction of metastables returning to the trapped $\ket{1S_d}$
ground state.  The branching ratio is found to be nearly 50\% for a typical
cold, dense sample.  Knowledge of $F_d$ is required for a detection efficiency
determination described in Ch.~\ref{ch:results}.  Results from the
radiation trapping simulations are applied to the analysis of
metastable decay data in the following chapters.


\chapter{Metastable Decay Measurements: Evidence for Two-Body Loss}
\label{ch:decay}
The decay behavior of metastable clouds was studied in hydrogen
samples spanning a wide range of densities, temperatures, and magnetic
trap shapes.  Measurement techniques were improved over
several experimental runs, culminating in the data which is the
focus of this chapter and the next.  Most of the data presented here was
taken over a three week period during which apparatus conditions were
relatively constant.  To improve the signal-to-noise ratio, the same
measurements were repeated multiple times in a few standard magnetic trap
configurations at specific points along an evaporation pathway.  These
traps are described in the first section below.

A main motivation for the metastable decay studies is to measure the
loss rate due to \twoStwoS\ inelastic collisions.  For such
measurements to be possible, there must be a discernible density
dependence of the metastable decay.  A simple first-pass analysis of
the data in Sec.~\ref{sec:decayrates} provides evidence that, in cold,
dense samples, the decay rate of the metastable cloud depends strongly
on the metastable density.  Furthermore, at high metastable
densities, the early part of a decay curve can be fit much better by a
model with a free two-body loss parameter and a fixed one-body
parameter than by a simple exponential decay model.  In
Ch.~\ref{ch:results} the model incorporating two-body loss is applied to
extract a two-body loss rate constant for metastable H.
 
\section{Summary of Traps}
\label{sec:summaryoftraps}
\begin{figure}
\centerline{\epsfig{file=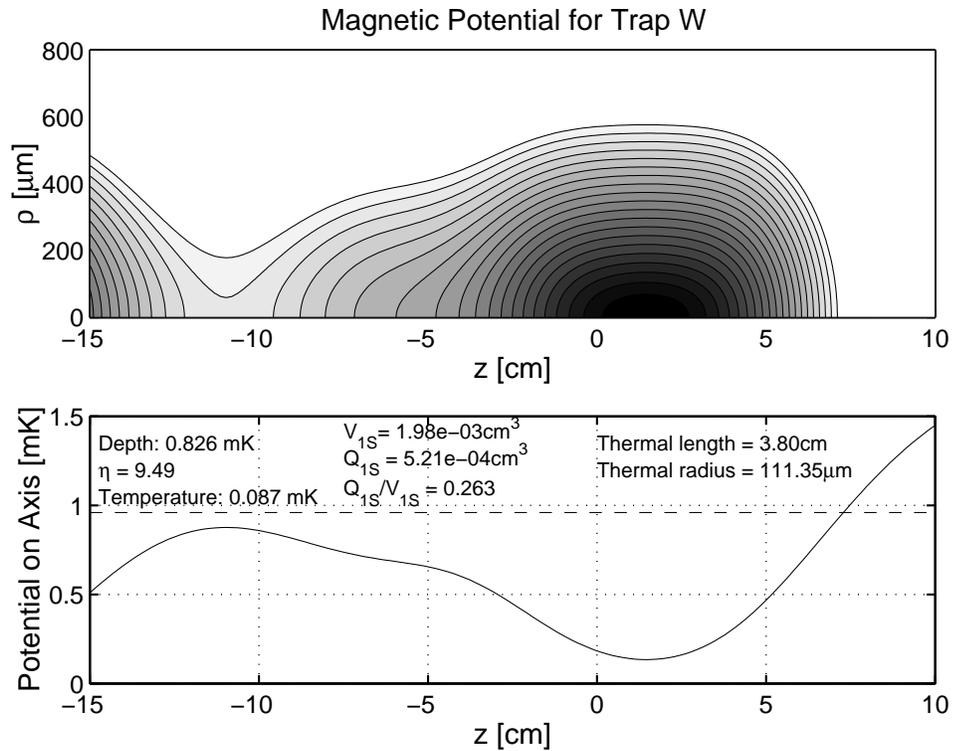,width=5in}}
\caption{Potential and characteristics of the $1S$ sample in Trap W.
The trap threshold is set by a 20~MHz rf field.  Although calculations
indicate that the magnetic trap threshold is below the rf threshold,
the probability of an atom escaping over the magnetic saddlepoint is
small compared to the probability of rf ejection, and the rf
frequency effectively determines the threshold energy.}
\label{fig:trapW}
\end{figure}
\begin{figure}
\centerline{\epsfig{file=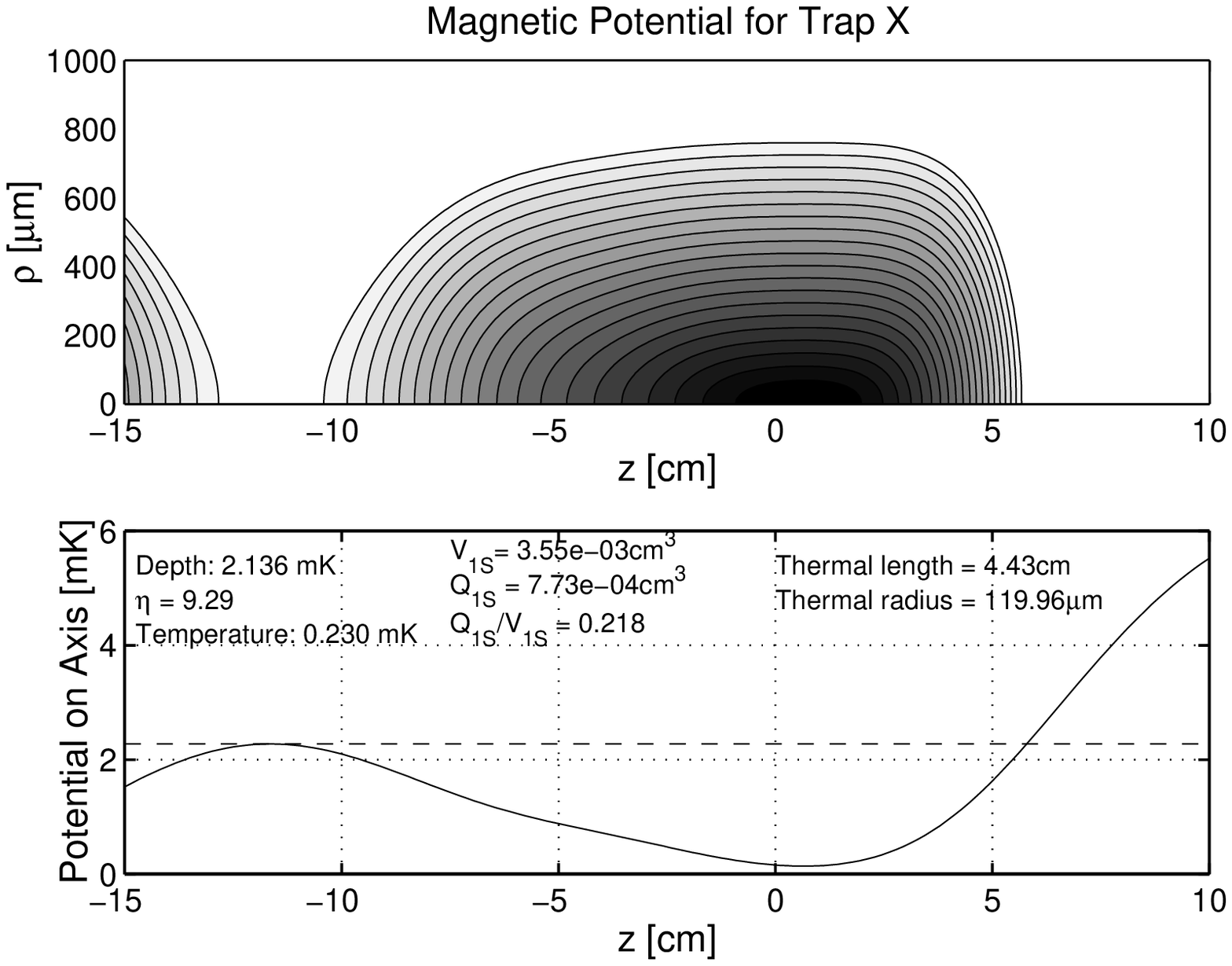,width=5in}}
\caption{Potential and characteristics of the $1S$ sample in Trap X.}
\label{fig:trapX}
\end{figure}
\begin{figure}
\centerline{\epsfig{file=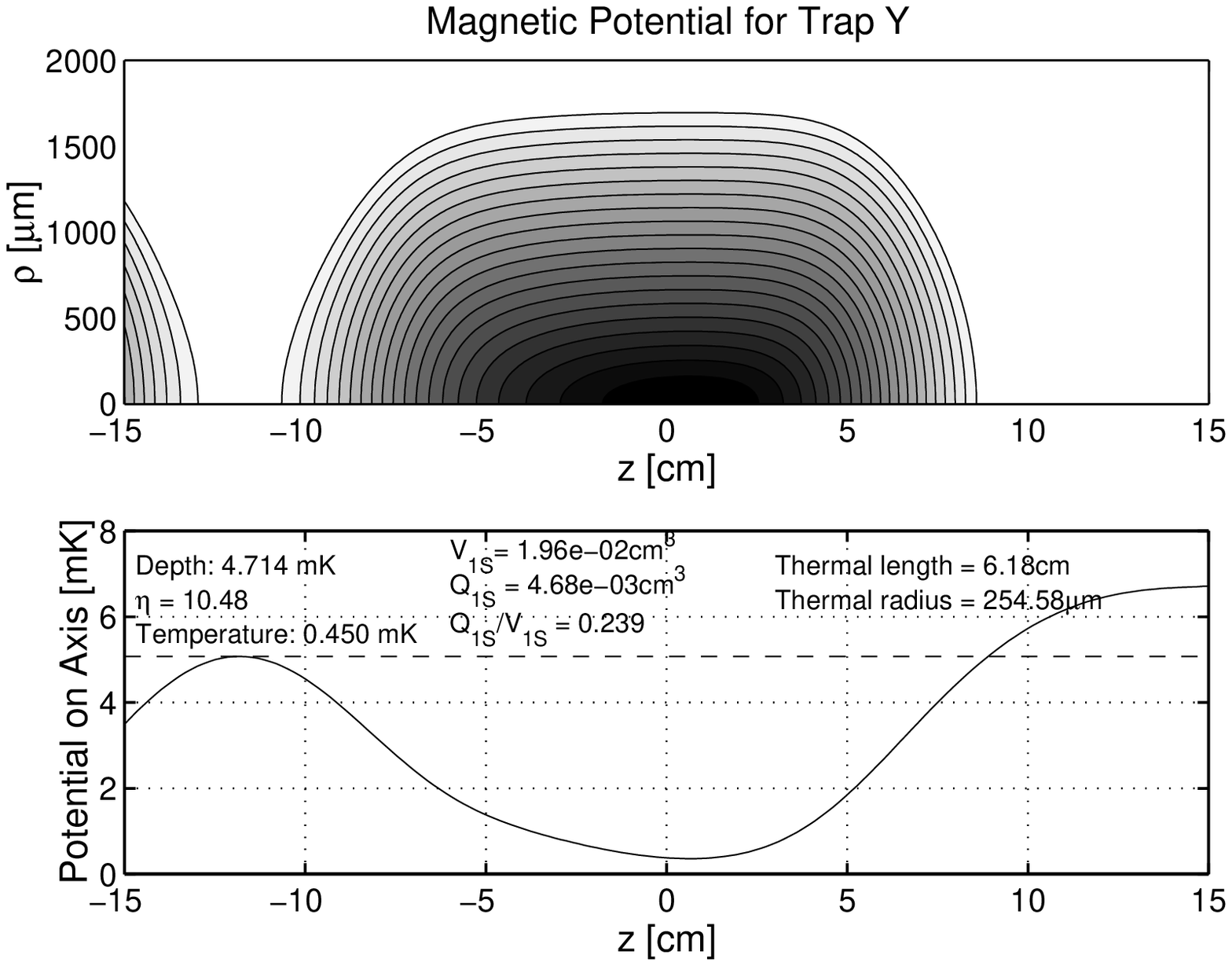,width=5in}}
\caption{Potential and characteristics of the $1S$ sample in Trap Y.}
\label{fig:trapY}
\end{figure}
\begin{figure}
\centerline{\epsfig{file=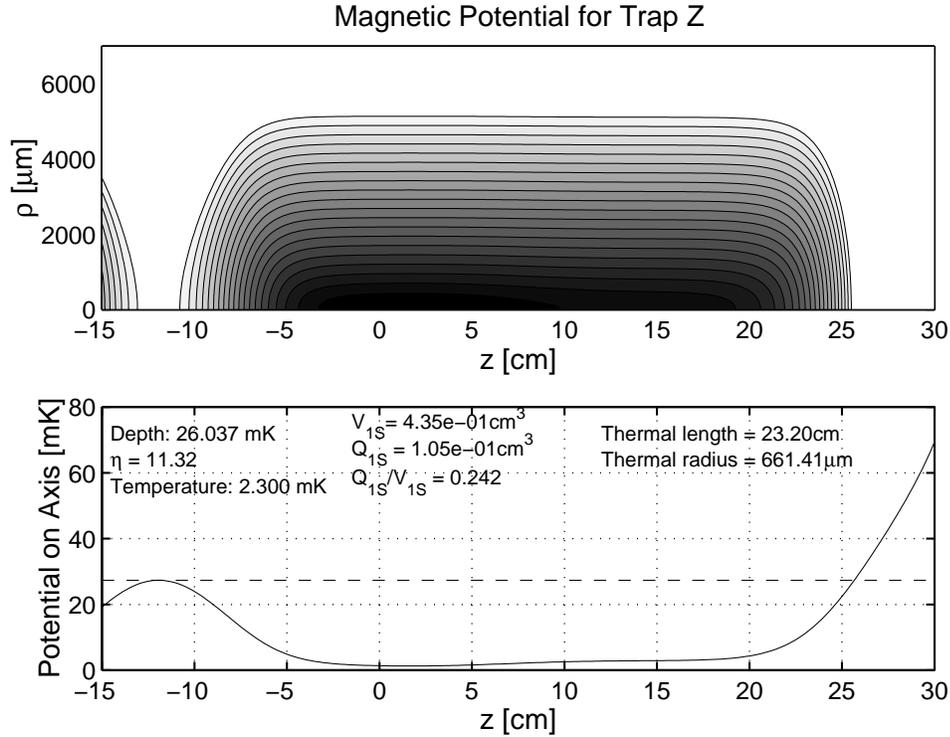,width=5in}}
\caption{Potential and characteristics of the $1S$ sample in Trap Z.}
\label{fig:trapZ}
\end{figure}

The four magnetic traps chosen for detailed studies are known as Traps
W, X, Y, and Z.  In each trap, a ground state sample was prepared
according to a consistent procedure, and metastable clouds were
excited in this sample.  The label ``Trap~W'' will sometimes be used
below to mean ``the sample in Trap~W.''  The trap shapes and some
parameters of the $1S$ cloud are shown in Figs.~\ref{fig:trapW},
\ref{fig:trapX}, \ref{fig:trapY}, and \ref{fig:trapZ}.

In order to determine collisional rate constants from the metastable
decay, it is important to know the $1S$ density and temperature for
each sample.  Both parameters are needed to calculate the spatial
distribution of the metastable cloud; these calculations will be
detailed in Ch.~\ref{ch:results}.  To determine the metastable density,
it is also necessary to know the detection efficiency for metastables.
The calibration of the detection efficiency presented in
Sec.~\ref{sec:deteff} requires knowledge of $N_{1S}$, the initial number
of atoms in the sample.  To find $N_{1S}$, the peak density $n_{1S}$ can
be multiplied by the numerically calculated effective volume $V_{1S}$.
The measured temperature and $1S$ density, together with $V_{1S}$ and
$N_{1S}$, are summarized for the four traps in
Table~\ref{tab:summaryoftraps}.

\begin{table}
\label{tab:summaryoftraps}
\begin{center}
\begin{tabular}{||c|c|c|c|c||}
\hline
Trap & $T$ (mK) & $n_{o,1S}\ (\times10^{13}$ \percc) & $V_{1S}\ ($cm$^3)$ &
$N_{1S}\ (\times10^{12})$ \\ \hline \hline
W & 0.087 & 6.5 & 0.0020 & 0.13 \\ \hline
X & 0.23  & 4.1 & 0.0036 & 0.15 \\ \hline
Y & 0.45  & 2.5 & 0.020  & 0.49 \\ \hline
Z & 2.3   & 1.3 & 0.44  & 10  \\ \hline  
\end{tabular}
\end{center}
\caption{Temperature, initial peak $1S$ density, effective $1S$
volume, and initial $1S$ atom number for the samples in which
metastable decay behavior was studied in detail.  The accuracy of
the values in this table is discussed in the text.}
\end{table}

\begin{figure}
\centerline{\epsfig{file=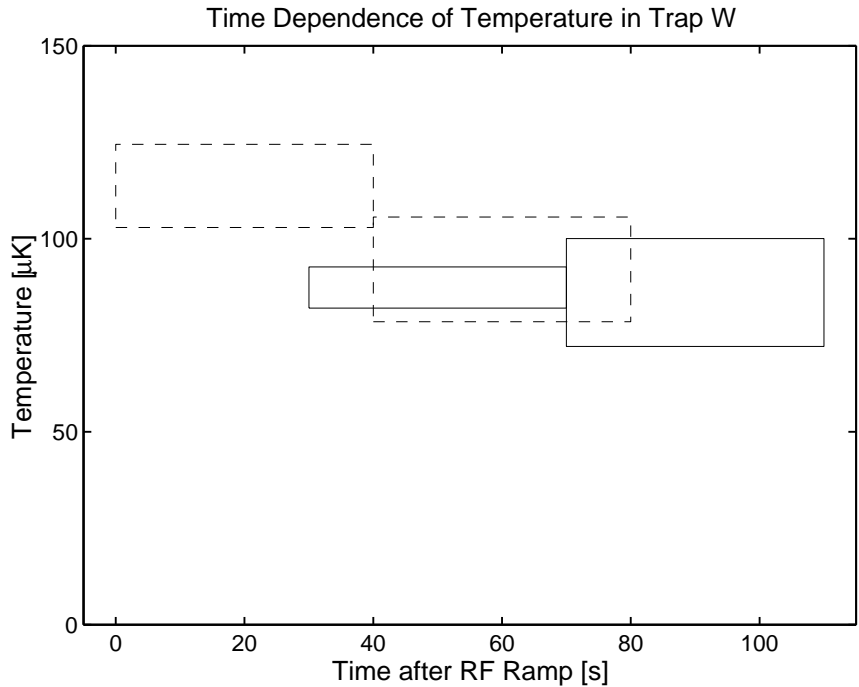,width=4.5in}}
\caption{Temperature determined from Doppler-sensitive scans in Trap W
for a sample probed immediately after the rf reached its final value
(dashed) and for a sample probed after the rf was already at its
final value for 30 s (solid).  The horizontal extents of the rectangles
indicate the boundaries of the 40 s time bin used for each temperature
determination, and the vertical extents indicate the error bars.
Considerable cooling occurs in the first 30 s after the rf ramp.  }
\label{fig:trapWcooling}
\end{figure}

When measuring the temperature or density of a sample, it is important
to ensure that the temperature has reached an equilibrium value which
is maintained during the measurement.  After the end of an evaporation
sequence in which the trap depth is ramped downward, the sample will
continue to cool for a time up to tens of seconds as the temperature
asymptotically approaches an equilibrium value.  This cooling has been
observed directly in Trap W by scanning the excitation laser frequency
repeatedly across the Doppler-sensitive line for about 1~minute after
the evaporation ramp (Fig.~\ref{fig:trapWcooling}).  The average
temperature during a given time interval was found by fitting the
lineshape accumulated during the interval to a Gaussian, as in
Fig.~\ref{fig:DStemperature}.  In the case of Traps W, X, and Y, the
standard practice was to hold the sample for 30~s after the
evaporation ramp to allow equilibrium to be reached.  At higher
temperatures, evaporation proceeds more efficiently, and the sample is
expected to be near equilibrium throughout the evaporation sequence.
For this reason, no hold time was employed for Trap~Z.

The temperatures of the equilibrated samples were determined by a
combination of methods.  For Trap~W, which involved the coldest
sample, the temperature was determined with an uncertainty of 5\% from
a Gaussian fit to the Doppler-sensitive \oneStwoS\ line
(Fig.~\ref{fig:DStemperature}).  In the case of Trap~X, the
temperature was found by comparing the width of the Doppler-free
\oneStwoS\ lines for low density samples in it and Trap~W (Fig.~\ref{fig:DFexamples}).  The temperature extracted in this way
agrees well with the trap depth divided by the equilibrium value of
$\eta$, determined self-consistently from numerical simulations of a
cylindrical quadrupole trap uniform in $z$
\cite{doy91}.  For higher temperature samples, the most reliable way
to determine temperature at present is to obtain $\eta$ from
simulations and use $kT = U_t/\eta$, where $U_t$ is the trap depth.
This was done for Traps Y and Z.  The estimated uncertainty in the
temperatures for Traps X, Y, and Z is 10\%.

Once the temperature of a sample is known, the $1S$ density can be
determined by the bolometric method described in
Sec.~\ref{subsec:densitymeasurements}.  The densities in
Table~\ref{tab:summaryoftraps} were determined this way.  They have an
estimated total uncertainty of 20\%.  Some of this uncertainty stems
from systematic uncertainty in the dipolar decay constant and affects
all of the density measurements in the same way; the measured
densities should be accurate to 10\% relative to each other.

\begin{figure}
\centerline{\epsfig{file=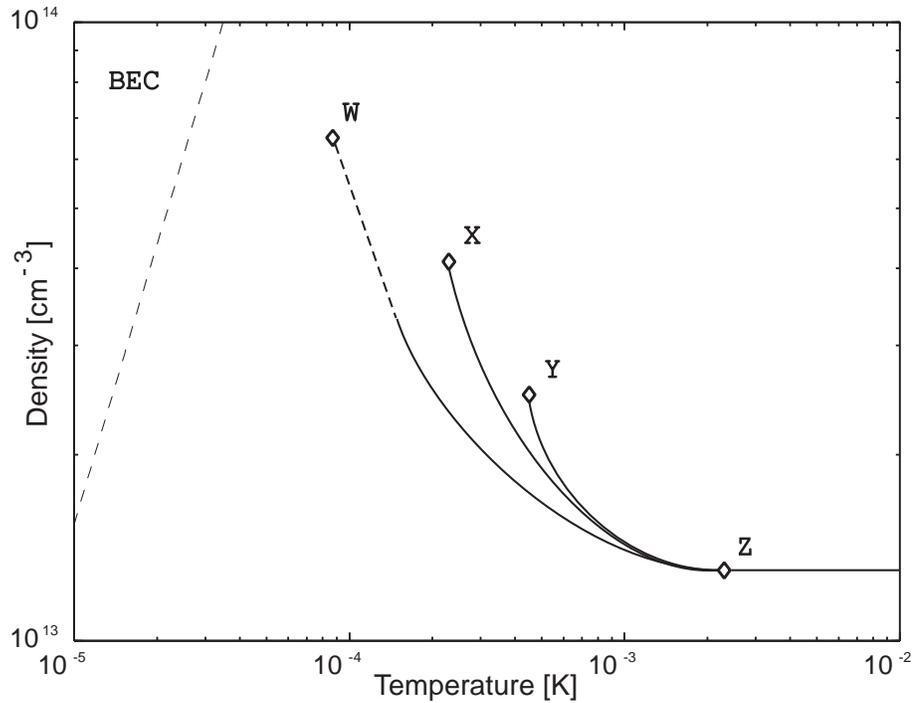,width=5in}}
\caption{Location of ground state samples in phase space.  The solid
lines roughly indicate the trajectory of the samples during magnetic
 saddlepoint evaporation, originating from higher temperatures at
 right.  A dashed line denotes the approximate rf evaporation
 trajectory of Trap W.  A longer dashed line marks the boundary between
 the BEC and normal phases of the gas.}
\label{fig:phasespace}
\end{figure}

Values for the effective volume of the $1S$ cloud were obtained by
numerical evaluation of $V_{1S}=\int$exp$(-U({\bf r})/kT)$ over the
trap region.  Since the gravity gradient (12~$\mu$K/cm) is not strong
enough to significantly distort the trapping potentials at these
temperatures, it was neglected in the calculations.  The uncertainty
in $V_{1S}$ is dominated by the temperature uncertainty.  For small
changes in $T$, numerical calculations show that $V_{1S}
\sim T^2$ in all of the traps.  Thus, $V_{1S}$ is accurate to about
10\% for Trap W and 20\% for the other traps.  The number of atoms
$N_{1S}$ in each trap was obtained by multiplying $n_{o,1S}$ by
$V_{1S}$.  Since the fractional uncertainties in $n_{o,1S}$ and
$V_{1S}$ are essentially independent, we can add them in quadrature to
find that the accuracy of the $N_{1S}$ values is 20-30\%, closer to
20\% for Trap W, and closer to 30\% for the others.

A phase space representation of the four standard samples is shown in
Fig~\ref{fig:phasespace}.  To prepare the sample in Trap W, an rf
evaporation field was ramped from 35~MHz to 20~MHz and then held
constant at 20~MHz for the duration of measurements.  No rf
evaporation was employed in the other traps.  Although Trap W is
considerably colder than Trap X, it contains almost as many atoms.
This is because Trap W has less radial compression than Trap X.  For
Traps W, X, and Y, the energy of the trap minimum was determined by
slowly sweeping the rf frequency upward from near DC until atoms were
ejected from the sample and observed on the bolometer.  In each of
these traps, the minimum was found to be consistently 180-200~$\mu$K
lower than the value predicted by numerical calculations based on the
magnet currents.  This discrepancy is assumed to be due either to
uncertainty in the magnet geometry or to a stray field component
pointing along the trap axis.  For Trap Z, the trap minimum was
estimated by subtracting 200~$\mu$K from the numerically calculated
trap minimum.

\section{Density-Dependent Decay Rates}
\label{sec:decayrates}
To measure a two-body loss process in metastable H, it is important to
have close control of not only the one-body decay rate and but also
the $2S$ density.  A first goal for our experiments was to see a
systematic effect on metastable decay due to density.  In this
section, it is shown how the effective decay rate varied in each trap
over the achievable $2S$ density range.  More specifically, an
effective decay rate is calculated as a function of the metastable signal
in a decay curve.  To a first approximation, the signal is simply
proportional to the $2S$ density in the trap.
 
The density of metastables excited by a UV laser pulse varies during
the excitation phase for several reasons.  First, the ground state
sample density decreases over many seconds due to dipolar decay.
Second, the laser frequency is generally scanned back and forth across
the center of the Doppler-free \oneStwoS\ resonance.  This allows the
line center frequency relative to the laser's reference cavity to be
established in each trap cycle.  The relative frequency drifts
typically $\sim20$~kHz per hour, measured at 243~nm.  Since the laser
frequency is stepped after each decay curve measurement, the detuning
is stepped, and the excitation rate changes.  Finally,
there are sources of noise operating on a time scale which is short
compared to the time required to make a decay measurement.  These
include fluctuations in the laser power and the overlap of the
counterpropagating UV beams.

In order to obtain enough signal-to-noise for a useful analysis, the
decay curves from the data set for a given trap must be grouped or
``binned'' together.  To look for an effect dependent on the number of
metastables excited, the decay curves can be binned according to the
total number of metastable signal counts in each decay curve.  This
means that decay curves corresponding to approximately the same
initial density are averaged together.  Next, the average decay curve
for each signal bin is fit with a model function.  For a preliminary
analysis, we use a simple exponential decay function
\be
s(t)=A_o e^{-\alpha t}
\label{eq:simpleexponential}
\ee
as a model, where $s(t)$ is the number of signal counts at a time $t$
following the laser pulse, and the free parameters are an effective decay rate
$\alpha$ and initial signal amplitude $A_o$.  Even when the decay
curve is distinctly nonexponential, the quantity $1/\alpha$
derived from a fit provides a useful measure of the lifetime of the
metastable cloud.  For cases where the decay is well described by
Eq.~\ref{eq:simpleexponential}, $\alpha$ can be interpreted as the
total rate of one-body metastable loss processes.

\begin{figure}
\centerline{\epsfig{file=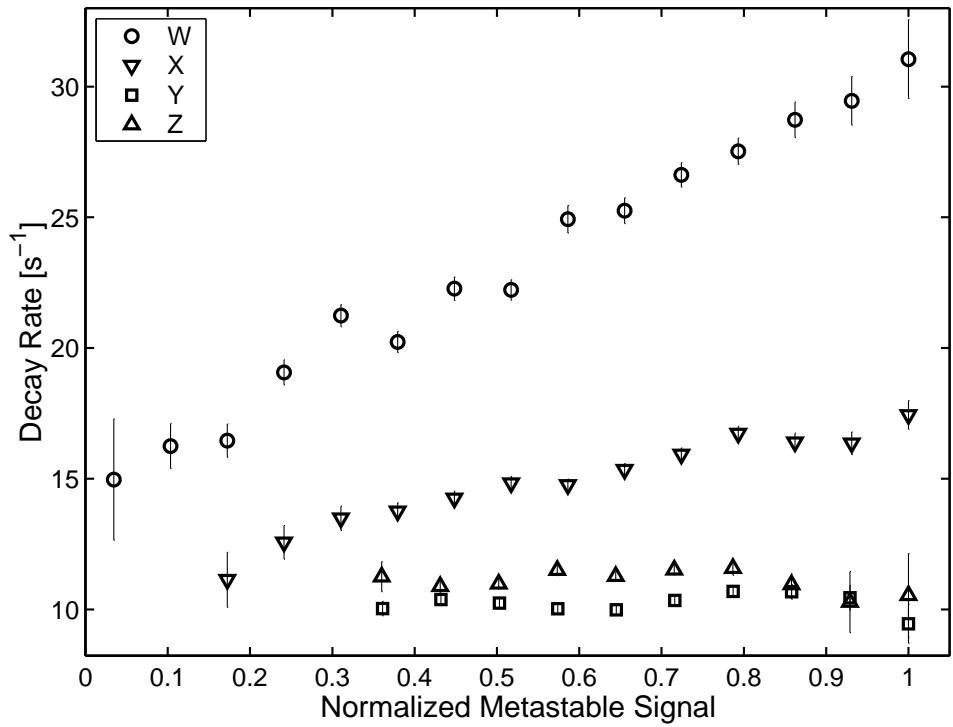,width=5in}}
\caption{Effective decay rate $\alpha$ determined by fitting
Eq.~\ref{eq:simpleexponential} to decay curves recorded in Traps W, X,
Y, and Z.  The curves were binned according to the total metastable
signal in each decay curve.  Error bars reflect only the statistical
uncertainty in the data.  The lower horizontal edge of the plot
corresponds to the natural $2S$ decay rate of 8.2 s$^{-1}$.  For this
figure, the metastable signal has been normalized to the peak signal
observed in each trap.  The peak metastable density corresponding to
the peak signal of Trap~W is about $2\times10^9$~\percc.  The peak
densities for Traps X, Y, and Z are roughly 10, 40, and 80 times
larger, respectively.}
\label{fig:decayratesWXYZ}
\end{figure}

\begin{figure}
\centerline{\epsfig{file=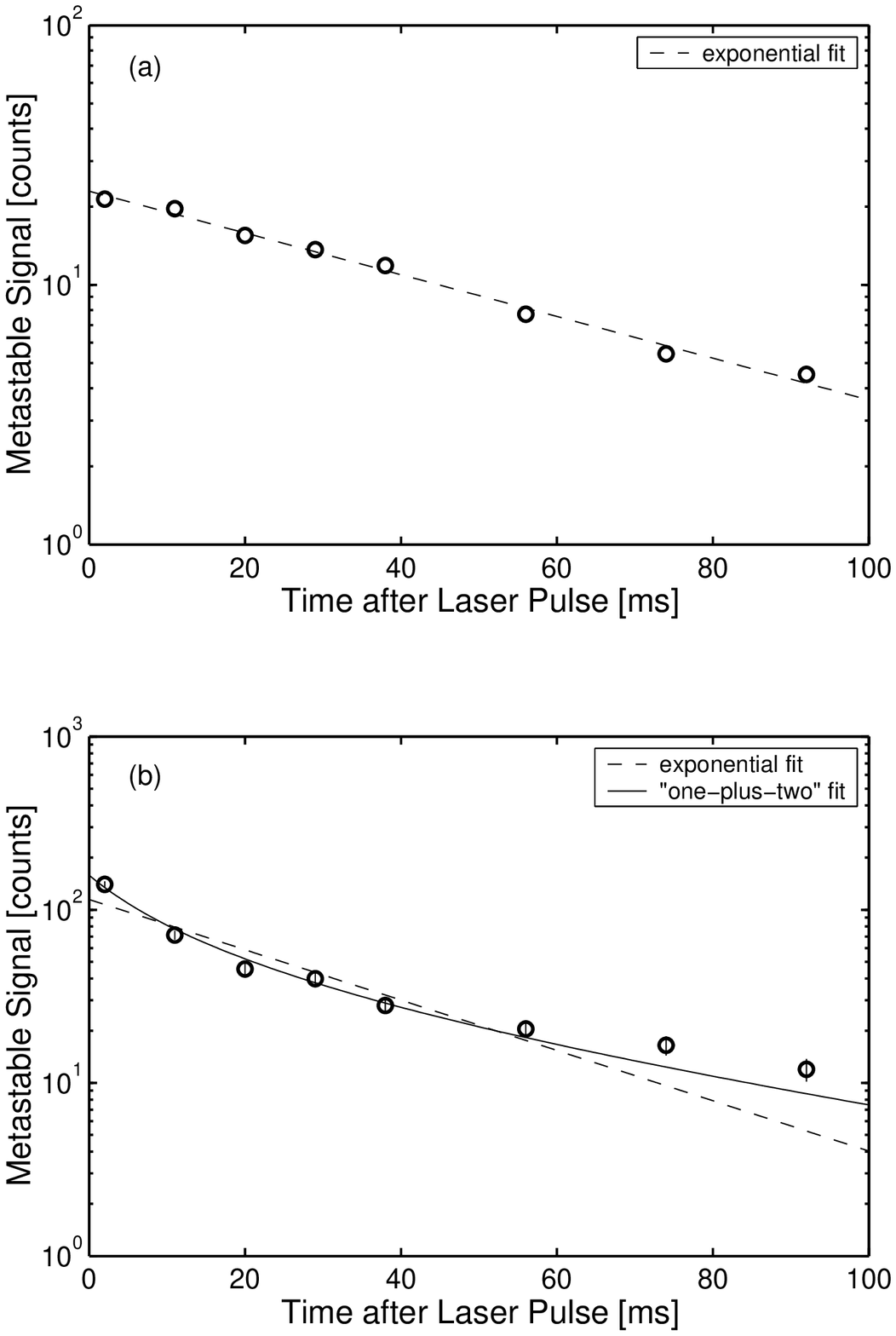,width=4in}}
\caption{Average decay curves at two different signal levels in Trap
W.  The parent data set was gathered over five trap cycles, and the
error bars account for only statistical error.  For the curve in (a),
more than 60 individual decay measurements have been averaged, and the
statistical errors are smaller than the markers; non-statistical
scatter is evident.  For (b), the curve is the average of the two
decay measurements in the data set with the largest signal.  In case
(b), where the initial metastable number is more than 6 times larger,
a ``one-plus-two'' model fits the data better than a simple
exponential.  The one-plus-two model (see Sec.~\ref{sec:oneplustwo})
incorporates terms corresponding to both one- and two-body loss, and
it assumes the shape of the $2S$ cloud is static.  In reality,
the $2S$ distribution flattens somewhat as the cloud decays, and it is
not surprising that the model deviates from the data at long wait times
(Sec.~\ref{sec:dynamicsimulation}).}
\label{fig:comparesingles}
\end{figure}

Figure~\ref{fig:decayratesWXYZ} summarizes the results for decay rates
determined from fits to a simple exponential.  The most dense
metastable cloud is excited in Trap W, which contains the coldest,
most dense ground state sample.  In this trap, there is a clear
monotonic increase of the effective decay rate with the number of
metastables excited.  For Trap X, which is warmer and less dense, the
maximum decay rates are not as large, but the same trend with
metastable signal is present.  In the warmest and least dense samples,
there is only a small increase in decay rate at the largest metastable
population achievable in those traps.

It should be mentioned that the observed dependence of decay rate on
metastable number is not a consequence of correlated variations in
ground state sample density.  To generate
Fig.~\ref{fig:decayratesWXYZ}, all metastable decay curves with
approximately the same initial metastable signal were binned together
without regard to the ground state density, which decayed over the
course of each laser scan from the initial values in
Table~\ref{tab:summaryoftraps}.  For each ground state density, a
range of metastable populations was generated by varying the laser
detuning.

Since $2S$ density increases with $2S$ signal in a given trap, the
data indicate the presence of metastable loss processes which depend
on the metastable density.  The effect is strongest in Trap W, where
the largest metastable densities were achieved.  As will be explained
in Sec.~\ref{sec:spatialdist}, a numerical simulation can be used to
calculate the spatial distribution of $2S$ atoms for a given trap and
laser geometry.  This distribution can be integrated to obtain the
effective volume of the metastable cloud.  For resonant excitation,
the peak number of metastables excited in a laser pulse is 50-70\%
larger in Trap W than in Trap Z, while the volume occupied by the $2S$
cloud in Trap W is $\sim60$ times smaller (assuming perfect overlap of the
counterpropagating UV beams).  This means that the peak metastable
density achieved in Trap W may be as much as 100 times larger than in Trap
Z.

Qualitatively, the increase in peak decay rate going from Z to W is
consistent with the existence of two-body inelastic processes which
become increasingly significant at higher $2S$ densities.  The data
could also be consistent with an inelastic rate constant which
increases with decreasing temperature.  However, theoretical
predictions for the relevant inelastic processes indicate that the
rate coefficients should be nearly independent of temperature in the
temperature range of these experiments (Fig~\ref{fig:forreyrates}).

In Traps W and X, the decay curves were observed to be exponential for
low $2S$ signal and distinctly nonexponential at high $2S$ signal.
This is illustrated in the semi-logarithmic plots of
Fig.~\ref{fig:comparesingles}.  In Fig.~\ref{fig:comparesingles}(a),
the data is well fit by a simple exponential.  For the decay curve
starting at high signal in Fig.~\ref{fig:comparesingles}(b), there is
a clear departure from an exponential.  The data is better fit with a
model which has a fixed one-body decay parameter and a free two-body
parameter (``one-plus-two model'').  The procedure for fitting to this
model and extracting a \twoStwoS\ inelastic rate constant is described
in the following chapter.


\chapter{Analysis of Metastable Loss Dynamics}
\label{ch:results}

In order to determine the loss rate constant $K_2$ associated with
\twoStwoS\ inelastic collisions, it is convenient to first assume that
the metastable cloud shape does not change as it decays.  This allows
the decay data to be fit by a simple model incorporating both one-body
and two-body loss terms.  The one-body parameter of the model can be
fixed at the exponential decay rate inferred from the data for the
limit of zero metastable density.  With the one-body parameter fixed,
the two-body parameter is easily determined from fits to decay curves.
To find $K_2$, it is also necessary to know the geometry of the
metastable cloud and the detection efficiency for metastables; these
pieces of information establish the link between the $2S$ density and
the observed signal.  As will be explained in this chapter, the
geometry and detection efficiency are difficult to determine
precisely, and this gives rise to much of the uncertainty in the final
values for $K_2$.  Some more uncertainty arises because the $2S$
spatial distribution flattens significantly during a decay, and it is
necessary to apply a correction factor based on a dynamic simulation
of the cloud.  Nevertheless, it is possible to determine $K_2$ to
within a factor of about 2 with our data, and the results are
compatible with the current state of theory.  It is also possible with
our data to set limits on the loss rates due to \oneStwoS\ inelastic
collisions.  The details of the analysis follow.

\section{One-Plus-Two Model}
\label{sec:oneplustwo}
The extraction of a rate constant from the decay data is
straightforward if we assume that (1) loss occurs solely due to a
combination of one- and two-body processes (``one-plus-two model'')
and (2) the shape of the metastable cloud is constant in time
(``static approximation'').  Three-body loss is not expected to play a
role at the metastable densities achieved to date ($\sim
10^{11}$~\percc\ or less).  If the $2S$ distribution in the trap and
the detection efficiency are known, then a total two-body loss
constant in terms of density can be computed from the two-body
parameter in the model.

In the static approximation of the one-plus-two model, the decay of
the metastable signal is described by
\be
\dot{s}=-\alpha_1 s - \alpha_2 s^2,
\label{eq:oneplustwodiffeq}
\ee
where $\alpha_1$ and $\alpha_2$ are, respectively, the one- and
two-body fit parameters.  The solution to
Eq.~\ref{eq:oneplustwodiffeq} is
\be
s(t) = \frac{\alpha_1 A_o e^{-\alpha_1 t}}
{\alpha_1 + \alpha_2 A_o (1 - e^{-\alpha_1 t})}. 
\label{eq:oneplustwo}
\ee

Since $\alpha_1$ and $\alpha_2$ are highly correlated, it is not
useful to vary both of these fit parameters simulataneously.
Fortunately, $\alpha_1$ can be determined by extrapolation of
effective decay rates $\alpha$ to the limit of very low metastable
density; this will be explained more fully in
Sec.~\ref{sec:experimentalresults}.  The parameter $\alpha_1$
represents the background one-body decay rate which occurs in a given
trap regardless of the metastable density.  With the fixed value for
$\alpha_1$, the decay curves for all initial $2S$ densities can
subsequently be fit with Eq.~\ref{eq:oneplustwo}, using $\alpha_2$ and
$A_o$ as free parameters.  This was the procedure used to determine
the solid curve in Fig.~\ref{fig:comparesingles}(b).
 
\section{Two-Body Loss Constant in the Static Approximation}
As discussed in Sec.~\ref{sec:clouddynamics}, the relative spatial
distribution of the metastable cloud evolves during its lifetime due
to diffusion and two-body loss.  For the purposes of extracting a
value for the two-body loss constant from the decay data, however, we
start with the simplifying assumption that the relative $2S$ distribution is
static in time.  The effect of metastable cloud dynamics will be
considered afterwards in Sec.~\ref{sec:dynamicsimulation}.

The desired two-body loss constant, $K_2$, is defined so that it gives the
total local rate of $2S$ density change due to two-body processes:
\be
\Bigl. \dot{n}_{2S}({\bf r}) \Bigr|_{\rm two-body} = - K_{2} n_{2S}^2({\bf r}).
\ee
In terms of the normalized metastable distribution $f_{2S}({\bf r}) =
n_{2S}({\bf r})/n_{o,2S}$, where $n_{o,2S}$ is the peak density, the
two-body loss from the entire trap is
\be
\Bigl. \dot{N}_{2S} \Bigr|_{\rm two-body} = - K_{2} n_{o,2S}^2 \int
f_{2S}^2({\bf r})\,d^3{\bf r},
\label{eq:entiretraploss}
\ee
where $N_{2S}$ is the metastable population.  Since the peak density
cannot be measured directly in the current experiments, it is useful
to rewrite the right-hand-side of Eq.~\ref{eq:entiretraploss} in terms
of $N_{2S}$.  Making use of the effective volume definitions
\be
V_{2S} = \int f_{2S}({\bf r})\,d^3{\bf r} = \frac{N_{2S}}{n_{o,2S}}
\ee
and
\be
Q_{2S} = \int f_{2S}^2({\bf r})\,d^3{\bf r},
\ee 
we have
\be
\Bigl. \dot{N}_{2S} \Bigr|_{\rm two-body} = - \frac{K_{2}
Q_{2S}}{V_{2S}^2} N_{2S}^2 = - \frac{K_{2}}{\zeta} N_{2S}^2,
\label{eq:twobodylossvszeta}
\ee
where $\zeta = V_{2S}^2/Q_{2S}$ is a quantity having units of volume and
encapsulating all necessary information about the $2S$ cloud
geometry.  The quantity $\zeta$ is a measure of the spatial extent of
the metastable distribution; $\zeta$ decreases as $f_{2S}(r)$ becomes
more sharply peaked in space.

The quantity measured experimentally is $s(t)$, the number of
metastable signal counts observed when quenching at time $t$ after
excitation.  The signal is proportional to the number of $2S$ atoms in the
trap by the detection efficiency $\epsilon$: 
\be
s = \epsilon N_{2S}.
\ee
Since the signal decay due to two-body loss is given by
\be
\Bigl. \dot{s} \Bigr|_{\rm two-body} = - \alpha_2 s^2,
\ee
it immediately follows that
\be
K_2 = \zeta \epsilon \alpha_2.
\label{eq:staticK2}
\ee
The method for determining $\alpha_2$ from the data has already been
described.  We turn our attention next to the calculations and
measurements required to obtain $\zeta$ and $\epsilon$.

\section{Spatial Distribution of Metastables}
\label{sec:spatialdist}

\boldmath
\subsection{Monte Carlo Calculation of $2S$ Distribution}
\unboldmath

A lineshape simulation was developed by L.~Willmann of the MIT
Ultracold Hydrogen Group to calculate the \oneStwoS\ excitation
spectrum for realistic experimental conditions.  The simulation can be
used to calculate the spectrum due to atoms excited at different $z$
positions.  If contributions for enough different $z$ positions are
calculated, the distribution of metastables immediately following
excitation can be extracted from the lineshape results.

The input parameters to the simulation include the laser field
geometry, the trap shape, the sample temperature, and the chemical
potential, which is related to the sample density by the Bose-Einstein
distribution.  The simulation is based on the principle that the
contribution to the excitation spectrum by a single atom is
proportional to the Fourier transform of the electric field
experienced by the atom during the excitation pulse.  Moreover, the
cold-collision shift is incorporated by allowing the resonance
frequency to vary during the pulse according to the trajectory of the
atom in the trap; if ${\bf r}$(t) is the atom trajectory, the resonance
frequency as a function of time is $\nu_o - \chi n_{1S}({\bf r}(t))$, where
$\nu_o$ is the unperturbed resonance frequency and $\chi$ is the
cold-collision shift per unit density.  When the laser field intensity
and instantaneous resonance frequency are sampled at a sufficient
number of points along the trajectory, a Fast Fourier Transform (FFT)
algorithm accurately produces the single-atom lineshape contribution.
  
In each simulation, a Monte Carlo routine first generates a density of
trajectory states table with trajectories classified according to $z$,
the maximum distance from trap center $\rho_{\rm max}$ during a
complete ``orbit'', and the orbital angular momentum $L$.  The orbits
are considered to occur at constant $z$, since the distance traveled
by the atoms along $z$ during a laser pulse is small compared to the
trap length scale.  Momentum-changing collisions are neglected since
typical collision times are longer than the laser pulse.  Next, the
FFT's of a prescribed number of trajectories are computed in a second
Monte Carlo procedure.  The contribution of each FFT to the final
spectrum is weighted according to the density of trajectory states and
the Bose-Einstein distribution.

The lineshape simulation has produced spectra in good agreement with
experimental spectra for a wide range of trap shapes, laser
geometries, sample densities, and temperatures.  For the purposes of
the analysis presented here, sufficient accuracy in the $2S$
distribution is obtained for a given set of input parameters after
about one hour of CPU time on a computer which performs the FFT
of a 1024-point trajectory every $\sim50$~ms.  

\subsection{Results of Spatial Distribution Calculations}

The Monte Carlo simulation provides the axial distribution of
metastables in the trap.  Under the assumption that a thermal
distribution is established quickly in the radial direction, the
distribution function $f_{2S}({\bf r})$ is easily calculated for the
entire trap.  Using the photoionization correction factor $C(z,t)$
described in Sec.~\ref{sec:photoionization}, the distribution function
is corrected for photoionization losses during the excitation pulse.
Numerical integration can then be used to determine the geometry
factor $\zeta$.

By varying the input parameters to the simulation, the sensitivity of
$\zeta$ to various experimental uncertainties can be determined.  The
most important uncertainty by far is found in how well the focuses of
the counterprograting UV beams are overlapped along $z$.  The
alignment of the beams to the trap axis can be done accurately using
the imaging technique of Fig.~\ref{fig:lateralscan}, and the concave
cell retromirror effectively fixes the centroid of the incoming and
return focuses.  However, it is difficult to ensure that no separation
exists between the focuses.  The present method of minimizing the
separation is to match the sizes of the overlapped incoming and return
beam modes in Telescope 2 of Fig.~\ref{fig:uvlayout}.  This is
accomplished by adjusting the length of the telescope.  The estimated
typical discrepancy in mode sizes for the decay measurements presented
here is 10\% or less, which translates into a separation of 1.8~cm or
less between the two beam waists.

\begin{figure}
\centerline{\epsfig{file=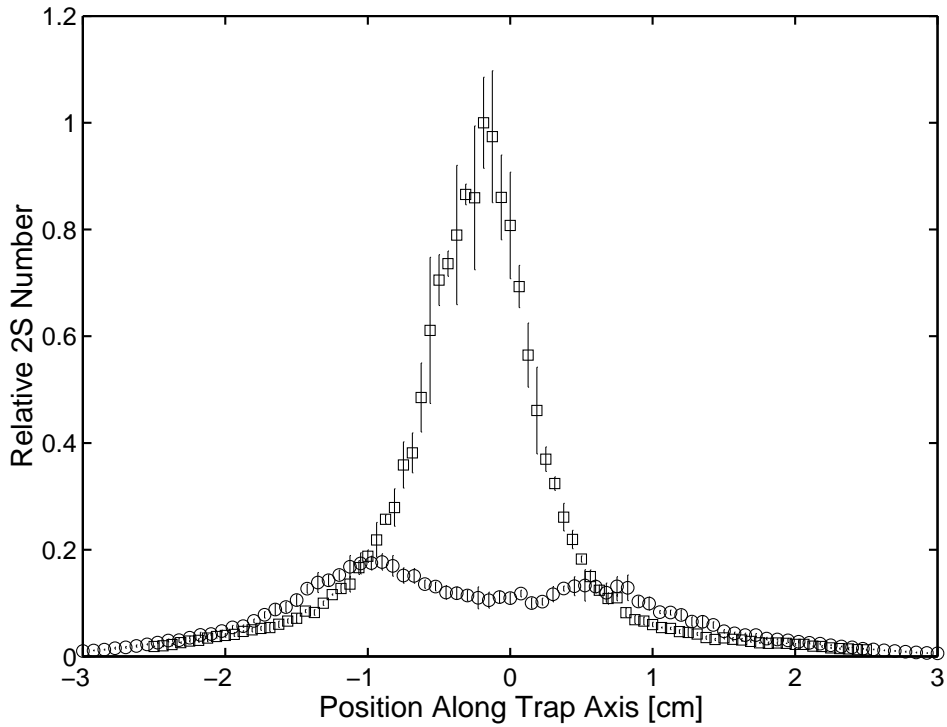,width=5in}}
\caption{Relative number of metastables as
a function of position relative to the trap minimum in Trap W.  At
each position, the Monte Carlo lineshape simulation was used to
calculate the number of atoms excited assuming the laser focuses were
perfectly overlapped (squares) and separated by 1.8~cm (circles) along
$z$.  The detuning of the laser from the \oneStwoS\ resonance is zero,
and the centroid of the focuses is located at -0.2~cm.  Error bars
indicate the scatter in the simulation results.}
\label{fig:axial2SdistW}
\end{figure}

The dependence of the $2S$ spatial distribution on the separation of
the laser focuses is illustrated for Trap~W in
Fig.~\ref{fig:axial2SdistW}.  A separation of 1.8~cm produces a
significant change since the local Doppler-free excitation rate
depends on the product of incoming and return beam intensities, and
the Rayleigh range is only 0.6~cm for the 21~$\mu$m waist radius.
Values of $\zeta$ determined in this and other traps for zero laser
detuning are summarized in Table~\ref{tab:zetaWXYZ}.

\begin{table}
\begin{center}
\begin{tabular}{||c|c|c||}
\hline
 & $\zeta\,(10^{-3}$~cm$^3)$, & $\zeta\,(10^{-3}$~cm$^3)$,
\\ 
Trap & $d_{\rm sep}=0$~cm & $d_{\rm sep}=1.8$~cm \\ \hline \hline
W & 2.0 & 4.3 \\ \hline
X & 3.3  & 6.4 \\ \hline
Y & 16  & 27 \\ \hline
Z & 150 & 250 \\ \hline  
\end{tabular}
\end{center}
\caption{Results for the geometry factor $\zeta$ at zero laser
detuning, calculated by numerical integration of $f_{2S}({\bf
r})$.  The spatial distribution function along $z$ was generated using
the Monte Carlo lineshape simulation, and the radial distribution was
assumed to be Maxwell-Boltzmann.  The assumed separation of incoming and
return laser focuses is denoted by $d_{\rm sep}$. }
\label{tab:zetaWXYZ}
\end{table}

\begin{figure}
\centerline{\epsfig{file=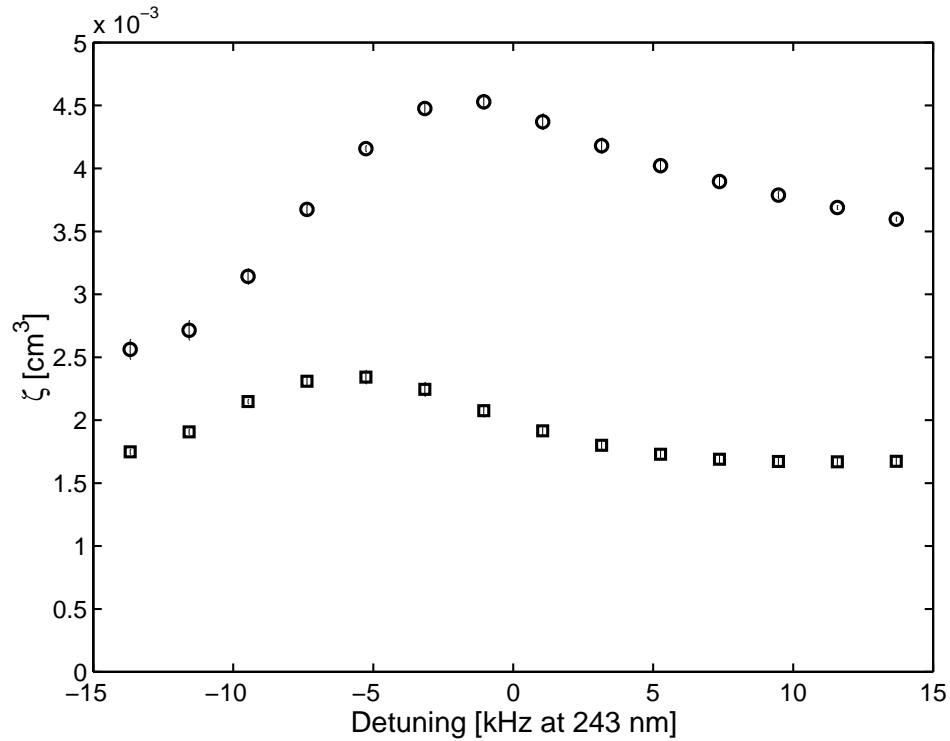,width=5in}}
\caption{Values of $\zeta$ in Trap W as a function of laser detuning.
Circles are for the case where the two laser focuses are separated
along the trap axis by 1.8~cm, and squares indicate the case of
perfect overlap.  The two cases provide approximate upper and lower
bounds for the true value of $\zeta$ in the experiment. }
\label{fig:zetabydetW}
\end{figure}

Though the laser field geometry is the most important factor in
determining the shape of the metastable cloud, the shape also depends
on laser detuning.  This is a consequence of the cold-collision shift.
The atomic resonance frequency of an atom passing in and out of a
dense region of the ground state cloud is chirped by the changing
cold-collision shift, which is proportional to the local density.  In
this way, the excitation spectrum of a single atom is both broadened
and shifted.  At red detunings (laser frequencies less than the
unperturbed \oneStwoS\ resonance frequency), the $2S$ distribution is
slightly more concentrated near the trap minimum in $z$ than for zero
detuning.  This is because the high density near the trap minimum
shifts the atomic resonances to lower frequency.  At blue detunings
the relative metastable distribution will generally again become more
concentrated about the trap minimum since most of the excited atoms
are ones whose resonances have been sufficiently chirped in a high
density region.

Figure~\ref{fig:zetabydetW} shows the variation of $\zeta$ in Trap W over the
range of detunings used for most $2S$ decay measurements.  The
dependence on detuning is stronger when the laser focuses are
separated.  In traps where peak densities are smaller and the ground
state cloud extents are larger relative to the laser depth of focus,
the dependence on detuning is less.  For example, in Trap Z, $\zeta$
varies by less than 20\% over the detuning range used in
Fig.~\ref{fig:zetabydetW}, even for focuses separated by 1.8~cm.

\section{Measurement of Detection Efficiency}
\label{sec:deteff}
To obtain $K_2$ from $\alpha_2$ in the static approximation, it is
crucial to know not only $\zeta$ but also the detection efficiency
$\epsilon$.  The detection efficiency completes the connection between
the metastable signal, which we observe, and the metastable density.

While it is difficult to measure $\epsilon$ precisely with the current
apparatus, it can be calibrated to better than a factor of 2 by
depleting the ground state sample as quickly as possible through
\oneStwoS\ excitation.  For an ideal measurement by this approach, the
only significant loss mechanism would be excitation, and the detection
efficiency would be given simply by the total number of signal counts
observed divided by the initial number of atoms in the trap,
$N_{1S}(0)$.  However, even in cases where the depletion rate due to
excitation is high, there can also be significant loss due to dipolar
decay and due to laser heating of the trapping cell, which causes
helium atoms to be evaporated from the cell walls.  Another
complicating factor in a real experiment is that a considerable
fraction of the metastables relax back to the trapped $d$ ground state
and are subsequently excited again (see Sec.~\ref{sec:radtrapsim}).
Finally, since the metastables are quenched after some wait time
following the laser pulse, some correction needs to be made for the
signal ``missing'' due to inelastic collisions of metastables during
the wait time. The following paragraphs outline a method for
extracting the detection efficiency which takes all these factors into
account.

\begin{figure}
\centerline{\epsfig{file=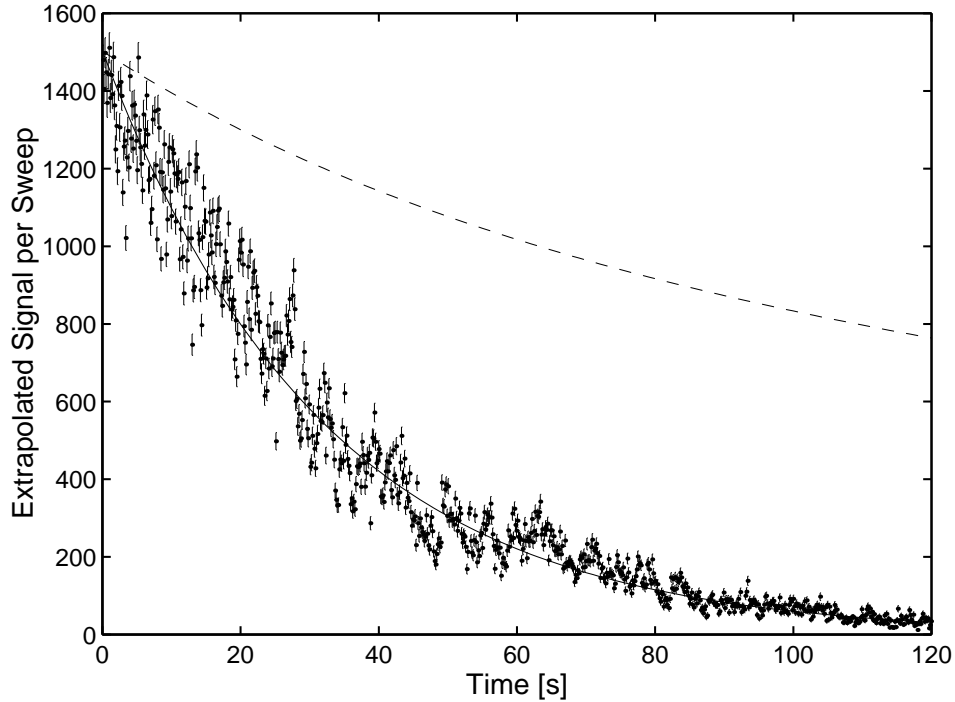,width=5in}}
\caption{Decay of $1S$ cloud in Trap X undergoing resonant
excitation.  Each data point is the total number of signal counts in a
sweep across resonance.  The signal has been corrected for counts
which were missed because of the delay time between excitation and
quench pulses.  Shown for comparison is the theoretical signal which
would be observed in the absence of any laser-induced loss (dashed
curve).  Error bars indicate only the uncertainty due to counting
statistics; significant non-statistical fluctuation in the signal is
evident, believed due to optical alignment jitter which affects the
overlap of incoming and return UV focuses.  The average detection
efficiency for this trap is obtained by integrating the fitted (solid)
curve and dividing by the total number of atoms excited to the $2S$
state.}
\label{fig:NDLXthesis}
\end{figure}

Figure~\ref{fig:NDLXthesis} shows the results of a ground state
depletion experiment in Trap X.  The 243~nm laser frequency is swept
back and forth across the central 10~kHz of the \oneStwoS\ line,
completing a sweep in one direction every 150~ms.  Since the laser
frequency remains near the center of the line, the sample is depleted
quickly; after 120~s, only a few percent of $N_{1S}(0)$ remain in the trap.
To analyze the data, the number of signal counts observed after each
laser pulse was first corrected for the loss which occurs during the
5.4~ms wait time between excitation and quench pulses.  This was done
by extrapolating backwards in time using Eq.~\ref{eq:oneplustwo} with
experimentally determined values of $\alpha_1$ and $\alpha_2$.  Next,
the extrapolated signal was summed across each laser sweep, and the
results were fit with a smooth curve $y(t)$, shown in
Fig.~\ref{fig:NDLXthesis}, where $t=0$ is the beginning of the
depletion experiment.  The quantity $y(t)$ serves as a
distribution function of the \oneStwoS\ signal in time.

To understand how the detection efficiency is extracted from $y(t)$,
we consider its relationship to $W_{1S-2S}(t)$, the excitation rate
at time $t$ averaged over a laser sweep:
\be
\epsilon(t)=\frac{y(t)}{W_{1S-2S}(t)}.
\ee
We neglect here the small number of metastables lost during the
excitation pulse itself.  The detection efficiency is an increasing
function of time, because the number of fluorescence photons ``lost''
due to radiation trapping decreases as the $1S$ density decreases
(Sec.~\ref{sec:radtrapsim}).  It is useful to calculate the average
detection efficiency $\bar{\epsilon}$ per metastable during the
depletion experiment,
\be
\bar{\epsilon}=\frac{\int W_{1S-2S}(t) \epsilon(t)\,dt}{\int
W_{1S-2S}(t)\,dt} = \frac{\int y(t)\,dt}{\int
\left[y(t)/\epsilon(t)\right] dt}\,.
\label{eq:epsilonbar0}
\ee
The integration is to be taken over the duration of the experiment.
Equation~\ref{eq:epsilonbar0} can be written in terms of the initial
detection efficiency $\epsilon(0)$ using
\be
\epsilon(t)=\frac{1-F(t)}{1-F(0)}\,\epsilon(0),
\ee
where $F(t)$ is the fraction of fluorescence photons in the detection
solid angle which are absorbed by the $1S$ cloud.  The radiation
trapping simulation described in Sec.~\ref{sec:radtrapsim} predicts
$F(0)=0.22$ in Trap X with an uncertainty of 20\% stemming primarily
from uncertainty in the initial ground state density.  This number is
relatively insensitive to the sample temperature, the laser detuning,
and the separation of the laser focuses.  Results from simulations at
several sample densities in Trap X were used to show $F(t)\simeq
[N_{1S}(t)/N_{1S}(0)]F(0)$.  The fraction of atoms remaining,
$N_{1S}(t)/N_{1S}(0)$, can be obtained by taking $y(t)/y(0)$ and
making a correction for the changing width of the
\oneStwoS\ line due to decreasing cold-collision shift.  Substituting
into Eq.~\ref{eq:epsilonbar0}, we obtain
\be
\bar{\epsilon}=\left( \frac{\epsilon(0)}{1-F(0)} \right) \frac{\int
y(t)\,dt}{\int [y(t)/(1-F(t))]\,dt}\,. 
\label{eq:epsilonbar1}
\ee
Now, Eq.~\ref{eq:epsilonbar0} can be interpreted as equating the
average detection efficiency with the total number of signal counts
divided by the total number of atoms excited to the $2S$ state.  We
can write the total number of atoms excited as
\be
\int W_{1S-2S}(t)\,dt = N_{1S}(0) - N_d - N_l + N_r,
\label{eq:intW}
\ee 
where $N_d$ is the total number of atoms lost to dipolar decay, $N_l$ is the
number lost due to laser heating in the trapping cell, and $N_r$ is
the number of relaxations back to the trapped $1S$ state when
metastables are quenched.  The small number of atoms remaining in the
trap at the end of the depletion measurement are neglected.
Substituting Eq.~\ref{eq:intW} into Eq.~\ref{eq:epsilonbar0}
yields an alternative expression for the average detection efficiency:
\be
\bar{\epsilon}=\frac{\int y(t)\,dt}{N_{1S}(0) - N_d - N_l + N_r}\,.
\label{eq:epsilonbar2}
\ee

The rate of relaxations back to the trapped sample is approximately given by
$d_rW_{1S-2S}(t)$, where $d_r$ is the spatially-averaged branching
ratio to $\ket{1S_d}$ for metastables in the quenching electric
field.  Thus,
\be
N_r \simeq d_r \int{\frac{y(t)}{\epsilon(t)}\,dt} =
\frac{d_r[1-F(0)]}{\epsilon(0)}
\,\int{\frac{y(t)}{1-F(t)}\,dt}\,.
\label{eq:Nr}
\ee  
For Trap X, Monte Carlo simulations predict that $d_r \simeq 0.43$,
virtually independent of sample density, temperature, laser detuning,
and separation of the focuses.\footnote{In the simulations, the
electric field was uniform and pointing in a direction perpendicular
to the trap axis.  This should be a good approximation to the
actual quenching field.  See the electrode arrangement in
Sec.~\ref{sec:strayfield}.}  Equation~\ref{eq:Nr} is not exact
because the branching ratio is unknown for those metastables decaying
before the quench field is applied.  The resulting uncertainty is
small, however, relative to other uncertainties in the detection
efficiency determination.

The number of atoms lost to dipolar decay
and the number lost to laser heating are calculated numerically from
$N_{1S}(t)$, using
\be
N_d=\frac{g Q_{1S}}{V_{1S}^2}\int N_{1S}^2(t)\,dt,
\label{eq:Nd}
\ee
where $g$ is the dipolar decay constant, and
\be
N_l=\frac{1}{\tau}\int N_{1S}(t)\,dt,
\label{eq:Nl}
\ee
where $\tau$ is the time constant for sample decay due to laser
heating.  At the high laser power ($\sim20$~mW) used to obtain the
data in Fig.~\ref{fig:NDLXthesis}, $\tau$ was determined to be about
200~s by tuning the laser far off resonance and observing the
depletion of the sample.  Taking $N_{1S}(0)$ from
Table~\ref{tab:summaryoftraps}, Eqs.~\ref{eq:Nd} and
\ref{eq:Nl} evaluate to $N_d=2.6\times10^{10}$ and
$N_l=2.3\times10^{10}$ for this data.

By combining Eqs.~\ref{eq:epsilonbar1}, \ref{eq:epsilonbar2}, and
\ref{eq:Nr}, the initial detection efficiency in Trap X is determined
to be
\be
\epsilon(0)=1.9^{+1.1}_{-0.6} \times 10^{-6}.
\label{eq:deteffX}
\ee
The errors are due predominantly to uncertainties in temperature and
initial density.  By comparing radiation trapping calculations for
Traps W, X, Y, and Z, the detection efficiency can be predicted for
the other traps as well (see Table~\ref{tab:deteffs}).  It turns out
that $1-F(0)$ (and thus $\epsilon(0))$ is roughly constant across the
traps.  As the sample becomes cooler and denser along the evaporation
pathway, it also becomes shorter along the trap axis, and the optical
density seen by Lyman-$\alpha$ photons traversing the axis towards the
detector remains nearly the same.  We note that the detection
efficiency for metastables at zero $1S$ density is simply
$\epsilon(0)[1-F(0)]^{-1}=2.4\times10^{-6}$, with relative
uncertainties similar to those in Eq.~\ref{eq:deteffX}.

\begin{table}
\begin{center}
\begin{tabular}{||c|c|c||}
\hline
Trap & $F$ & $\epsilon(0) (\times 10^{-6})$ \\ \hline \hline
W & 0.28 & 1.8 \\ \hline
X & 0.22 & 1.9 \\ \hline
Y & 0.26 & 1.8 \\ \hline
Z & 0.32 & 1.7 \\ \hline  
\end{tabular}
\end{center}
\caption{Summary of initial detection efficiencies for the various
samples.  For Trap X, $\epsilon(0)$ was determined by the method
described in the text.  For the other traps, $\epsilon(0)$ has been
scaled according to $(1-F)$, where $F$ is the fraction of fluorescence
photons in the detection solid angle which are absorbed by the $1S$
cloud.  While the absolute accuracy of the $\epsilon(0)$ values
derives from the uncertainty of the Trap X measurement (lower relative
error $\simeq30$\%, upper relative error $\simeq60$\%), the
ratios of the values for different traps are estimated to have an
uncertainty of only about 10\%.}
\label{tab:deteffs}
\end{table}

With knowledge of the detection efficiency, the population of a
metastable cloud can be inferred from the number of signal photons
detected during the quench pulse.  In Trap W, for example, as many as
150 counts were observed for a single laser shot, implying
$N_{2S}\simeq8\times10^7$.  Depending on the separation of the laser
focuses, the effective volume of the $2S$ cloud in Trap W lies between
$4\times10^{-4}$ and $1.2\times10^{-3}$~cm$^3$.  This means that peak
metastable densities as high as $\sim 10^{11}$~\percc\ have been achieved.

\section{Two-Body Loss: Experimental Results}
\label{sec:experimentalresults}
\subsection{$K_2$ in the Static Approximation}

With values for $\zeta$ and $\epsilon$ in hand, we can finally turn
our attention again to $\alpha_2$ and the determination of $K_2$.

\begin{figure}
\centerline{\epsfig{file=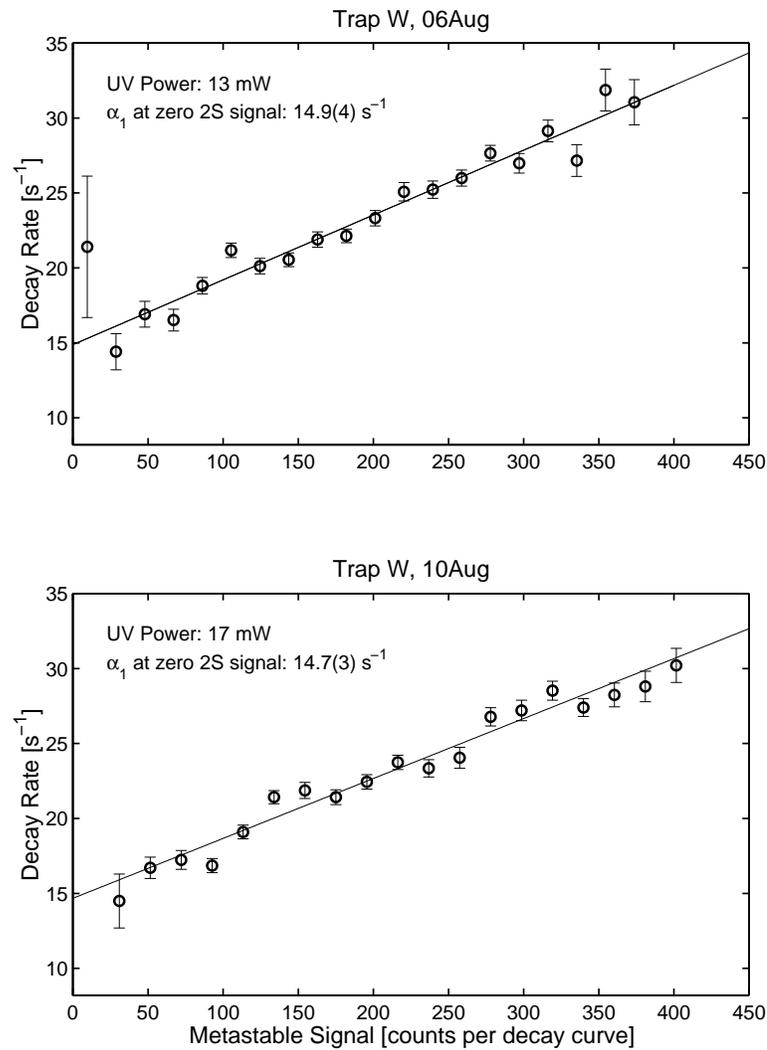,width=4in}}
\caption{Determination of average $\alpha_1$ by extrapolation of exponential
decay rates to zero metastable density for two complete data sets in
Trap W.  The data sets were taken four days apart under identical
conditions, except that the excitation laser power was 30\% larger in
the lower plot.  The values for $\alpha_1$ are in good agreement,
indicating that systematics associated with laser power are
insignificant.  The fact that the peak signal increases by less than
10\% for a 30\% increase in UV power indicates that these experiments
are in a saturation regime for metastable production.}
\label{fig:TrapWalpha1}
\end{figure}

Before fitting the decay curves of a given data set to extract
$\alpha_2$, the fixed value for $\alpha_1$, the ``background''
one-body loss rate, must be chosen.  This is accomplished by fitting
decay curves at different metastable signal levels with simple
exponentials and extrapolating the corresponding decay rates to zero
metastable signal.  Plots of the best-fitting exponential decay rate
$\alpha$ against metastable signal show, to a good approximation, that
decay rates are linear with metastable signal, suggesting that linear
extrapolation is a reasonable way to obtain $\alpha_1$.

\begin{figure}
\centerline{\epsfig{file=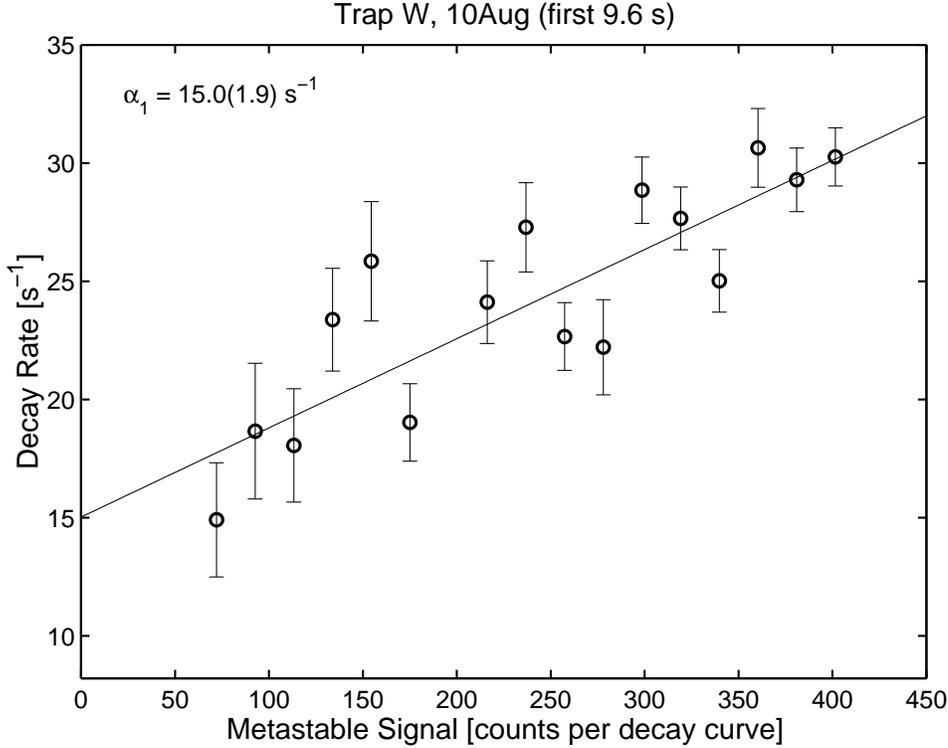,width=5in}}
\caption{An extrapolation similar to those in
Fig.~\ref{fig:TrapWalpha1}, but using decay curves from only the first
2 sweeps across the \oneStwoS\ resonance, during which time the ground
state density is known and relatively constant.  Error bars on the
decay rates are derived assuming only statistical errors in the decay
curves, which means they are smaller than the true uncertainties.  To
avoid underestimating the error on $\alpha_1$, the uncertainty derived
from the linear fit has been multiplied by the square root of the
$\chi^2$ statistic of the fit.}
\label{fig:alpha1from10Aug}
\end{figure}

Figure~\ref{fig:TrapWalpha1} shows linear extrapolations of the
one-body rates to zero signal for two different data sets in Trap~W.
The resulting values of $\alpha_1$, employing decay curves from the
entire data sets, are in good agreement.  Since the data sets were
taken on different days with different laser powers, the agreement
provides evidence that (1) trap conditions, including stray electric
fields, were relatively constant over a period of several days and (2)
systematic effects associated with changes in the laser power are
small.

The decay data compiled to create Fig.~\ref{fig:TrapWalpha1} was taken
continuously over the first 120~s after sample preparation at a rate
of one decay curve every 0.8~s.  The ground state sample depletes
substantially during this time due to dipolar decay.  In Trap W, for
example, the peak ground state density decreases by two-thirds.  To
minimize the role of systematics associated with changes in the ground
state sample density, only decay curves taken within the first 10-15~s
after sample preparation were considered in the determination of
$K_2$.  In all of the traps, the $1S$ density decreased by less than
20\% in the time required to acquire this data subset.  Furthermore,
since the laser duty cycle was only 2\%, ground state losses due to
laser heating and excitation were insignificant.  Since $\alpha_1$ can
vary as the ground state density changes (see
Sec.~\ref{sec:onebodyloss}), the same 10-15~s subset of decay curves
was employed to determine a value of $\alpha_1$ to use when fitting
to the one-plus-two model.  As an example,
Fig.~\ref{fig:alpha1from10Aug} shows the linear extrapolation used to
find $\alpha_1$ for the first 9.6~s of a Trap W data set.

\begin{figure}
\centerline{\epsfig{file=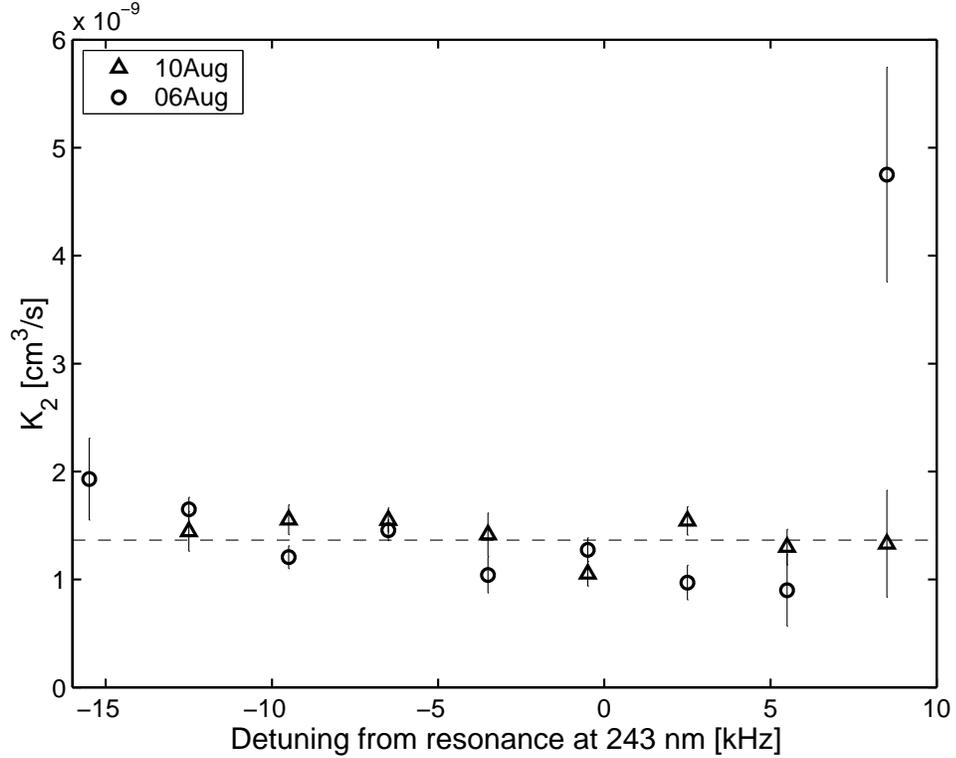,width=5in}}
\caption{Experimental results for the two-body loss constant $K_2$ in
Trap~W, derived in the static approximation.  The decay data has been
binned by laser detuning.  A value of $1.8\times10^{-6}$ has been used
for $\epsilon$, and $\zeta$ has been calculated for each laser
detuning assuming perfect overlap of the laser focuses (see
Fig.~\ref{fig:zetabydetW}).  Results from two data sets with different
laser powers are in reasonable agreement.  The error bars on
individual points derive from the fits to the one-plus-two model which
include only statistical uncertainties.  The weighted average (dashed
line) of all points is $1.4\times10^{-9}$~cm$^3$/s.  The error bars on
$K_2$ are discussed in the text.}
\label{fig:K2bydetW}
\end{figure}

After $\alpha_1$ has been determined, the decay data is fit to the
one-plus-two model.  To accumulate enough statistics, a typical data
set encompasses five consecutive trap cycles.  The decay curves taken
during the first two or three sweeps across resonance in each trap
cycle are binned according to laser detuning.  The resulting curves
are fit with Eq.~\ref{eq:oneplustwo}, using only $\alpha_2$ and $A_o$
as free parameters.  The values for $\alpha_2$ at different laser
detunings are substituted into Eq.~\ref{eq:staticK2} to obtain values
for $K_2$ in the static approximation.  Since the $1S$ density and the
cold-collision shift do not change much in 10-15~s, it is a good
approximation to assume that the value of $\zeta$ calculated for each detuning
is constant.  As shown in Fig.~\ref{fig:K2bydetW}, the $K_2$ values
are relatively independent of detuning, indicating that the variations
in metastable cloud shape with detuning are reasonably accounted for
by the spatial distribution calculations.  The experimental value of
$K_2$ for a given trap is taken to be the weighted average of the
values determined at different detunings.

\begin{figure}
\centerline{\epsfig{file=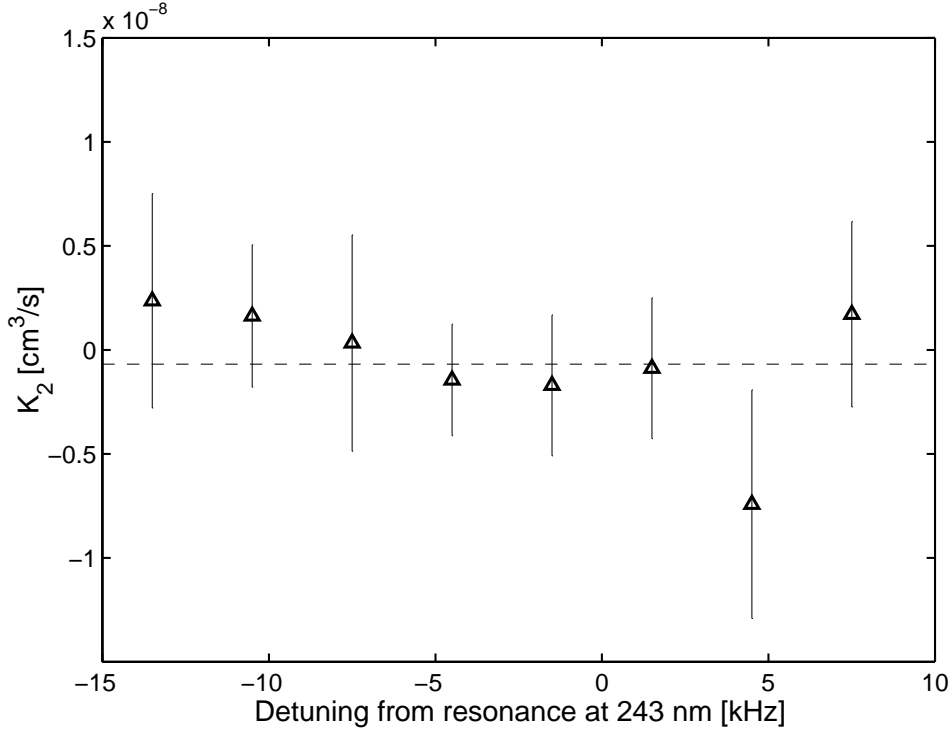,width=5in}}
\caption{Determination of $K_2$ in Trap~Z assuming perfect overlap of the
laser focuses and a detection efficiency $\epsilon =
1.7\times10^{-6}$.  The weighted average (dashed line) is consistent
with zero within the statistical uncertainties.  By summing the
systematic uncertainties, an upper limit for $K_2$ of $7 \times
10^{-9}$~cm$^3$/s is obtained.}
\label{fig:K2bydetZ}
\end{figure}

Table~\ref{tab:SAresults} summarizes the results for $K_2$ obtained in
the static approximation.  The asymmetric error bars for Traps W and X
derive mostly from systematic uncertainties, which are summarized in
Sec.~\ref{sec:uncertainties}.  In traps Y and Z, the weighted average
of $K_2$ is consistent with zero (Fig.~\ref{fig:K2bydetZ}).
Nevertheless, an upper limit for $K_2$ has been established by linear
addition of the uncertainties in $\alpha_2$, $\epsilon$, and $\zeta$
appropriate for these traps.  If $K_2$ maintains the same order of
magnitude across the temperature range spanned by Traps W, X, Y, and
Z, it is not surprising that two-body loss is difficult to detect in
Traps Y and Z, given the much lower metastable densities
and the sizeable systematics.

At red detunings larger than 10~kHz and blue detunings larger than
5~kHz, the metastable cloud in Trap W appeared sometimes to decay very
rapidly in the first 10-20~ms after excitation, implying values of
$K_2$ far above the wighted average.  For example, note the outlying
point in the most blue-detuned bin of Fig.~\ref{fig:K2bydetW}.  The
origin of these anomalously steep decay curves is not yet understood.
Perhaps at large detunings either (1) the metastable distributions
calculated by Monte Carlo simulation are inaccurate or (2) transient
redistribution of the $2S$ cloud in the first $\sim10$~ms somehow
leads to an accelerated decay of the observed fluorescence.  The
statistical weight of the anomalously fast decay curves is small,
however, and their inclusion in the weighted average has little effect
on the final results.

\begin{table}
\begin{center}
\begin{tabular}{||c|c|c||}
\hline
Trap & $T$~(mK) & $K_2$~$(10^{-9}$~cm$^3$/s) \\ \hline \hline
& & \\
W & 0.087 &$1.4^{+1.3}_{-0.5}$ \\ \hline
& & \\
X & 0.23 & $0.74^{+0.70}_{-0.35}$ \\ \hline
& & \\
Y & 0.45 & $<2$ \\ \hline
& & \\
Z & 2.3 & $<7$ \\ \hline
\end{tabular}
\end{center}
\caption{Results for $K_2$ in the static approximation.  The sources
of uncertainty will be discussed in Sec.~\ref{sec:uncertainties}.}
\label{tab:SAresults}
\end{table}

A check was made to see if the results for $K_2$ were biased by the
fact that only the first few time points in the decay curve are
sensitive to two-body decay.  For example, consider that during
the first 30~ms following excitation of a large metastable population
in Trap W, the $2S$ number decays by a factor of 4.  The two-body loss
rate, which depends locally on the square of the metastable density,
decreases dramatically during this time.  For the remaining
$\sim60$~ms of the measured decay curve, one-body loss dominates.  The
check consisted in repeating the measurements in Trap~W with eight
wait times concentrated in the first 30~ms of the decay.  The
resulting value for $K_2$ agreed to within 10\% of the previous
measurements made with eight time points distributed across 92~ms.
This confirms that the results were not significantly biased by the
chosen distribution of wait times.

\subsection{Uncertainties: Static Approximation}
\label{sec:uncertainties}
The uncertainties relevant to determination of $K_2$ under the
assumption of a static $2S$ cloud shape are summarized for Trap W in
Table~\ref{tab:uncertainties}.  Most of the uncertainties are
systematic in nature, stemming from imprecise knowledge of the
excitation geometry and also of the sample temperature and density.
The uncertainties which enter through $\zeta$ have been estimated by
varying the inputs to the numerical calculation of the $2S$
distribution.  The overall error bars are asymmetric due to asymmetric
uncertainty in the detection efficiency and the fact that laser
misalignments generally lead to an increase in $\zeta$.

\begin{table}
\begin{center}
\begin{tabular}{||l|c|c||}
\hline
Source of Uncertainty & Est. Contribution to & Est. Contribution to \\
       & Lower Rel. Error (\%) & Upper Rel. Error (\%) \\
\hline \hline
$\alpha_2$ & 12 & 12 \\ \hline
$\epsilon$ & 32 & 58 \\ \hline
$\zeta$ & & \\
\ \ overlap of waists & 0 & 75 \\
\ \ size of waists & 16 & 16 \\
\ \ trap shape & 5 & 5 \\
\ \ cold-collision shift & 4 & 4 \\
\ \ trap/laser focus offset & 0 & 3 \\ 
\ \ $n_{1S}$ & 3 & 3 \\
\ \ $T$ & 2 & 2 \\ 
\ \ laser linewidth & $< 1$ & $< 1$ \\
\ \ photoionization & $< 1$ & $< 1$ \\
\hline
Total Relative Error (\%) & 38 & 97 \\
\hline
\end{tabular}
\end{center}
\caption{Uncertainties for the static approximation determination of
$K_2$ in Trap W.  To obtain the total relative error, the
values in the table are added in quadrature, excluding the $\zeta$
contributions from ground state density $n_{1S}$ and the temperature
$T$ (see explanation in text).  The ``overlap of waists'' refers to
the separation of incoming and return laser focuses along $z$, while
the ``trap/laser focus offset'' refers to the location of the laser
focuses in $z$ with respect to the trap minimum.  The contribution labeled
``cold-collision shift'' is due to uncertainty in the parameter which
relates the shift of the \oneStwoS\ transition to the $1S$ density.}
\label{tab:uncertainties}
\end{table}

The uncertainty budget for Trap X is similar to
Table~\ref{tab:uncertainties}.  For Trap X, however, the uncertainty
in $\alpha_2$ is 30\%, and the upper relative error associated with
overlap of the focuses is 65\%.  The smaller relative errors
associated with $\zeta$ have not been explicitly calculated for Trap
X, but are assumed not to differ significantly from those in Trap W.
In both of these traps, the uncertainty in $\alpha_2$ stems mainly
from the imprecise extrapolation of $\alpha_1$; the parameters
$\alpha_1$ and $\alpha_2$ are highly correlated in the one-plus-two
model.

With a couple exceptions, the uncertainties in
Table~\ref{tab:uncertainties} are assumed to be uncorrelated.  The
errors in ground state temperature and density are responsible for
most of the uncertainty in $\epsilon$.  However, the $T$ and $n_{1S}$
contributions to $\epsilon$ are somewhat anti-correlated with the
corresponding contributions to $\zeta$.  In other words, an error in
$T$ (or $n_{1S}$) leads to errors in $\epsilon$ and $\zeta$ with
opposite signs.  To be conservative, the possibility of small partial
cancellations is ignored by neglecting the $n_{1S}$ and $T$
contributions to $\zeta$.  The estimated total relative error is found
by adding all other uncertainties in quadrature.

\subsection{Effects of the Changing Metastable Cloud Shape}
\label{sec:dynamicsimulation}
As acknowledged before, the shape of the metastable cloud changes
while it is decaying.  Although thermal equilibrium is maintained in the radial
direction, the shape of the cloud evolves in the axial direction due
to diffusion and to two-body loss, which occurs preferentially where
metastable density is highest.  To estimate the corrections which must
be made to the static approximation analysis in order to account for these
dynamics, a numerical simulation of the decaying metastable cloud was
developed.

The simulation begins with a distribution $f_{2S}({\bf r})$, calculated
for a particular trap, excitation geometery, and laser detuning
by the Monte Carlo method of Sec.~\ref{sec:spatialdist}.  Other inputs
to the simulation include the peak ground state density $n_{o,1S}$, a putative
value for $K_2$, a uniform one-body loss rate $\alpha_1$, a
one-dimensional diffusion parameter, the detection efficiency
$\epsilon$, and $s_o$, the number of signal counts which would be
experimentally observed when quenching immediately after the
excitation pulse.  

The diffusion parameter is the proportionality
constant $C_{\rm diff}$ in the relation
\be
\sigma_z(n_{a,1S},t) = C_{\rm diff}\left(\frac{t}{n_{a,1S}}\right)^{1/2},
\ee
where $\sigma_z$ is the rms diffusion length of $2S$ atoms along $z$
in a background $1S$ gas, $t$ is the corresponding diffusion time, and
$n_{a,1S}$ is the $1S$ density on the trap axis.  Using the Monte Carlo
diffusion simulation described in Sec.~\ref{sec:diffsim}, $C_{\rm
diff}$ was calculated to be $4.3\times10^5$~(cm ms)$^{-1/2}$ at the
minimum of Trap W.  This constant has some dependence on the local
radial curvature of the trap.  However, for the purpose of simulating
the evolution of the metastable cloud in the presence of large
two-body losses, it is a good approximation to neglect the variation of $C_{\rm
diff}$ with $z$.

\begin{figure}
\centerline{\epsfig{file=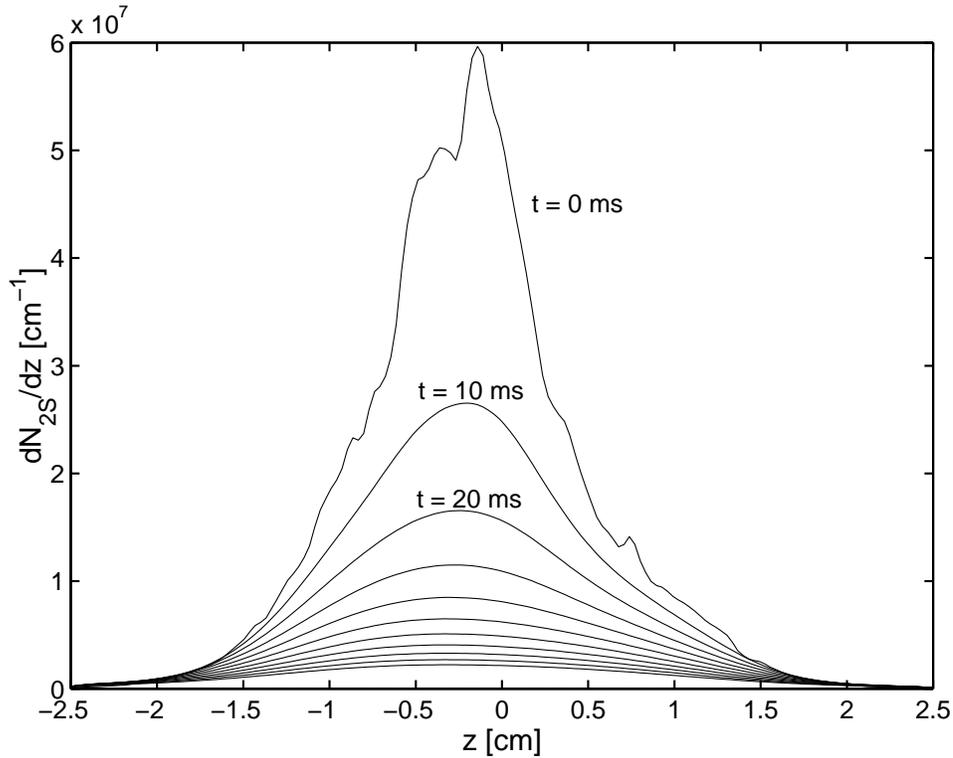,width=5in}}
\caption{Results from a dynamic simulation of the time evolution of a
metastable cloud in Trap W.  The plot depicts the distribution of $2S$
atoms, $dN_{2S}/dz$, in the $z$ direction as the population decays;
the distribution is shown at 10~ms intervals from $t=0$~ms to
$t=100$~ms following the excitation pulse.  The ripples on the initial
curve are due to the finite number of trajectories used in the Monte
Carlo calculation of the $2S$ spatial distribution.  Diffusion and
loss processes in the simulation quickly smooth out the ripples.  In
this example, the laser focuses were assumed to be perfectly
overlapped, and the initial number of $2S$ atoms was $7\times10^7$,
which approximates the largest number of metastables excited in the
decay measurements. }
\label{fig:2Sdistvstime}
\end{figure}

In the simulation, the initial number of metastables $s_o/\epsilon$ is
used to establish an initial absolute $2S$ distribution.  At any given
time, if the number distribution $dN_{2S}(z)/dz$ is known, the density
distribution $n_{2S}({\bf r})$ can be calculated everywhere assuming
that the radial distribution is Maxwell-Boltzmann.  For each step in
time $\Delta t$, the total number of atoms lost to one- and two-body
processes is numerically integrated as a function of $z$, and
$dN_{2S}(z)/dz$ is adjusted accordingly.  To account for diffusion
during the time step, the distribution in $z$ is also convoluted with
a Gaussian of rms width $\sigma_z(z,\Delta t)$, whose $z$ dependence
comes from $n_{a,1S}(z)$.  The time evolution of $dN_{2S}(z)/dz$ for
an example simulation in Trap W is shown in
Fig.~\ref{fig:2Sdistvstime}.

\begin{figure}
\centerline{\epsfig{file=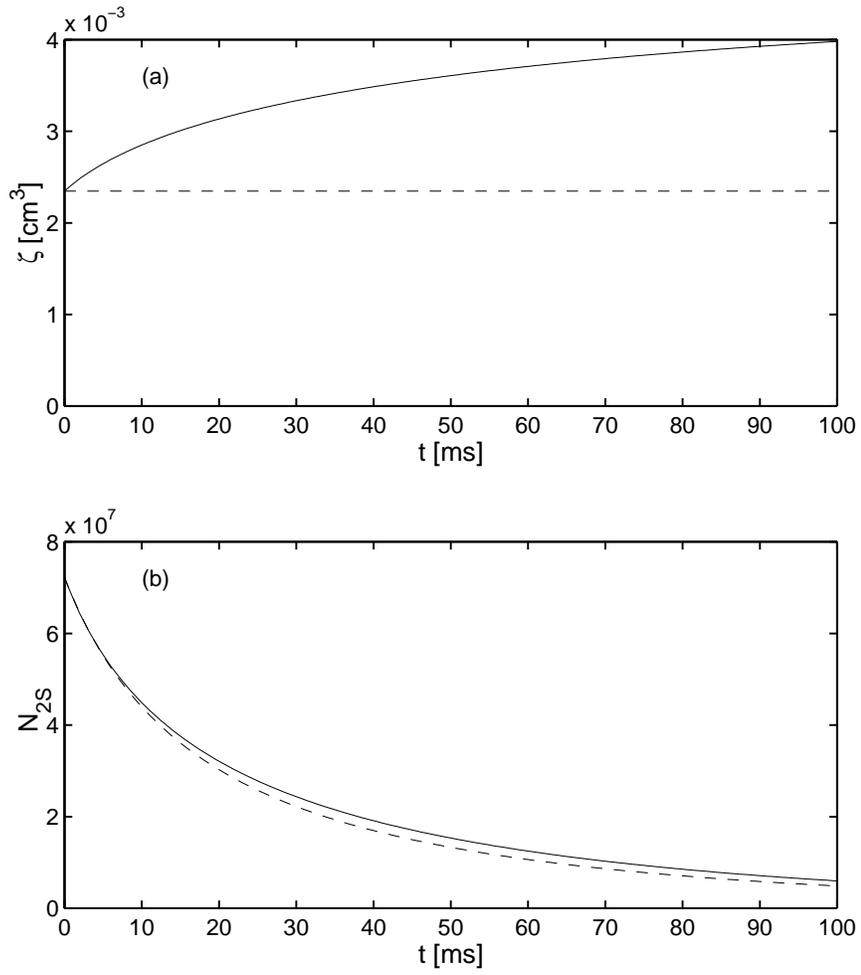,width=4.5in}}
\caption{Simulation results for $\zeta$ and $N_{2S}$ in dynamic
(solid curves) and static (dashed curves) evolution of the metastable
cloud.  The initial conditions are the same as those of
Fig.~\ref{fig:2Sdistvstime}. }
\label{fig:zetaN2Svstime}
\end{figure}

To compare the cases of ``static'' and ``dynamic'' evolution of the
metastable cloud, the dynamic simulation also evolves the number in a
static cloud starting from the same initial conditions.  At each time
step, loss is calculated in the static cloud in the same way as in the
dynamic cloud, but only the total $2S$ number is adjusted and not the
relative distribution.  The static and dynamic evolutions for the
initial conditions of Fig.~\ref{fig:2Sdistvstime} are compared in
Fig.~\ref{fig:zetaN2Svstime}.  After each step in time, the geometry factor
$\zeta$ is calculated for both static and dynamic cases.  Since diffusion and
two-body loss cause $f_{2S}({\bf r})$ to become less sharply peaked,
$\zeta$ increases with time in the dynamic simulation.  The rate of
increase slows as the peak $2S$ density drops and two-body loss
becomes less important relative to one-body loss.  The effect of the
increase in $\zeta$ in the dynamic case is to decrease the
two-body loss rate relative to the static case
(Eq.~\ref{eq:twobodylossvszeta}).  This effect is only important,
however, during the early part of the decay when two-body loss dominates.

\begin{figure}
\centerline{\epsfig{file=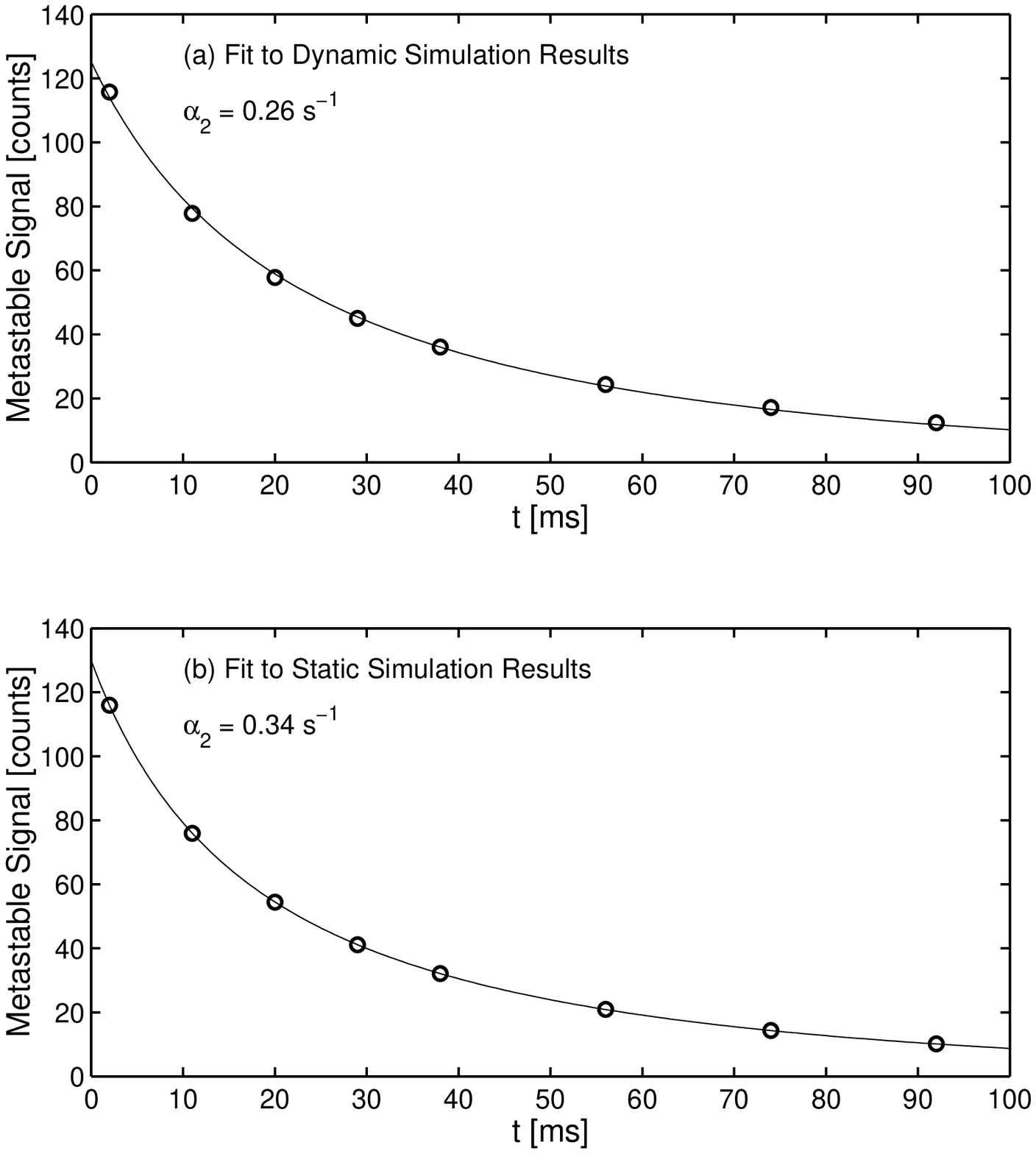,width=4.5in}}
\caption{Fits of the one-plus-two model to simulated decay data
derived from the $N_{2S}(t)$ curves of Fig.~\ref{fig:zetaN2Svstime}.  In
both cases, the one-body constant $\alpha_1$ has been fixed at an
experimental value of 15~s$^{-1}$, and the one-plus-two model is an
excellent fit to the simulation results.  The ratio of $\alpha_2$ values
determined in the two cases provides a correction factor for
$\alpha_2$ in the static approximation.}
\label{fig:fitstostatdyn}
\end{figure}

As illustrated in Fig.~\ref{fig:fitstostatdyn}, the results of the
simulation can be used to calculate a dynamic correction for $\alpha_2$
in the one-plus-two model (Sec.~\ref{sec:oneplustwo}).  In the figure, the
simulated $N_{2S}(t)$ curves have been multiplied by $\epsilon$ at the
experimental wait times to obtain decay curves similar to those in the
experiment.  Applying statistical error bars and using an
experimentally determined value for $\alpha_1$, the simulated static
and dynamic decay curves are fit to the one-plus-two model
to find $\alpha_2$ in each case.  A
dynamic correction factor $f_{\rm dc}$ can be defined as follows:
\be
f_{\rm dc} = \frac{\alpha_{\rm 2, static}}{\alpha_{\rm 2, dynamic}}.
\ee
The values of $\alpha_2$ extracted from fits of the one-plus-two model
to experimental decay curves can be multiplied by $f_{\rm dc}$ to
account for the flattening of the metastable distribution as the cloud
decays.

The simulation was executed many times with the parameters of Traps W
and X to find $f_{\rm dc}$ as a function of laser detuning.  These
results were used to correct the static approximation values for
$K_2$, and a new weighted average $K_2$ was determined for each trap.
The input parameters to the simulation were varied to estimate the
uncertainty in the overall correction to $K_2$.  It was found that the
static approximation values of $K_2$ must be multiplied by $1.35$ and
$1.30$, respectively, to correct for dynamics in Traps W and X.  The
uncertainty in these factors is estimated to be 10\%.  For Traps Y and
Z, a conservative upper limit for the correction factor is 2, and the
upper limits for $K_2$ in the static approximation are multiplied
accordingly.  The corrected experimental values for $K_2$ are listed
in Table~\ref{tab:correctedK2}.

In principle, the dynamic simulation could be used to accurately
``fit'' the experimental decay curves.  However, the simulation has
too many weakly constrained parameters for a more accurate determination
of $K_2$ than what has already been accomplished by correcting the
static approximation.  In addition, since many iterations of the
simulation are required to find a convergence of the simulation
results, the CPU time required is enormously larger than when
fitting to the simple one-plus-two model. 

It should be noted that the one-body loss rate was assumed in the
simulation to be uniform over the entire sample.  This is a good
assumption if the one-body loss rate is mostly due to natural decay
and quenching by stray fields.  In Trap W, however, it is possible
that there is significant one-body loss due to inelastic
\oneStwoS\ collisions (see Sec.~\ref{sec:onebodyloss}).  Since the
contribution to the one-body loss rate due to \oneStwoS\ collisions is
proportional to the ground state density, the total one-body loss rate
in this case will be appreciably higher where the ground state sample
is most dense, and there will be additional flattening of the $2S$
spatial distribution.  This additional flattening is not likely to
perturb the results for $K_2$ by more than a few percent, however.

\begin{table}
\begin{center}
\begin{tabular}{||c|c|c||}
\hline
Trap & $T$~(mK) & $K_2$~$(10^{-9}$~cm$^3$/s) \\ \hline \hline
& & \\
W & 0.087 &$1.8^{+1.8}_{-0.7}$ \\ \hline
& & \\
X & 0.23 & $1.0^{+0.9}_{-0.5}$ \\ \hline
& & \\
Y & 0.45 & $<4$ \\ \hline
& & \\
Z & 2.3 & $<14$ \\ \hline
\end{tabular}
\end{center}
\caption{Results for $K_2$ including dynamic corrections.}
\label{tab:correctedK2}
\end{table}

The effects of the rf evaporation field (present only in Trap W) on
the metastable cloud were neglected in the simulation.  In principle,
since the hyperfine structure of the $2S$ atoms is the same as the
$1S$ atoms at low fields, the rf ``surface of death'' can eject $2S$
atoms with the same probability as $1S$ atoms.  Experimentally,
however, this additional source of loss appears not to be significant
for the rf field strengths in our apparatus.  Decay measurements were
made with a cold, dense sample where the rf field was switched off
during the wait time before each quench pulse.  The results were
compared to the case where the rf field was on throughout the decay
measurements.  Within uncertainties, no difference in decay behavior
was observed.

Evaporation of metastables over the magnetic saddlepoint threshold
should not play a role in the evolution of the metastable cloud in any of
the traps. The $2S$ atoms do not have time to diffuse from the region
of excitation to the magnetic barrier.

\subsection{Summary of Two-Body Loss Results}

\begin{figure}
\centerline{\epsfig{file=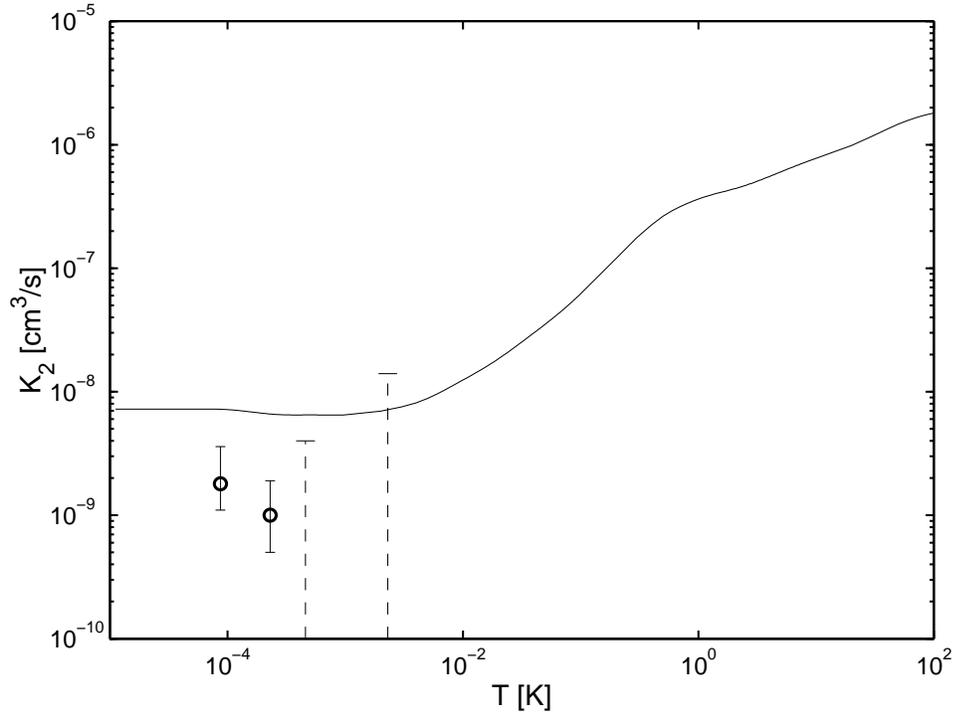,width=5in}}
\caption{Theoretical and experimental values for $K_2$ as a
function of temperature.  The solid line is the sum of theoretical
loss rate constants due to \twoStwoS\ excitation transfer and
ionization processes, calculated by Forrey and collaborators
(Sec.~\ref{sec:theoreticalrateconstants}).  Experimental values are
indicated by circles.  At the temperatures where only
upper limits were determined, dashed lines indicate the possible range
of $K_2$.}
\label{fig:K2theoryandexp}
\end{figure}

The final experimental values for the two-body loss constant of
metastable H are plotted in Fig.~\ref{fig:K2theoryandexp} along with a
theoretical curve based on Eq.~\ref{eq:K2fromtheory}.  There are at least
two features noteworthy in the comparison.  

First, in its current state of development, the theory of Forrey \etal\
predicts a somewhat larger value of $K_2$ than what was measured at
the lowest temperatures.  However, considering that the theory does
not yet include hyperfine structure or the effects of a trapping
magnetic field (Sec.~\ref{sec:theoreticalrateconstants}), corrections
which could shift the theoretical results by as much as an order of
magnitude, we conclude that the theory and experiment are in range of
one other.

Second, the two experimental data points suggest that there is a
significant temperature dependence of the inelastic rates between
87~$\mu$K and 230~$\mu$K.  Although the uncertainty ranges on the
absolute values of $K_2$ are large enough to overlap, the error bars
are mostly due to systematic uncertainties which affect both points in
approximately the same way.  When the correlated errors are removed,
our analysis leads to the conclusion that $K_2$ increases by a factor
of $1.9\pm0.4$ as the temperature decreases from 230~$\mu$K to
87~$\mu$K.  This temperature dependence is considerably larger than
predicted by theory.  It will be interesting to see whether further
theoretical refinements resolve this discrepancy.

\boldmath
\section{One-Body Loss and Inelastic \oneStwoS\ Collisions}
\unboldmath
\label{sec:onebodyloss}
Taking advantage of the fact that metastable decay data
is taken continuously while the $1S$ sample decays, we can look for a
dependence of decay rates on the $1S$ density.  This will lead to
quantitative conclusions about the prevalence of loss caused by
inelastic \oneStwoS\ collisions in the trap.  

The total one-body decay rate $\alpha_1$ for a metastable cloud is the
sum of several rates, including the natural decay rate, the quenching
rate due to stray electric fields, the loss rate due to the generation
of helium vapor by laser heating, and any \oneStwoS\ inelastic
collision rates.  Ejection of metastables by an rf evaporation field
would also contribute to the total one-body loss rate; as mentioned in
Sec.~\ref{sec:dynamicsimulation}, however, loss due to rf has not been
detected in our apparatus.  The natural decay rate is well known:
$\gamma_{\rm nat} = 8.2$~s$^{-1}$.  Loss due to residual electric
fields is difficult to measure directly, but field compensation
measurements (Sec.~\ref{sec:strayfield}) repeated several times in a
single data-taking session showed that the rate varies by less than
1~s$^{-1}$ over several hours.  In the decay experiments reported
here, the laser heating loss is expected to be very small due to the
small laser duty cycle.  If there are inelastic \oneStwoS\ collisions
causing measurable metastable loss, then the one-body decay rate
should vary with the ground state density.  Experimentally, the
parameter $\alpha_1$, found by extrapolating fitted exponential decay
rates to zero metastable density, ranges between 10 and 17~s$^{-1}$ in
the initially prepared samples of Traps W, X, Y, and Z.  There is not
an obvious trend with $1S$ density, though, which may be due of the
fact that these samples are at different temperatures.

\begin{figure}
\centerline{\epsfig{file=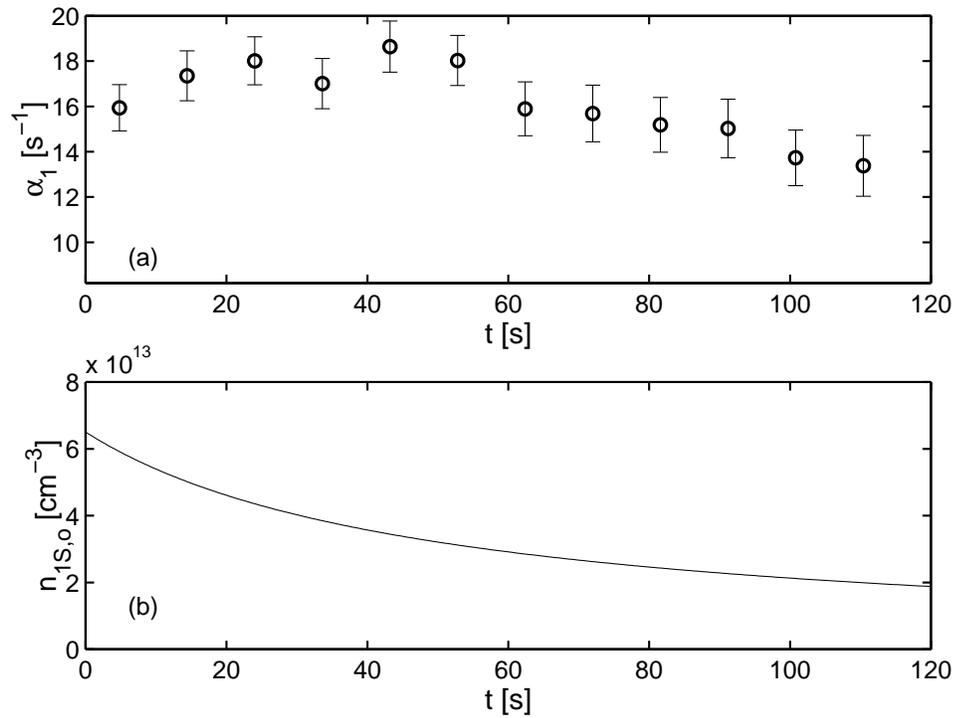,width=5in}}
\caption{(a) Weighted average of $\alpha_1$ determined by
extrapolation to zero metastable signal in Trap W as a function of
time after sample preparation.  Each point represents a value of
$\alpha_1$ extracted from 9.6~s of $2S$ decay data.  The one-body
metastable loss rate appears to decrease as the ground state sample
decays, suggesting that some of the one-body loss may be due to \oneStwoS\
inelastic collisions.  The bottom of the plot corresponds to the
natural decay rate, and error bars include only statistical
uncertainties.  (b) Peak ground state density calculated as a
function of time in Trap W assuming that the only significant $1S$
loss mechanism is dipolar decay.}
\label{fig:alpha1vstW}
\end{figure}

By dividing the decay data for a single trap into time bins
corresponding to a fixed number of sweeps across resonance, we can
examine whether the one-body loss rate changes as the ground state
sample decays at a constant temperature.
Figure~\ref{fig:alpha1vstW}(a) shows values for $\alpha_1$ determined
in consecutive time bins for Trap W.  The putative ground state
density is shown for the same time interval in
Fig.~\ref{fig:alpha1vstW}(b).  Although not conclusive, the data suggests
that the total one-body loss rate decreases as the $1S$ density
decreases.  This decrease in $\alpha_1$ may be due to a decline in
inelastic \oneStwoS\ collision rates, which are proportional to the
ground state density.  In principle, the points of
Fig.~\ref{fig:alpha1vstW}(a) could be fit to a model which includes
constant one-body decay terms and a term proportional to the $1S$
density.  The uncertainties in both the determinations of $\alpha_1$
and the density as a function of time are too large, however, to
extract a meaningful value for the total \oneStwoS\ inelastic loss
rate without prior knowledge of the stray electric field contribution.

On the other hand, it is possible to establish a conservative upper
limit on the total inelastic contribution $\alpha_{1,{\rm inel}}$ to
the one-body decay rate by subtracting the natural decay rate from $\alpha_1$:
\be
\alpha_{1,{\rm inel}} < \alpha_1 - \gamma_{\rm nat}.
\label{eq:alphainellimit} 
\ee
In the static approximation, $\alpha_{1,{\rm inel}}$ can be simply related
to the total metastable loss rate constant for inelastic \oneStwoS\
collisions $K_{12}$, defined by
\be
\Bigl. \dot{n}_{2S}({\bf r}) \Bigr|_{1S-2S} = - K_{12}\,n_{1S}({\bf
r}) n_{2S}({\bf r}).
\label{eq:K12definition}
\ee
Note that $K_{12}$ only encompasses collisions resulting in loss of a $2S$
atom from the trap; for example, collisions in which only the $1S$
atom changes state are not included.  Integrating both sides of
Eq.~\ref{eq:K12definition} over all space, we obtain
\begin{eqnarray}
\Bigl. \dot{N}_{2S} \Bigr|_{1S-2S} & = & - K_{12} \int n_{1S}({\bf
r}) n_{2S}({\bf r})\,d^3{\bf r} \\
& = & -K_{12} n_{1S,o} n_{2S,o} \int f_{1S}({\bf r}) f_{2S}({\bf r})\,d^3{\bf r}\\
& = & -\left( \frac{K_{12} n_{1S,o} Q_{12}}{V_{2S}}\right) N_{2S},
\label{eq:lossratefromK12}
\end{eqnarray}
where $Q_{12} = \int f_{1S}({\bf r}) f_{2S}({\bf r})\,d^3{\bf r}$ is
an overlap integral for the ground state and metastable clouds.
Since by definition
\be
\Bigl. \dot{N}_{2S} \Bigr|_{1S-2S} = -\alpha_{1,{\rm inel}} N_{2S},
\label{eq:lossratefromalphainel}
\ee
it immediately follows that
\be
K_{12} = \frac{\alpha_{1,{\rm inel}} V_{2S}}{n_{1S,o} Q_{12}}.
\ee

By combining an upper limit for the initial value of $\alpha_{1,{\rm
inel}}$ in a trap with an upper limit for the ratio $V_{2S}/Q_{12}$
and the known peak density $n_{1S,o}$, an upper limit for $K_{12}$ is
obtained.  An upper limit for $\alpha_{1,{\rm inel}}$ has been
estimated for each trap by first fitting a line to $\alpha_1$ as a
function of time and then subtracting $\gamma_{\rm nat}$ from the
initial value of the fitted line.  In Trap W, for example, the data
suggests that $\alpha_1$ may be as large as 18~s$^{-1}$, implying an
upper limit of 10~s$^{-1}$ for $\alpha_{1,{\rm inel}}$.  The ratio
$V_{2S}/Q_{12}$ can be computed numerically using the Monte Carlo $2S$
distribution results; it has a maximum value of approximately 3 for
all traps with all probable excitation geometries.  Upper limits for
$K_{12}$ are summarized in Table~\ref{tab:K12results}.

The experimental upper limits for $K_{12}$ are more than two orders of
magnitude larger than the rate constant $g=1.2\times10^{-15}$~cm$^3$/s
for $1S$-$1S$ dipolar decay.  If the true value of $K_{12}$ turns out
to be closer to the upper limits in Table~\ref{tab:K12results} than to
$g$ in this temperature range, this may imply that \oneStwoS\
excitation transfer collisions (Eq.~\ref{eq:1S2Sexcitationtransfer})
are much more likely than
\oneStwoS\ hyperfine-changing collisions
(Eq.~\ref{eq:1S2Shyperfinechanging}) mediated by magnetic dipole forces.
More precise metastable decay measurements are required to
establish lower bounds for $K_{12}$.

\begin{table}
\begin{center}
\begin{tabular}{||c|c|c||}
\hline
Trap & $T$~(mK) & $K_{12}$~$(10^{-13}$~cm$^3$/s) \\ \hline \hline
W & 0.087 &$<5$ \\ \hline
X & 0.23 & $<3$ \\ \hline
Y & 0.45 & $<2$ \\ \hline
Z & 2.3 & $<11$ \\ \hline
\end{tabular}
\end{center}
\caption{Experimental upper limits for $K_{12}$ in the static approximation.}
\label{tab:K12results}
\end{table}



\chapter{Outlook: Spectroscopy of Metastable Hydrogen}
\label{ch:outlook}
\begin{figure}
\centerline{\epsfig{file=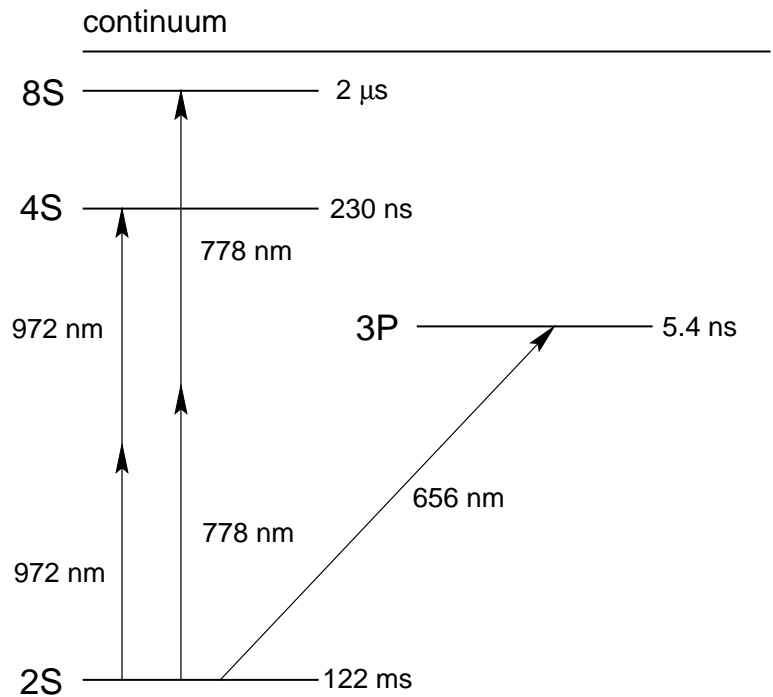,width=4in}}
\caption{Some transitions from the metastable state of hydrogen which can
be excited with readily available diode lasers.  Spectroscopy of
these transitions in trapped metastable hydrogen may provide new insight
into cold collisions and also fundamental physics.}
\label{fig:energylevels}
\end{figure}

As mentioned in the introduction to this thesis, the ability to
produce large, long-lived clouds of metastable H in a magnetic trap
opens the door for a variety of new experiments involving excited
states of hydrogen.  Of particular interest are spectroscopic
experiments involving transitions from the $2S$ state to higher-lying
states (Fig.~\ref{fig:energylevels}).  Spectroscopy of a strong
single-photon transition such as the $2S$-$3P$ Balmer-$\alpha$
transition will provide a more sensitive probe of the metastable
cloud.  Excitation on this transition may also allow the first direct
imaging of the metastable cloud.  From a fundamental physics
standpoint, the most exciting possibilities of all involve precision
spectroscopy of two-photon $2S$-$nS$ transitions \cite{wk00}.  Recent
breakthroughs in optical frequency metrology \cite{urh99} make it more
feasible than ever to measure these frequencies with unprecedented
accuracy, leading to improved values for the Lamb shift and Rydberg
constant \cite{sjd99}.

This chapter contains a brief discussion of experiments which could
feasibly be carried out in the next few years.  Much fruitful work
can be done yet with the current cryogenic apparatus.  However, to
fully exploit the promise of metastable H spectroscopy, a new
apparatus with more optical access and better detection efficiency is desired.

\section{Further Experiments with the Current Apparatus}
\subsection{Optical Refinements}
The principal sources of systematic noise and uncertainty for
experiments with metastables are currently related to the
UV excitation laser.  There are a number of possible refinements to the
optical apparatus which would dramatically reduce these systematics.

In the determination of $K_2$ described in Ch.~\ref{ch:results}, a
major source of uncertainty stemmed from uncertainty in the overlap of
the laser focuses along the trap axis.  This overlap can be done
empirically by changing the length of the telescope nearest the
cryostat (see Fig.~\ref{fig:uvlayout}) and observing the change in
metastable signal.  Assuming the centroid of the laser focuses is near
the bottom of the trap in $z$, the signal will be optimized when the
focuses are overlapped.  To perform this optimization accurately, the
telescope length should be scanned during a single trap cycle with
stable UV power.  This could be accomplished by mounting one of the
telescope lenses on a computer-controlled stepper motor.  Another
source of uncertainty is the spot size at the laser focuses.  It is
difficult to measure the beam radius at the UV waists directly, but
the radius could be inferred more accurately by measuring the beam
profile at several points outside of the cryostat.

On a time scale of seconds to minutes, fluctuations in the transverse
overlap of the laser focuses is probably the most important source of
non-statistical noise in our \oneStwoS\ spectroscopy.  Some of this
noise is apparently due to the pendulum motion of the 2.5~m-long
cryostat.  Greater efforts to mechanically isolate the cryostat from
low frequencies may alleviate this problem.  Improving the pointing
and transverse mode stability of the UV beam will also make the task
of actively stabilizing the beam overlap in the cryostat much easier.
Currently, much of the noise in the UV beam pointing and mode quality
is due to the instability of the 486~nm dye laser beam which pumps the
SHG cavity (Fig.~\ref{fig:uvlayout}).  The pointing and mode stability
of the dye laser output could be greatly enhanced by spatial
filtering, such as by passing the the beam through a short length of
fiber.  This would come at the expense of a significant amount of
power, but the trade-off of power for stability may be worthwhile for some
experiments.

With a more stable UV laser and less uncertainty in the laser
geometry, the decay measurements described in this thesis could be
performed with a better signal-to-noise ratio, and interpretation of
the data would be more straightforward.  In general, the geometry of
the metastable cloud could be predicted more accurately, thus reducing
a major uncertainty in the measurement of $K_2$.  Another important
source of uncertainty, the calibration of the detection efficiency,
would also benefit from a better-known and more stable laser geometry.
However, it may be difficult to significantly improve the measurement
of $K_2$ or other collisional parameters without a new method for
metastable detection.  One possibility is Balmer-$\alpha$
spectroscopy, discussed below in Sec.~\ref{sec:balmeralpha}.

\subsection{Excitation Studies}
\begin{figure}
\centerline{\epsfig{file=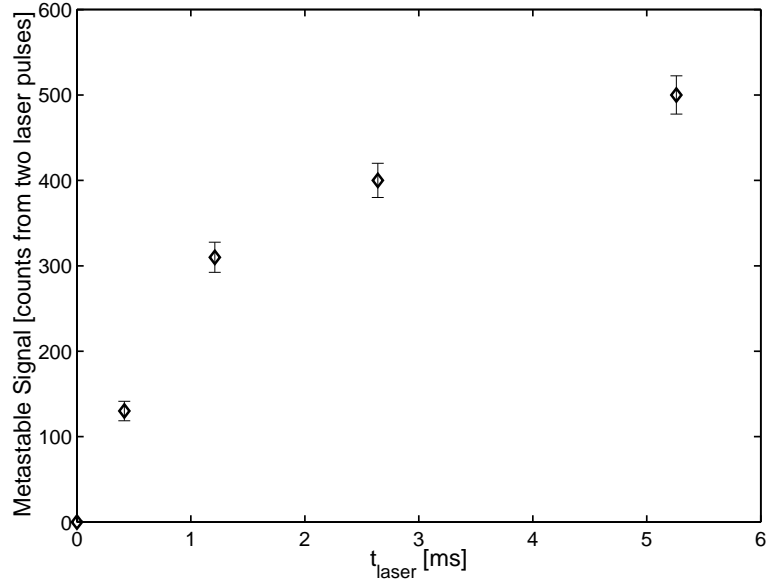,width=4in}}
\caption{Metastable signal observed for different excitation pulse
lengths when quenching at a fixed wait time of 5~ms after the
excitation pulse.  The sample was prepared at roughly 200~$\mu$K in a
magnetic trap similar to Trap W, and the laser was tuned to the peak
of the Doppler-free \oneStwoS\ line.  The number of metastables begins
to saturate at long pulse times due to photoionization and two-body
loss.}
\label{fig:saturation}
\end{figure}

Information about photoionization and two-body loss can be gleaned
from studies of the number of metastables excited as a function of the
excitation pulse length.  Results from a preliminary study of this
type can be seen in Fig.~\ref{fig:saturation}.  The number of $2S$
atoms generated begins to saturate after a few milliseconds.  This is
consistent with simulations of excitation behavior which incorporate
the theoretical value for the photoionization cross-section at 243~nm.
In cold, dense samples, \twoStwoS\ two-body loss can play an important
role in the saturation behavior as well.

Each point in Fig.~\ref{fig:saturation} was measured in a different
trap cycle.  The duty cycle of the mechanical beam chopper and the
timing generator were adjusted between cycles to produce different
excitation pulse lengths with a constant wait time between the end of
the excitation pulse and the quench pulse.  Such a study requires a
couple hours to execute and is sensitive to drift in mode quality,
power, and alignment of the UV beam over this time scale.  A better
way to look at the excitation behavior of metastables would be to use
different excitation pulse lengths in succession during a single trap
cycle.  This could be accomplished using a fast mechanical shutter or
a chopper wheel with several different-sized apertures.  An AOM could
also be used to chop the beam at the expense of about 40\% of the UV
power.

Experimental excitation curves measured in different traps could be
analyzed with a numerical model incorporating excitation,
photoionization, and collisional loss mechanisms.  In traps where the
metastable density is low enough so that two-body loss is not
important, photoionization will be the principal loss mechanism at
high UV powers.  If geometry and laser power are well known, the data
could be analyzed to verify the photoionization cross section.
Alternatively, it may be possible to use the theoretical cross section
to calibrate the laser intensity.  Finally, in situations where the
metastable density is very large, it may be possible to correct the
saturation curve for photoionization and extract a two-body loss rate.
Excitation studies may serve as a useful complement to the decay
measurements described in this thesis.

\subsection{Balmer-$\alpha$ Spectroscopy}
\label{sec:balmeralpha}
Preliminary work has been completed in the MIT Ultracold Hydrogen
Group towards probing the $2S$-$3P$ transition with a 656~nm
diode laser.  The retromirror of the current trapping cell has a
dichroic dieletric coating which is 99\% reflective at both 243~nm and
656~nm.  Thus, red Balmer-$\alpha$ light can be introduced into the
cryostat along the same path as the UV~beam, and the retro-reflected
red light can be detected using a beam splitter outside the cryostat.
This is potentially a highly sensitive probe of the metastable cloud.
Alternatively, the absorption of red light can be surmised from the
depletion of the metastable number, observed by quenching and
detecting Lyman-$\alpha$ as before.

Diode lasers with free-running wavelengths near 656~nm and output
powers up to 50~mW are commercially available for $\sim\$100$.  The
diodes can be induced to operate at a single frequency in a single
transverse mode by optical feedback from an external grating.  The
laser frequency can be actively stabilized by piezoelectric control of
the grating angle.  With a standard external-cavity grating-stabilized
diode laser design, a linewidth of a few hundred kilohertz is easily
achieved \cite{wh91,msw92}.  For Balmer-$\alpha$ spectroscopy, it is
convenient to use the $2S$-$3P$ saturated absorption lines of a
hydrogen discharge as the frequency reference.  The stabilized laser
linewidth is small compared to the 30~MHz natural linewidth of the
Balmer-$\alpha$ transition.  At 100~$\mu$K, the Doppler width is only
about 10\% of the natural linewidth, and Zeeman broadening will be
insignificant.  Thus, in experiments with trapped metastables, the
experimental Balmer-$\alpha$ linewidth will be approximately the
natural linewidth.

The Einstein coefficients relevant for atoms in the $3P$ state are
$A_{3P-1S}=1.64\times10^8$~s$^{-1}$ and
$A_{3P-2S}=0.22\times10^8$~s$^{-1}$ \cite{bs57p266}.  The excitation
rate $W_{2S-3P}$ from $2S$ to $3P$ in the presence of a resonant
656~nm laser can be written
\be
W_{2S-3P} = \sigma_o \left(\frac{I}{\hbar \omega}\right)
\ee
where $I$ is the intensity and the resonant absorption cross-section
is
\be
\sigma_o = \left( \frac{A_{3P-2S}}{A_{3P-1S}+A_{3P-2S}}\right)
6\pi\left(\frac{\lambda}{2\pi}\right)^2 = 2.4 \times 10^{-10}\,{\rm cm}^2.
\ee
For absorption experiments, the Balmer-$\alpha$ beam waist should be
larger than the thermal radius of the metastable cloud.  This will
facilitate alignment and allow saturation intensity to be achieved
over most of the atom cloud with a modest laser power.  Consider, for
example, a metastable cloud in Trap~W, which has a thermal radius of
approximately 110~$\mu$m.  If the laser radius is three times larger,
then the minimum power required for saturation at the center of the
trap is 24~$\mu$W.  In an absorption experiment, the
Balmer-$\alpha$ laser would be pulsed at a tiny duty cycle, and even
at powers far above saturation, the heating of the trapping cell and
cryostat would be negligible.

It should be possible to make absorption measurements which are nearly
shot noise limited.  To minimize shot noise, laser pulses can be used
which contain only as many photons as are necessary to excite most of
the metastables.  For the case of a saturation power beam with radius
three times larger than the thermal radius in Trap W, a $\sim200$~ns
pulse of $\sim10^7$ photons can excite most of a cloud of $10^7$
metastable atoms to the $3P$ state.  The relative error due to shot
noise for each laser pulse will only be $\sim10^{-7/2}$, which is less
than 0.1\%.  In practice, it will be more convenient to use somewhat
longer laser pulses and lower power, but similar shot noise levels
should be achievable.  This shot noise limit would be a dramatic
improvement from the $\sim10$\% statistical error on the $\sim100$
Lyman-$\alpha$ photons observed after a typical quench pulse.

By calibrating the number of photons removed from the Balmer-$\alpha$
beam, it will be possible to measure the number of metastables in the
trap.  This may lead to a better calibration of the detection
efficiency for metastables by the quenching method.  The relative
attenuation of the 656~nm beam will additionally give the column density of
the metastable cloud.  Used in combination with the shape of the
magnetic trap, this data may allow a better determination of the
spatial distribution of metastables.  Spectroscopy of the
Balmer-$\alpha$ line may also provide information about the
temperature of the sample.  For an accurate temperature determination,
the signal-to-noise ratio will have to be high, since the Doppler
broadening of typical samples is much smaller than the natural
$2S$-$3P$ linewidth.  Other experiments, such as photoassociation
spectroscopy on the  $2S$-$3P$ transition may become possible as well.

In the long term, imaging of the metastable cloud using
Balmer-$\alpha$ fluorescence may be the most useful diagnostic of all
for trapped metastable hydrogen.  The implementation of an imaging
detector, however, will require major changes to the apparatus (Sec.~\ref{sec:ideas}).

\subsection{Precision Spectroscopy and Frequency Metrology}
In the introduction to this thesis, it was described how the
measurement of a $2S$-$nS$ frequency interval, when combined with the
accurately known $1S$-$2S$ frequency, allows various contributions to
the structure of the hydrogen atom to be deconvolved.  With a sample
of cold metastable hydrogen, it should be possible to improve the
accuracy of $2S$-$nS$ frequencies by an order of magnitude over the
best measurements to date.  This will lead to a corresponding increase
in the accuracy of the Rydberg constant and Lamb shift.  If the
frequencies of several $2S$-$nS$ transitions can be measured
precisely, we will have stringent checks on the understanding of
systematic effects.

As a first demonstration of two-photon metastable H spectroscopy, the
MIT Ultracold Hydrogen Group plans to use light from a 972~nm diode
laser to excite the the $2S$-$4S$ transition.  To perform Doppler-free
two-photon spectroscopy, a standing wave is required, and most likely
the current retromirror will have to be replaced with one designed for
higher reflectivity at 972~nm.  The diode laser technology for this
and other $2S$-$nS$ transitions is similar to what was described above
for the Balmer-$\alpha$ frequency.  The $2S$-$4S$ interval
differs by less than 5~GHz from $1/4$ of the \oneStwoS\ interval;
frequency metrology can be accomplished by looking at the beat note
between the 486~nm dye laser, which is frequency-doubled to excite the
\oneStwoS\ transition, and the second harmonic of the 972~nm laser.
Because it has been calibrated precisely against a cesium standard,
the \oneStwoS\ transition serves as a {\em de facto} secondary
frequency standard \cite{nhr00}.

Stimulated by our need for convenient optical frequency metrology, a
new mode-locked femtosecond laser has recently been developed in the
group of F.~K\"{a}rtner at MIT \cite{smg02,emk01}.  The ``double-chirped''
mirrors of this laser allow it to produce an extraordinarily broad
comb of frequencies without dispersion-compensating prisms or fiber.
It will allow direct comparison of optical frequencies in the range
600-1100~nm.  Likely candidates for precision spectroscopy are the
$2S$-$8S$ (two photons at 778~nm) and $2S$-$10S$ (two photons at
759~nm) transitions, whose frequencies have been measured previously
in beam experiments.  In an experiment with cold metastable hydrogen,
these frequencies could be calibrated with the frequency comb and the
972~nm subharmonic of the $1S$-$2S$ transition.  An alternative
approach is to frequency-triple part of the comb for direct comparison
with the 243~nm frequency.

Based on our studies with metastable hydrogen so far, we can speculate
that the optimum metastable cloud for $2S$-$nS$ spectroscopy in the
current apparatus will be similar to the one excited in Trap X
(Sec.~\ref{sec:summaryoftraps}).  In this sample, the peak $2S$
density $(> 3\times10^{10}$~cm$^{-3})$ and number $(> 4\times10^7)$ are
sufficient to allow high $2S$-$nS$ excitation rates.  At the same time,
two-body loss will not play a significant role in limiting the signal
rate; $K_2$ appears to be roughly a factor of 2 smaller at 230~$\mu$K
as compared to 90~$\mu$K (Fig.~\ref{fig:K2theoryandexp}), and the
metastable density is somewhat smaller in the warmer trap as well.  The
moderate ground state density in Trap X may also be desired to reduce the
complicating effects of cold collision shifts and inelastic collisions
involving $1S$ atoms.

Before concluding this section, it is worth mentioning that the
mode-locked femtosecond laser will also permit a new determination of
the \oneStwoS\ absolute frequency.  Using an appropriate optical
frequency divider \cite{uhg97}, two optical frequencies with an
interval having a simple relationship to the \oneStwoS\ frequency can
be produced inside the range of the mode-locked laser's frequency
comb.  With the apparatus changes described in the next section, it
should be possible to perform \oneStwoS\ spectroscopy with much lower
ground state densities, minimizing the cold collision shift, which is
currently the most source of important systematic error.  Provided a
primary reference such as a cesium fountain clock, the transition
frequency can potentially be measured with an accuracy surpassing that
of H\"{a}nsch and collaborators \cite{nhr00}.

\section{A Cryogenic Trap Optimized for Optical Studies}
\label{sec:ideas}
A redesign of the magnet system and cryostat to improve optical access
and detection solid angle will have many benefits for spectroscopic
and other optical studies of trapped hydrogen.  These include better
signal-to-noise ratios, easier alignment, more possible geometries for
probe beams, and the potential to use imaging detectors.

\begin{figure}
\centerline{\epsfig{file=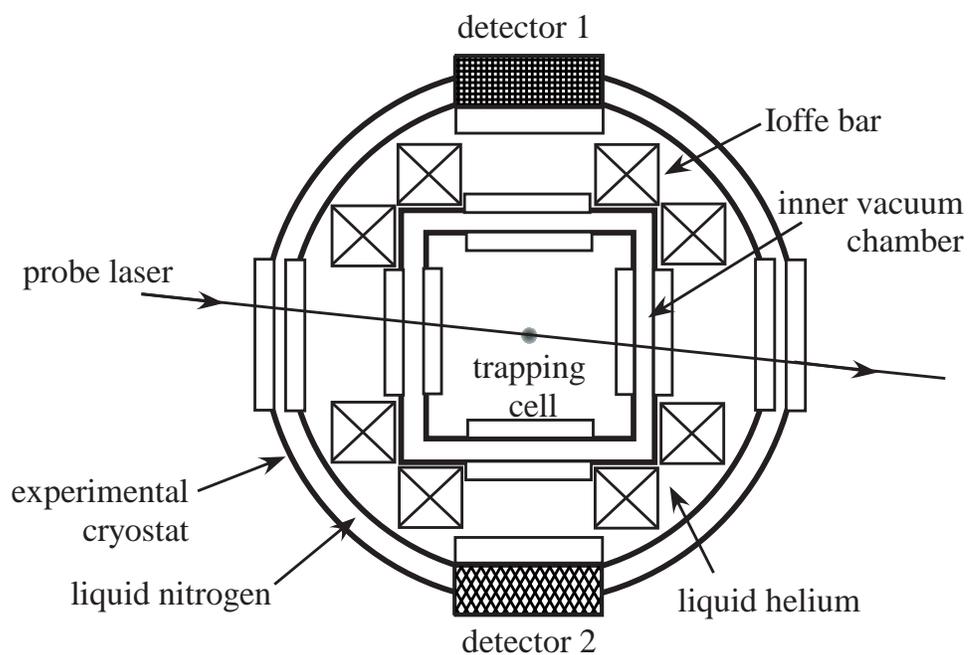,width=5in}}
\caption{Cross-sectional view of cryogenic trapping apparatus with
transverse optical access and large detection solid angle.  Metastable
hydrogen could be excited, as done currently, by a 243 nm laser along
the axis of the Ioffe-Pritchard trap (oriented into the page at the
center of the trapping cell).  One or more probe lasers could be
introduced transverse to the atom cloud.  Each of the detectors, anchored at
liquid nitrogen temperature, could be a CCD, PMT array, or MCP
detector.  The magnet system and detector mounts must be designed so
that magnetic fields are compatible with the detectors.
}
\label{fig:apparatusidea}
\end{figure}

Figure~\ref{fig:apparatusidea} is a schematic for a possible cryostat
and magnet system with a geometry favorable for optical experiments.
The cross section shows eight superconducting ``Ioffe bars'' which
give rise to a radial quadrupole field.  ``Pinch'' solenoids above and
below the plane of the page would provide the axial confinement.
Spaces between the Ioffe bars provide access for laser beams and
detectors.  As in the current apparatus, an additional window on the
axis of the trap could allow entry of a retro-reflected 243~nm beam
for generation of metastables.

The cryogenic apparatus can be greatly simplified if we sacrifice the
ability to do rf evaporation.  Our experiments have shown that large
numbers and high densities of metastables can be produced without rf
evaporation of the ground state sample (as in Trap X, for example).
In Fig.~\ref{fig:apparatusidea}, the trapping cell could be made from
a tube of non-magnetic metal with high thermal conductivity, such as
copper.  To incorporate four windows as shown, a square profile may be
most convenient.  Even more complexity can be avoided if the cryostat
is based on a $^3$He refrigerator rather than a dilution refrigerator.
The lowest practical temperature of a $^3$He refrigerator is about
300~mK, which in the present loading scheme is too warm for the atom
cloud to thermally disconnect from the cell walls
(Sec.~\ref{sec:thermaldisconnect}).  It may be possible, however, to
induce thermal disconnect by raising the magnetic fields at the walls
and forcing magnetic saddlepoint evaporation elsewhere.

It is desirable to reduce both the aspect ratio of the trap and the
length of the axial dimension of the entire apparatus.  With more
axial compression, the atom clouds would be more easily probed with
transverse laser beams, and the efficiency of magnetic saddlepoint
evaporation at low temperatures may be enhanced \cite{pmw98}.  If the
length of the whole apparatus can be significantly reduced, this will
facilitate reduction of fluctuations in the transverse overlap of the UV
beams, currently a major source of noise in our experiments.  With a
$^3$He refrigerator, it may even be possible to mount the entire
cryostat on an optical table.

Bringing the detectors close to the atoms and providing them with a
large numerical aperture will greatly enhance the detection
efficiency.  Fig.~\ref{fig:apparatusidea} depicts an arrangement where
the detectors sit in contact with a liquid nitrogen jacket and can be
easily swapped out, depending on the experiments being performed.
With a large detection solid angle, it should be possible to observe
the natural two-photon fluorescence decay of metastables without
applying a quench pulse.  With Lyman-$\alpha$ detectors, the
fluorescence occurring due to excitation transfer can be observed
during the decay of the metastable cloud; the relative importance of
ionization and excitation transfer processes could be unraveled.  With
a CCD camera and a Balmer-$\alpha$ source, the metastable cloud can be
imaged as in most other atom trapping experiments.  This diagnostic
would potentially yield a wealth of information about the $2S$ and,
indirectly, the $1S$ samples.

With a much improved detection efficiency, it will be possible to do
spectroscopy with significantly lower laser powers or significantly
lower ground state densities than at present.  If lower laser powers
are possible, then the UV retromirror could exist outside of the
cryostat, and it could be either vibrationally isolated or actively
stabilized to ensure a stable overlap of the counterpropagating beams.
Lower $1S$ densities are key for minimizing cold collision shifts.  In
particular, the combination of high detection efficiency and low
sample density is desired for pushing spectroscopy of the \oneStwoS\
transition to unprecedented resolution levels.

For spectroscopy of the metastable cloud, the benefits of close
optical access to the atoms on two sides are clear.  The necessary
beams would be easy to introduce and align.  For one-photon
transitions like $2S$-$3P$, single-pass absorption experiments would
be possible.  For the the two-photon $2S$-$nS$ transitions, the
retromirror needed to achieve a standing wave could be positioned
outside the cryostat.

The possibilities for the next generation spectroscopy apparatus are
so many, it may be difficult to choose which ones to pursue.  We can
be certain, however, that metastable H has many stories yet
to tell us about cold collisions and fundamental physics.





\bibliographystyle{unsrt}

\bibliography{cumulative}

\begin{thebibliography}{10}

\bibitem{kil99}
T.~C. Killian.
\newblock {\em 1S-2S Spectroscopy of Trapped Hydrogen: The Cold Collision
  Frequency Shift and Studies of BEC}.
\newblock PhD thesis, Massachusetts Institute of Technology, 1999.

\bibitem{kfw98}
Thomas~C. Killian, Dale~G. Fried, Lorenz Willmann, David Landhuis, Stephen~C.
  Moss, Daniel Kleppner, and Thomas~J. Greytak.
\newblock Cold-collision frequency shift of the {$1S$-$2S$} transition in
  hydrogen.
\newblock {\em Physical Review Letters}, 81:3807, 1998.

\bibitem{jdd96}
M.~J. Jamieson, A.~Dalgarno, and J.~M. Doyle.
\newblock Scattering lengths for collisions of ground state and metastable
  state hydrogen atoms.
\newblock {\em Molecular Physics}, 87:817, 1996.

\bibitem{osw99}
T.~Orlikowski, G.~Staszewska, and L.~Wolniewicz.
\newblock Long range adiabatic potentials and scattering lengths for the {EF},
  e and h states of the hydrogen molecule.
\newblock {\em Molecular Physics}, 96:1445, 1999.

\bibitem{pes01}
C.~J. Pethick and H.~T.~C. Stoof.
\newblock Collisional frequency shifts of absorption lines in an atomic
  hydrogen gas.
\newblock {\em Physical Review A}, 64:013618, 2001.

\bibitem{ucj92}
X.~Urbain, A.~Cornet, and J.~Jureta.
\newblock Associative ionization in collisions between metastable hydrogen
  atoms.
\newblock {\em Journal of Physics B: Atomic, Molecular, and Optical Phyics},
  25:L189, 1992.

\bibitem{fjf01}
R.~C.~Forrey, S.~Jonsell, P.~Froelich, A.~Saenz, and A.~Dalgarno, to be
  published.

\bibitem{lmm02}
D.~Landhuis, L.~Matos, S.~C.~Moss, K.~Vant, J.~K.~Steinberger, L.~Willmann,
  T.~J.~Greytak, and D.~Kleppner, in preparation.

\bibitem{fcd00}
R.~C. Forrey, R.~C\^{o}t\'{e}, A.~Dalgarno, S.~Jonsell, A.~Saenz, and
  P.~Froelich.
\newblock Collisions between metastable hydrogen atoms at thermal energies.
\newblock {\em Physical Review Letters}, 85:4245, 2000.

\bibitem{bfd98}
N.~Balakrishnan, R.~C. Forrey, and A.~Dalgarno.
\newblock Quenching of {H$_2$} vibrations in ultracold {$^3$He} and {$^4$He}
  collisions.
\newblock {\em Physical Review Letters}, 80:3224, 1998.

\bibitem{uhg97}
Th. Udem, A.~Huber, B.~Gross, J.~Reichert, M.~Prevedelli, M.~Weitz, and T.~W.
  H\"{a}nsch.
\newblock Phase-coherent measurement of the hydrogen {$1S$-$2S$} transition
  frequency with an optical frequency interval divider chain.
\newblock {\em Physical Review Letters}, 79:2646, 1997.

\bibitem{sjd99}
C.~Schwob, L.~Jozefowski, B.~de~Beauvoir, L.~Hilico, F.~Nez, L.~Julien,
  F.~Biraben, O.~Acef, J.-J. Zondy, and A.~Clairon.
\newblock Optical frequency measurement of the {$2S$-$12D$} transitions in
  hydrogen and deuterium: {R}ydberg constant and {L}amb shift determinations.
\newblock {\em Physical Review Letters}, 82:4960, 1999.

\bibitem{nhr00}
M.~Niering, R.~Holzwarth, J.~Reichert, P.~Pokasov, Th. Udem, M.~Weitz, T.~W.
  H\"{a}nsch, P.~Lemonde, G.~Santarelli, M.~Abgrall, P.~Laurent, C.~Salomon,
  and A.~Clairon.
\newblock Measurement of the hydrogen {$1S$-$2S$} transition frequency by phase
  coherent comparison with a microwave cesium fountain clock.
\newblock {\em Physical Review Letters}, 84:5496, 2000.

\bibitem{pac01}
K.~Pachucki.
\newblock Logarithmic two-loop corrections to the {L}amb shift in hydrogen.
\newblock {\em Physical Review A}, 63:042503--1, 2001.

\bibitem{whs95}
M.~Weitz, A.~Huber, F.~Schmidt-Kaler, D.~Leibfried, W.~Vassen, C.~Zimmermann,
  K.~Pachucki, T.~W. H\"{a}nsch, L.~Julien, and F.~Biraben.
\newblock Precision measurement of the {$1S$} ground-state {L}amb shift in
  atomic hydrogen and deuterium by frequency comparison.
\newblock {\em Physical Review A}, 52:2664, 1995.

\bibitem{bbn96}
S.~Bourzeix, B.~de~Beauvoir, F.~Nez, M.~D. Plimmer, F.~de~Tomasi, L.~Julien,
  F.~Biraben, and D.~N. Stacy.
\newblock High resolution spectroscopy of the hydrogen atom: determination of
  the {$1S$} {L}amb shift.
\newblock {\em Physical Review Letters}, 76:384, 1996.

\bibitem{bnj97}
B.~de~Beauvoir, F.~Nez, L.~Julien, B.~Cagnac, F.~Biraben, D.~Touahri,
  L.~Hilico, O.~Acef, A.~Clairon, and J.~J. Zondy.
\newblock Absolute freqeuncy measurement of the {$2S$-$8S/D$} transitions in
  hydrogen and deuterium: New determination of the {R}ydberg constant.
\newblock {\em Physical Review Letters}, 78:440, 1997.

\bibitem{urh99}
Th. Udem, J.~Reichert, R.~Holzwarth, and T.~W. H\"{a}nsch.
\newblock Accurate measurement of large optical frequency differences with a
  mode-locked laser.
\newblock {\em Optics Letters}, 24:881, 1999.

\bibitem{djy00}
S.~A. Diddams, D.~J. Jones, J.~Ye, S.~T. Cundiff, J.~L. Hall, J.~K. Ranka,
  R.~S. Windeler, R.~Holzwarth, Th. Udem, and T.~W. H\"{a}nsch.
\newblock Direct link between microwave and optical frequencies with a 300
  {THz} femtosecond laser comb.
\newblock {\em Physical Review Letters}, 84:5102, 2000.

\bibitem{udv01}
Th. Udem, S.~A. Diddams, K.~R. Vogel, C.~W. Oates, E.~A. Curtis, W.~D. Lee,
  W.~M. Itano, R.~E. Drullinger, J.~C. Bergquist, and L.~Hollberg.
\newblock Absolute frequency measurements of {Hg$^+$} and {Ca} optical clock
  transitions with a femtosecond laser.
\newblock {\em Physical Review Letters}, 86:4996, 2001.

\bibitem{wk00}
L.~Willmann and D.~Kleppner.
\newblock Ultracold hydrogen.
\newblock In S.~G. Karshenboim, F.~S. Pavone, F.~Bassani, M.~Inguscio, and
  T.~W. H\"{a}nsch, editors, {\em The Hydrogen Atom: Precision Physics of
  Simple Atomic Systems}, pages 42--56. Springer-Verlag, 2001.

\bibitem{ys96}
M.~Yasuda and F.~Shimizu.
\newblock Observation of two-atom correlation of an ultracold neon atomic beam.
\newblock {\em Physical Review Letters}, 77:3090, 1996.

\bibitem{myk96}
M.~Morinaga, M.~Yasuda, T.~Kishimoto, F.~Shimizu, J.~Fujita, and S.~Matsui.
\newblock Holographic manipulation of a cold atomic beam.
\newblock {\em Physical Review Letters}, 77:802, 1996.

\bibitem{rsb01}
A.~Robert, O.~Sirjean, A.~Browaeys, J.~Poupard, S.~Nowak, D.~Boiron, C.~I.
  Westbrook, and A.~Aspect.
\newblock A {B}ose-{E}instein condensate of metastable atoms.
\newblock {\em Science}, 292:461, 2001.

\bibitem{kdh99}
M.~Kozuma, L.~Deng, E.~W. Hagley, J.~Wen, R.~Lutwak, K.~Helmerson, S.~L.
  Rolston, and W.~D. Phillips.
\newblock Coherent splitting of {Bose}-{E}instein condensed atoms with
  optically induced {B}ragg diffraction.
\newblock {\em Physical Review Letters}, 82:871--875, 1999.

\bibitem{npm96}
S.~Nowak, T.~Pfau, and J.~Mlynek.
\newblock Nanolithography with metastable helium.
\newblock {\em Applied Physics B}, 63:203, 1996.

\bibitem{jtd98}
K.~S. Johnson, J.~H. Thywissen, N.~H. Dekker, K.~K. Berggren, A.~P. Chu,
  R.~Younkin, and M.~Prentiss.
\newblock Localization of metastable atom beams with optical standing waves:
  Nanolithography at the {H}eisenberg limit.
\newblock {\em Science}, 280:1583, 1998.

\bibitem{hkd87}
H.~F. Hess, G.~P. Kochanski, J.~M. Doyle, N.~Masuhara, D.~Kleppner, and T.~J.
  Greytak.
\newblock Magnetic trapping of spin-polarized atomic hydrogen.
\newblock {\em Physical Review Letters}, 59:672, 1987.

\bibitem{doy91}
J.~M. Doyle.
\newblock {\em Energy Distribution Measurements of Magnetically Trapped Spin
  Polarized Atomic Hydrogen: Evaporative Cooling and Surface Sticking}.
\newblock PhD thesis, Massachusetts Institute of Technology, 1991.

\bibitem{cfk96}
C.~L. Cesar, D.~G. Fried, T.~C. Killian, A.~D. Polcyn, J.~C. Sandberg, I.~A.
  Yu, T.~J. Greytak, D.~Kleppner, and J.~M. Doyle.
\newblock Two-photon spectroscopy of trapped atomic hydrogen.
\newblock {\em Physical Review Letters}, 77:255, 1996.

\bibitem{bs77p214}
H.~A. Bethe and E.~E. Salpeter.
\newblock {\em Quantum Mechanics of One- and Two-Electron Atoms}, page 214.
\newblock Plenum Publishing Corp., 1977.

\bibitem{pri83}
D.~E. Pritchard.
\newblock Cooling neutral atoms in a magnetic trap for precision spectroscopy.
\newblock {\em Physical Review Letters}, 51:1336, 1983.

\bibitem{bbc79}
F.~Biraben, M.~Bassini, and B.~Cagnac.
\newblock {Line-shapes in Doppler-free two-photon spectroscopy. The effect of
  finite transit time}.
\newblock {\em Le Journal de Physique}, 40:445, 1979.

\bibitem{fkw98}
Dale~G. Fried, Thomas~C. Killian, Lorenz Willmann, David Landhuis, Stephen~C.
  Moss, Daniel Kleppner, and Thomas~J. Greytak.
\newblock Bose-{E}instein condensation of atomic hydrogen.
\newblock {\em Physical Review Letters}, 81:3811, 1998.

\bibitem{bea86}
R.~G. Beausoleil.
\newblock {\em Continuous-Wave Measurement of the 1S-2S Transition Frequency in
  Atomic Hydrogen: The 1S Lamb Shift}.
\newblock PhD thesis, Stanford University, 1986.

\bibitem{san93}
J.~C. Sandberg.
\newblock {\em Research Toward Laser Spectroscopy of Trapped Atomic Hydrogen}.
\newblock PhD thesis, Massachusetts Institute of Technology, 1993.

\bibitem{ck99}
C.~L. Cesar and D.~Kleppner.
\newblock Two-photon {D}oppler-free spectroscopy of trapped atoms.
\newblock {\em Physical Review A}, 59:4564, 1999.

\bibitem{ces95}
C.~L. Cesar.
\newblock {\em Two-Photon Spectroscopy of Trapped Atomic Hydrogen}.
\newblock PhD thesis, Massachusetts Institute of Technology, 1995.

\bibitem{fri99}
D.~G. Fried.
\newblock {\em Bose-Einstein Condensation of Atomic Hydrogen}.
\newblock PhD thesis, Massachusetts Institute of Technology, 1999.

\bibitem{skv88}
H.~T.~C. Stoof, J.~M. V.~A. Koelman, and B.~J. Verhaar.
\newblock Spin-exchange and dipole relaxation rates in atomic hydrogen:
  Rigorous and simplified calculations.
\newblock {\em Physical Review B}, 38:4688, 1988.

\bibitem{cew99}
E.~A. Cornell, J.~R. Ensher, and C.~E. Wieman.
\newblock Experiments in dilute atomic {B}ose-{E}instein condensates.
\newblock In {M. Inguscio, et al.}, editor, {\em Bose-Einstein Condensation in
  Atomic Gases}. Proceedings of the International School of Physics ``Enrico
  Fermi'', Course CXL, IOS Press Ohmsha, 1999.

\bibitem{die01}
K.~Dieckmann.
\newblock {\em Bose-Einstein Condensation with High Atom Number in a Deep
  Magnetic Trap}.
\newblock PhD thesis, University of Amsterdam, 2001.

\bibitem{mos01}
S.~C. Moss.
\newblock {\em Formation and Decay of a Bose-Einstein Condensate in Atomic
  Hydrogen}.
\newblock PhD thesis, Massachusetts Institute of Technology, 2002.

\bibitem{Yu93}
I.~A. Yu.
\newblock {\em Ultracold Surface Collisions: Sticking Probability of Atomic
  Hydrogen on Superfluid $^4$He}.
\newblock PhD thesis, Massachusetts Institute of Technology, 1993.

\bibitem{valpeyfisher}
Valpey Fisher Corporation.

\bibitem{1266}
Emerson \& Cumming Inc., Canton, Massachusetts.

\bibitem{aerodag}
Acheson Colloids Company, Port Huron, Michigan.

\bibitem{gtb81}
J.~M. Greben, A.~W. Thomas, and A.~J. Berlinsky.
\newblock Quantum theory of hydrogen recombination.
\newblock {\em Canadian Journal of Physics}, 59:945, 1981.

\bibitem{sto01}
H.~T.~C. Stoof, private communication.

\bibitem{gre01}
T.~J.~Greytak, unpublished.

\bibitem{lrw96}
O.~J. Luiten, M.~W. Reynolds, and J.~T.~M. Walraven.
\newblock Kinetic theory of evaporative cooling of a trapped gas.
\newblock {\em Physical Review A}, 53:381, 1996.

\bibitem{bhk86}
D.~A. Bell, H.~F. Hess, G.~P. Kochanski, S.~Buchman, L.~Pollack, Y.~M. Xiao,
  D.~Kleppner, and T.~J. Greytak.
\newblock Relaxation and recombination in spin-polarized atomic hydrogen.
\newblock {\em Physical Review B}, 34:7670, 1986.

\bibitem{svk01}
A.~I. Safonov, S.~A. Vasilyev, A.~A. Kharitonov, I.~I.~Lukashevich
  S.~T.~Boldarev, and S.~Jaakola.
\newblock Adsorption and two-body recombination of atomic hydrogen on
  $^3${H}e-$^4${H}e mixture films.
\newblock {\em Physical Review Letters}, 86:15, 2001.

\bibitem{kzm89}
R.~Kallenbach, C.~Zimmermann, D.~H. McIntyre, and T.W. H\"{a}nsch.
\newblock A blue dye laser with sub-kilohertz stability.
\newblock {\em Optics Communications}, 70:56, 1989.

\bibitem{coherent}
Coherent Laser Group, Santa Clara, California.

\bibitem{radiantdyes}
Radiant Dyes GmbH, Wermelskirchen, Germany.

\bibitem{dhk83}
R.~W.~P. Drever, J.~L. Hall, F.~V. Kowalski, J.~Hough, G.~M. Ford, A.~J.
  Munley, and H.~Ward.
\newblock Laser phase and frequency stabilization using an optical resonator.
\newblock {\em Applied Physics B}, 31:97, 1983.

\bibitem{mci87}
D.~H. McIntyre.
\newblock {\em High Resolution Laser Spectroscopy of Tellurium and Hydrogen: A
  Measurement of the Rydberg Constant}.
\newblock PhD thesis, Stanford University, 1987.

\bibitem{hac80}
T.~W. H\"{a}nsch and B.~Couillaud.
\newblock {Laser frequency stabilization by polarization spectroscopy of a
  reflecting reference cavity}.
\newblock {\em Optics Communications}, 35:441, 1980.

\bibitem{thorlabspzt}
Part number AE0203D04, Thorlabs, Inc., Newton, New Jersey.

\bibitem{bs77p287}
H.~A. Bethe and E.~E. Salpeter.
\newblock {\em Quantum Mechanics of One- and Two-Electron Atoms}, page 287.
\newblock Plenum Publishing Corp., 1977.

\bibitem{jsp02}
S.~Jonsell, A.~Saenz, P.~Froelich, R.~C. Forrey, R.~C\^{o}t\'{e}, and
  A.~Dalgarno.
\newblock Long-range interactions between two $2s$ excited hydrogen atoms.
\newblock {\em Physical Review A}, 65:042501, 2002.

\bibitem{for01}
R.~C.~Forrey, private communication.

\bibitem{rbj88}
R.~van Roijen, J.~J. Berkhout, S.~Jaakkola, and J.~T.~M. Walraven.
\newblock Experiments with atomic hydrogen in a magnetic trapping field.
\newblock {\em Physical Review Letters}, 61:931, 1988.

\bibitem{wig48}
E.~P. Wigner.
\newblock On the behavior of cross sections near thresholds.
\newblock {\em Physical Review}, 73:1002, 1948.

\bibitem{wbz99}
J.~Weiner, V.~S. Bagnato, S.~Zilio, and P.~S. Julienne.
\newblock Experiments and theory in cold and ultracold collisions.
\newblock {\em Reviews of Modern Physics}, 71:1, 1999.

\bibitem{wol01}
L.~Wolniewicz, private communication.

\bibitem{jdk95}
M.~J. Jamieson, A.~Dalgarno, and M.~Kimura.
\newblock Scattering lengths and effective ranges for {He}-{He} and
  spin-polarized {H-H} and {D-D} scattering.
\newblock {\em Physical Review A}, 51:2626, 1995.

\bibitem{okk02}
M.~\"{O}. Oktel, T.~C. Killian, D.~Kleppner, and L.~S. Levitov.
\newblock Sum rule for the optical spectrum of a trapped gas.
\newblock {\em Physical Review A}, 65:033617, 2002.

\bibitem{mlm02}
S.~C.~Moss, D.~Landhuis, L.~Matos, K.~Vant, J.~K.~Steinberger, L.~Willmann,
  T.~J.~Greytak, and D.~Kleppner, in preparation.

\bibitem{bur65}
A.~Burgess.
\newblock Tables of hydrogenic photoionization cross-sections and
  recombinations coefficients.
\newblock {\em Memoirs of the Royal Astronomical Society}, 69:1, 1965.

\bibitem{sak94}
J.~J. Sakurai.
\newblock {\em Modern Quantum Mechanics}, pages 339--341.
\newblock Addison-Wesley, revised edition, 1994.

\bibitem{sw89}
D.~W. Snoke and J.~P. Wolf.
\newblock Population dynamics of a {Bose} gas near saturation.
\newblock {\em Physical Review B}, 39:4030, 1989.

\bibitem{oha89}
J.~F. O'Hanlon.
\newblock {\em A User's Guide to Vacuum Technology}, page~20.
\newblock Wiley, 2nd edition, 1989.

\bibitem{chc70}
S.~Chapman and T.~G. Cowling.
\newblock {\em The Mathematical Theory of Non-Uniform Gases}, page 195.
\newblock Cambridge U.~P., 3rd edition, 1970.

\bibitem{dal01}
A.~Dalgarno, private communication.

\bibitem{bj83}
B.~H. Bransden and C.~J. Joachain.
\newblock {\em Physics of Atoms and Molecules}, pages 212--214.
\newblock Longman, 1983.

\bibitem{wh91}
C.~E. Wieman and L.~Hollberg.
\newblock Using diode lasers for atomic physics.
\newblock {\em Review of Scientific Instruments}, 62:1, 1991.

\bibitem{msw92}
K.~B. MacAdam, A.~Steinbach, and C.~Wieman.
\newblock A narrow-band tunable diode laser system with grating feedback, and a
  saturated absorption spectrometer for {C}s and {R}b.
\newblock {\em American Journal of Physics}, 60:1098, 1992.

\bibitem{bs57p266}
H.~A. Bethe and E.~E. Salpeter.
\newblock {\em Quantum Mechanics of One- and Two-Electron Atoms}, page 266.
\newblock Springer-Verlag, 1957.

\bibitem{smg02}
T.~R.~Schibli, L.~M.~Matos, F.~J.~Grawert, J.~G.~Fujimoto, and
  F.~X.~K\"{a}rtner, in preparation.

\bibitem{emk01}
R.~Ell, U.~Morgener, F.~X. K\"{a}rtner, J.~G. Fujimoto, E.~P. Ippen,
  V.~Scheuer, G.~Angelow, T.~Tschudi, M.~J. Lederer, A.~Boiko, and
  B.~Luther-Davies.
\newblock Generation of 5-fs pulses and octave-spanning spectra directly from a
  {T}i:sapphire laser.
\newblock {\em Optics Letters}, 26:373, 2001.

\bibitem{pmw98}
P.~W.~H. Pinkse, A.~Mosk, M.~Weidem\"{u}ller, M.~W. Reynolds, and T.~W.
  Hijmans.
\newblock One-dimensional evaporative cooling of magnetically trapped atomic
  hydrogen.
\newblock {\em Physical Review A}, 57:4747, 1998.

\end{thebibliography}

\end{document}